\documentclass[11pt,DIV=16]{scrartcl}
\addtokomafont{disposition}{\rmfamily}
\pdfoutput=1
\makeatletter
\DeclareOldFontCommand{\rm}{\normalfont\rmfamily}{\mathrm}
\DeclareOldFontCommand{\sf}{\normalfont\sffamily}{\mathsf}
\DeclareOldFontCommand{\tt}{\normalfont\ttfamily}{\mathtt}
\DeclareOldFontCommand{\bf}{\normalfont\bfseries}{\mathbf}
\DeclareOldFontCommand{\it}{\normalfont\itshape}{\mathit}
\DeclareOldFontCommand{\sl}{\normalfont\slshape}{\@nomath\sl}
\DeclareOldFontCommand{\sc}{\normalfont\scshape}{\@nomath\sc}
\makeatother

\usepackage{epsf,amsmath,amssymb,graphicx,scalefnt,rotating,enumitem,fancyhdr}
\usepackage{cite}
\usepackage{filemod}
\usepackage[affil-it]{authblk}
\usepackage[usenames,dvipsnames]{color}
\usepackage{multirow}
\usepackage{appendix}
\usepackage[bottom]{footmisc}
\usepackage[utf8]{inputenc}
\usepackage{color,soul}
\usepackage{hyperref}
\usepackage[all]{hypcap}
\usepackage{cancel}
\usepackage[utf8]{inputenc}

\newcommand{\abbrev}{\scalefont{.9}}
\newcommand{\EulerGamma}{\gamma_\text{E}}
\newcommand{\ep}{\epsilon}

\newcommand{\eqn}[1]{eq.\,(\ref{#1})}

\newcommand{\mssm}{{\abbrev MSSM}}

\newcommand{\msbar}{\ensuremath{\overline{\text{\abbrev MS}}}}
\newcommand{\drbar}{\ensuremath{\overline{\text{\abbrev DR}}}}
\newcommand{\mdrbar}{\ensuremath{\overline{\text{\abbrev MDR}}}}

\newcommand{\mgl}{M_3}
\newcommand{\Mt}{M_{t}} 

\newcommand{\MW}{M_W}
\newcommand{\MZ}{M_Z}

\newcommand{\Mh}{M_h}
\newcommand{\MH}{M_H}
\newcommand{\MA}{M_A}
\newcommand{\MHp}{M_{H^\pm}}
\newcommand{\MS}{M_S}
\newcommand{\ovms}{{\cal O}(v^2/\MS^2)}
\newcommand{\thdm}{2HDM}
\newcommand{\smallSM}{{\scriptscriptstyle{\rm SM}}}

\newcommand{\smallHET}{{\scriptscriptstyle{\rm HET}}}
\newcommand{\lambdaSM}{\lambda_{\smallSM}}
\newcommand{\Qewsb}{Q_{{\scriptscriptstyle{\rm EWSB}}}}
\newcommand{\Qin}{Q_{{\scriptscriptstyle{\rm in}}}}
\newcommand{\QSM}{Q_\smallSM}

\newcommand{\beq}{\begin{equation}}
\newcommand{\eeq}{\end{equation}}
\newcommand{\bea}{\begin{eqnarray}}
\newcommand{\eea}{\end{eqnarray}}

\def\at{\alpha_t}

\def\as{\alpha_s}

\newcommand{\Rpar}{$R$-parity}

\newcommand{\FH}{\texttt{FeynHiggs}}
\newcommand{\FS}{\texttt{FlexibleSUSY}}
\newcommand{\FE}{\texttt{FlexibleEFTHiggs}}
\newcommand{\HIM}{\texttt{Himalaya}}
\newcommand{\SA}{\texttt{SARAH}}
\newcommand{\SP}{\texttt{SPheno}}
\newcommand{\Su}{\texttt{SuSpect}}
\newcommand{\SH}{\texttt{SusyHD}}
\newcommand{\ME}{\texttt{MhEFT}}
\newcommand{\HS}{\texttt{HSSUSY}}
\newcommand{\SSY}{\texttt{SOFTSUSY}}
\newcommand{\Htm}{\texttt{H3m}}
\newcommand{\NT}{\texttt{NMSSMTools}}
\newcommand{\NC}{\texttt{NMSSMCALC}}
\newcommand{\SF}{\texttt{SuSeFLAV}}
\newcommand{\HB}{\texttt{HiggsBounds}}
\newcommand{\HSi}{\texttt{HiggsSignals}}
\newcommand{\munu}{\texttt{munuSSM}}

\def\kutsnr{eleven} 

\fancypagestyle{firstpage}
{
  
  \fancyhead[R]{{\tt Draft --- \filemodprint{\jobname}
    \\\footnotesize[latex compiled on \today]}}
}

\begin{document}
\pagenumbering{roman}
\thispagestyle{empty}

\vspace*{-1.8cm}
\begin{flushright}
{\tt DESY 20-229},~~
{\tt IFT-UAM/CSIC-20-184},~~
{\tt FR-PHENO-2020-021},\\
{\tt KA-TP-23-2020},~~
{\tt MPP-2020-235},~~
{\tt P3H-20-086},~~
{\tt TTK-20-53}~~
\end{flushright}

\vspace*{2cm}

\begin{center}
{\huge \bf Higgs-mass predictions in the \mssm \\[0.2cm]
and beyond}

\vspace{1cm}

{\large
  
  P.~Slavich$^a$ and S.~Heinemeyer$^{b,c,d}$~(eds.),\\[3mm]

  E.~Bagnaschi$^e$,
  H.~Bahl$^f$,
  M.~Goodsell$^a$,
  H.E.~Haber$^g$,
  T.~Hahn$^h$,
  R.~Harlander$^i$,\\[2mm]
  W.~Hollik$^h$,
  G.~Lee$^{j,k,l}$,
  M.~M\"uhlleitner$^{m}$,
  S.~Pa\ss ehr$^i$,
  H.~Rzehak$^{n}$,
  D.~St\"ockinger$^o$,\\[2mm]
  A.~Voigt$^p$,
  C.E.M.~Wagner$^{q,r,s}$ and
  G.~Weiglein$^f$,\\[3mm]

  B.C.~Allanach$^t$,
  T.~Biek\"otter$^f$,
  S.~Borowka$^{u\,\ddag}$,
  J.~Braathen$^f$,
  M.~Carena$^{r,s,v}$,\\[1.5mm]
  T.N.~Dao$^w$,
  G.~Degrassi$^x$,
  F.~Domingo$^y$,
  P.~Drechsel$^{f\,\ddag}$,
  U.~Ellwanger$^z$,
  M.~Gabelmann$^m$,\\[1.5mm]
  R.~Gr\"ober$^{aa}$,
  J.~Klappert$^i$,
  T.~Kwasnitza$^o$,
  D.~Meuser$^f$,
  L.~Mihaila$^{bb\,\ddag}$,
  N.~Murphy$^{cc\,\ddag}$,\\[1.5mm]
  K.~Nickel$^{y\,\ddag}$,
  W.~Porod$^{dd}$,
  E.A.~Reyes~Rojas$^{ee}$,
  I.~Sobolev$^f$ and
  F.~Staub$^{m\,\ddag}$
}  
\end{center}

\vspace{2cm}

\begin{abstract}
  Predictions for the Higgs masses are a distinctive feature of
supersymmetric extensions of the Standard Model, where they play a
crucial role in constraining the parameter space. The discovery of a
Higgs boson and the remarkably precise measurement of its mass at the
LHC have spurred new efforts aimed at improving the accuracy of the
theoretical predictions for the Higgs masses in supersymmetric
models. The {\em ``Precision SUSY Higgs Mass Calculation Initiative”}
(KUTS) was launched in 2014 to provide a forum for discussions between
the different groups involved in these efforts. This report aims to
present a comprehensive overview of the current status of Higgs-mass
calculations in supersymmetric models, to document the many advances
that were achieved in recent years and were discussed during the KUTS
meetings, and to outline the prospects for future improvements in
these calculations.

\end{abstract}

\newpage
\section*{List of affiliations}

{\small
${}^a$ Sorbonne Universit\'e, CNRS, Laboratoire de Physique Th\'eorique
et Hautes \'Energies, LPTHE,\\ \mbox{}~~F-75005 Paris, France\\[1.5mm]
${}^b$ Instituto de F\'isica Te\'orica, (UAM/CSIC),
Universidad Aut\'onoma de Madrid,\\
\mbox{}~~Cantoblanco, E-28049 Madrid, Spain\\[1.5mm]
${}^c$ Campus of International Excellence UAM+CSIC, Cantoblanco,
E-28049, Madrid, Spain\\[1.5mm]
${}^d$ Instituto de F\'isica de Cantabria (CSIC-UC), E-39005 Santander,
Spain\\[1.5mm]
${}^e$ Paul Scherrer Institut, CH-5232 Villigen, Switzerland\\[1.5mm]
${}^f$ DESY, Notkestra{\ss}e 85, D-22607 Hamburg, Germany\\[1.5mm]
${}^g$ Santa Cruz Institute for Particle Physics, University of California,
Santa Cruz, CA 95064, USA\\[1.5mm]
${}^h$ Max-Planck Institut f\"ur Physik, D-80805 M\"unchen, Germany\\[1.5mm]
${}^i$ Institute for Theoretical Particle Physics and Cosmology,
RWTH Aachen University,\\ \mbox{}~~D-52074 Aachen, Germany\\[1.5mm]
${}^j$ Department of Physics, Korea University, KR-136-713 Seoul,
Korea\\[1.5mm]
${}^k$ Department of Physics, LEPP, Cornell University, Ithaca, NY-14853, USA
\\[1.5mm]
${}^l$ Department of Physics, University of Toronto, Toronto, ON, Canada
\\[1.5mm]
${}^m$ Institute for Theoretical Physics (ITP), Karlsruhe Institute of
Technology, D-76131 Karlsruhe, Germany\\[1.5mm]
${}^n$ Albert-Ludwigs-Universit\"at Freiburg, Physikalisches Institut,
D-79104 Freiburg, Germany\\[1.5mm]
${}^o$ Institut f\"ur Kern- und Teilchenphysik, TU Dresden,
D-01069 Dresden, Germany\\[1.5mm]
${}^p$ Fachbereich Energie und Biotechnologie, Hochschule
Flensburg, D-24943 Flensburg, Germany\\[1.5mm]
${}^q$ High Energy Physics Division, Argonne National Laboratory,
Argonne, IL-60439, USA\\[1.5mm]
${}^r$ Enrico Fermi Institute, University of Chicago,
Chicago, IL-60637, USA \\[1.5mm]
${}^s$ Kavli Institute for Cosmological Physics, University of Chicago,
Chicago, IL-60637, USA\\[1.5mm]
${}^t$ DAMTP, University of Cambridge, CB30WA Cambridge, UK\\[1.5mm]
${}^u$ Theoretical Physics Department, CERN, CH-1211 Geneva 23,
Switzerland\\[1.5mm]
${}^v$ Fermi National Accelerator Laboratory, Batavia, IL 60510, USA\\[1.5mm]
${}^w$ Institute for Interdisciplinary Research in Science and Education, ICISE,
       590000 Quy Nhon, Vietnam.\\[1.5mm]
${}^x$ Dipartimento di Matematica e Fisica, Universit\`a degli Studi
       Roma Tre, I-00146 Roma, Italy\\[1.5mm]
${}^y$ Bethe Center for Theoretical Physics \& Physikalisches Institut
der Universit\"at Bonn, D-53115 Bonn, Germany\\[1.5mm]
${}^z$ University Paris-Saclay, CNRS/IN2P3, IJCLab, F-91405 Orsay,
France\\[1.5mm]
${}^{aa}$ Dipartimento di Fisica e Astronomia ``G. Galilei'',
Universit\`a di Padova \& INFN, Sezione di Padova,\\
\mbox{}~~~~I-35131 Padova, Italy\\[1.5mm]
${}^{bb}$ Institute for Theoretical Physics,
University of Heidelberg, D-69120
Heidelberg, Germany\\[1.5mm]
${}^{cc}$ CP3-Origins, University of Southern Denmark,
DK-5230 Odense M, Denmark\\[1.5mm]
${}^{dd}$ Institute for Theoretical Physics and Astrophysics, 
Julius-Maximilians-Universit{\"a}t W{\"u}rzburg,\\
\mbox{}~~~~D-97074 W{\"u}rzburg, Germany\\[1.5mm]
${}^{ee}$ Universidad de Pamplona (UDP), Pamplona -- Norte de Santander,
Colombia\\[10mm]
${}^{\ddag}$ (Former academic affiliation)

}

\newpage
\tableofcontents


\newpage
\pagenumbering{arabic}
\setcounter{page}{1}

\section{Introduction}
\label{sec:intro}

The spectacular discovery of a scalar particle with mass around
$125$~GeV by the ATLAS and CMS collaborations~\cite{Aad:2012tfa,
Chatrchyan:2012xdj, Aad:2015zhl} at CERN constitutes a milestone in
the quest for understanding the physics of electroweak symmetry
breaking (EWSB). While the properties of the observed particle are
compatible with those predicted for the Higgs boson of the Standard
Model (SM) within the present experimental and theoretical
uncertainties~\cite{Khachatryan:2016vau}, they are also in agreement
with the predictions of many models of physics beyond the SM (BSM).
{For the latter}, the requirement that the model under consideration
include a state that can be identified with the observed particle can
translate into important constraints on the model's parameter space.

One of the prime candidates for BSM physics is supersymmetry (SUSY),
which predicts scalar partners for all SM fermions, as well as
fermionic partners for all bosons. A remarkable feature of SUSY
extensions of the SM is the requirement of an extended Higgs sector,
with additional neutral and charged bosons. In such models the
couplings of the Higgs bosons to matter fermions and to gauge bosons
can differ significantly from those of the SM Higgs. Moreover, in
contrast to the case of the SM, the masses of the Higgs bosons are not
free parameters, as SUSY requires all quartic scalar couplings to be
related to the gauge and Yukawa couplings.
For example, in the minimal SUSY extension of the SM, the MSSM, the
tree-level mass of the lighter neutral CP-even scalar in the Higgs
sector is bounded from above by the mass of the $Z$ boson. However,
radiative corrections involving loops of SM particles and their SUSY
partners alter the tree-level predictions for the Higgs masses,
introducing a dependence on all of the SUSY-particle masses and
couplings. The most relevant corrections to the Higgs masses in the
MSSM are those controlled by the top Yukawa coupling, which involve
the top quark and its scalar partners, the stops. These corrections
are enhanced by logarithms of the ratio between stop and top masses,
and also show a significant dependence on the value of the left--right
stop mixing parameter $X_t$. In particular, a SM-like Higgs boson with
mass around $125$~GeV can be obtained in the MSSM with an average stop
mass $\MS$ of about $2$~TeV when $|X_t/\MS| \approx 2$, whereas for
vanishing~$X_t$ the stops need to be heavier than $10$~TeV.

Non-minimal SUSY extensions of the SM have also been considered: in
the {next-to-minimal extension, the} NMSSM, the Higgs sector of the
MSSM is augmented with a gauge-singlet complex scalar and its
fermionic partner; the {so-called ``$\mu$-from-$\nu$'' extension, or}
$\mu\nu$SSM{,} includes right-handed neutrinos and their scalar
partners; models with Dirac gauginos include additional scalars and
fermions in the adjoint representation of their gauge groups.  {Most}
of these models feature additional tree-level contributions to the
quartic Higgs couplings, as well as additional contributions to the
radiative corrections to the Higgs masses. As a consequence, the
tree-level bound on the mass of the lightest neutral scalar is {in
general} relaxed, and a SM-like Higgs boson of suitable mass can be
obtained with smaller stop masses than in the case of the MSSM.

Since the first realization of the importance of the radiative
corrections in the early 1990s, an impressive theoretical effort has
been devoted to the precise calculation of the Higgs masses in SUSY
{extensions} of the SM. This effort focused initially on the simplest
realization of the MSSM, assuming the conservation of CP symmetry,
$R$-parity and flavor, but it eventually grew to include the most
general MSSM Lagrangian, as well as non-minimal SUSY extensions of the
SM such as those mentioned above. The discovery of a Higgs boson in
2012 at CERN has given new impetus to the quest for high-precision
predictions for the Higgs masses in SUSY models. First of all, the
need for multi-TeV stop masses ({namely, $\MS \gtrsim 2$~TeV,} at
least in the MSSM) to ensure a Higgs mass of about $125$~GeV implies
that the logarithmically-enhanced corrections controlled by the top
Yukawa coupling are particularly large. To obtain a reliable
prediction for the mass of the SM-like Higgs boson, these corrections
should be resummed to all orders in the perturbative expansion by
means of an effective field theory (EFT) approach. More generally, for
the available information on the Higgs mass to be most effectively
exploited when constraining the parameter space of SUSY models, the
uncertainty of the theory prediction should ideally be below the
experimental precision of the measurement, which has already reached
the per-mille level.

The uncertainty of the theory prediction for the Higgs masses in a
given SUSY model {should be estimated 
{for} each considered point of  {the
SUSY} parameter space. The uncertainty} has a ``parametric''
component, arising from the experimental uncertainty of the SM input
parameters, and a proper ``theory'' component, arising from unknown
corrections that are of higher order with respect to the accuracy of
the calculation. While the former can be straightforwardly estimated,
and is currently dominated by the uncertainty of the top mass, the
latter, which we are henceforth denoting as ``theory uncertainty'',
requires a fair amount of guesswork on the expected size of the
uncomputed corrections.  Rule-of-thumb estimates of about $3$~GeV for
the theory uncertainty of the prediction for the SM-like Higgs mass in
the MSSM were provided in the early 2000s, based on the accuracy of
the then-available calculations and on the expectation that the SUSY
scale would be at most of the order of $1$~TeV. While the progress in
the Higgs-mass calculations in SUSY models should have naturally
entailed an overall reduction of the theory uncertainty, further
studies made it clear that the latter can still vary substantially,
depending on the specific region of the SUSY parameter space that is
being considered. Even in the most favorable scenarios, the theory
uncertainty of the prediction for a SM-like Higgs mass in the MSSM
remains at the percent level, i.e.~one order of magnitude larger than
the experimental precision of the measurement. For SUSY models beyond
the MSSM  {only a few} studies of the
theory uncertainty of the Higgs-mass predictions 
{have been performed} so far. The presence of additional
particles and interactions contributing to the radiative corrections
generally increases the theory uncertainty in these models compared to
the MSSM. For specific regions in the parameter space, however, the
radiative corrections required to obtain a Higgs mass of 125 GeV, and
thus the associated uncertainty, can be smaller than is typically the
case in the MSSM.

\bigskip

In order to address this situation and bring the theory uncertainty of
the Higgs-mass predictions in SUSY models closer to the experimental
precision, the {\em ``Precision SUSY Higgs Mass Calculation
Initiative''} (KUTS)\,\footnote{The acronym KUTS originates as an
inside joke, which the authors will explain on request.}  was launched
in 2014. The initiative aims to provide a forum for discussions
between the different groups involved in the calculation of the Higgs
masses in SUSY models. Since its inception, \kutsnr\ KUTS meetings
have taken place,\footnote{The programs of the KUTS meetings 
are available at the web page~ {\tt https://sites.google.com/site/kutsmh/}}
discussing the advances achieved over the
years. These included: new fixed-order (FO) calculations of the Higgs
masses in the MSSM and other SUSY models; new EFT calculations for the
all-order resummation of large logarithmic effects; the improved
combination of FO and EFT calculations; efforts to provide a reliable
estimate of the theory accuracy as a function of the SUSY parameters;
new or improved computer codes for a state-of-the-art numerical
evaluation of SUSY Higgs boson masses.

The purpose of this report is two-fold. On the one hand, we aim to
provide a comprehensive overview of the status of Higgs-mass
calculations in SUSY models. On the other hand, we document the
specific advances that were achieved in recent years and were
discussed during the KUTS meetings, and we outline the prospects for
future improvements in these calculations. The report is written as a
non-technical review, in which the interested reader is guided to the
literature where detailed accounts of the different topics can be
found. In section~\ref{sec:THoverview} we provide a general
introduction\footnote{Due to the general nature of many of the
concepts discussed in section~\ref{sec:THoverview}, we {shall} omit
all citations there; {see} ref.~\cite{Draper:2016pys} for a
pedagogical review of Higgs-mass calculations in SUSY models {and an
accompanying list of references. However, we aim to provide a
comprehensive bibliography in the sections that follow.}}  to
high-precision predictions for the Higgs masses in SUSY models; in
sections~\ref{sec:fo} to \ref{sec:hybrid} we discuss in detail the
recent advances in the FO, EFT and ``hybrid'' calculations,
respectively; section~\ref{sec:uncertainties} concerns the estimation
of the theory uncertainty; section~\ref{sec:outlook} provides an
outlook.
Finally, we include in the appendix a survey of the existing public
codes for Higgs-mass calculations in SUSY models.

\newpage
\section{Calculating the Higgs masses in SUSY extensions of the SM}
\label{sec:THoverview}
\subsection{Radiative corrections to the Higgs masses}
\label{sec:general:radcor}

The squared physical masses of a set of $n$ scalar fields that mix
with each other are the real parts of the solutions for $p^2$ of the
equation
\beq 
{\rm det} \left[ \Gamma_{ij}(p^2)
\right] = 0~,
\label{eq:det}
\eeq
where $\Gamma_{ij}(p^2)$ denotes the $n\!\times\!n$ inverse-propagator
matrix, $p$ being the external momentum flowing into the scalar
self-energies. We can decompose $\Gamma_{ij}(p^2)$ as
\beq
\label{eq:gamma}
-i\,\Gamma_{ij}(p^2) ~=~ p^2\,\delta_{ij} ~-~ {\cal M}_{ij,\,0}^2 ~-~
\Delta {\cal M}_{ij}^2(p^2)~,
\eeq
where ${\cal M}_{ij,\,0}^2$ denotes the tree-level mass matrix written
in terms of renormalized parameters, and $\Delta {\cal M}_{ij}^2(p^2)$
collectively denotes the radiative corrections to the mass matrix. 

The entries of ${\cal M}_{ij,\,0}^2$ are in general combinations of
mass parameters ($m^2_{ij}$) for the scalar fields and of products of
trilinear ($a_{ijk}$) and quartic ($\lambda_{ijkl}$) scalar couplings
with appropriate powers of the vacuum expectation values (vevs) of the
scalar fields, $v_i$. The minimum conditions of the scalar potential
relate the mass parameters in the tree-level mass matrix to
combinations of other mass parameters, couplings and vevs. In SUSY
models, the quartic scalar couplings are not free parameters, but are
related to combinations of the gauge couplings (via the $D$-term
contributions to the scalar potential) and of the Yukawa couplings
(via the $F$-term contributions). This leads to non-trivial relations
among the scalar masses and the other parameters of the model. For
example, in the MSSM -- whose Higgs sector consists of two $SU(2)$
doublets $H_1$ and $H_2$ with opposite hypercharge -- one finds the
well-known tree-level formula for the masses of the lighter and
heavier CP-even Higgs bosons, denoted as $h$ and $H$, respectively
\begin{equation}
  M_{h/H}^2=\frac{1}{2}\left(\MA^2+\MZ^2
  \mp\sqrt{(\MA^2-\MZ^2)^2+4\MA^2\MZ^2\sin^22\beta}\right)\,,
    \label{eq:treeMSSM}
\end{equation}
where $\MA$ is the mass of the CP-odd Higgs boson, $\MZ$ is the mass
of the $Z$ boson, and $\tan\beta\equiv v_2/v_1$ is the ratio of the
vevs of the two doublets. This leads to the tree-level bound
$\Mh<\MZ|\cos2\beta|$, which is saturated for $\MA \gg \MZ$.

At each order in the perturbative expansion, the radiative corrections
$\Delta {\cal M}_{ij}^2(p^2)$ in \eqn{eq:gamma} include:
momentum-dependent contributions of the scalar self-energies,
$\Sigma_{ij}(p^2)$; ``tadpole'' contributions, $T_i$, arising if the
minimum conditions of the potential have been used to simplify the
tree-level matrix; contributions of the renormalization constants of
the scalar fields, $\delta Z_{ij}$; finally, counterterm contributions
arising from the renormalization of the parameters that enter the
lower-order parts of $\Gamma_{ij}(p^2)$.
In a pure FO approach, the radiative corrections to the scalar masses
are obtained by evaluating $\Delta {\cal M}_{ij}^2(p^2)$ as a power
series in the various coupling constants up to a certain order in the
perturbative expansion. For example, the numerically dominant one-loop
corrections to the Higgs mass matrix in the MSSM, i.e.~those involving
top and stop loops and controlled only by the top Yukawa coupling
$y_t$, are proportional to $y_t^4\,v^2/(16\pi^2)$, where $v^2\equiv
v_1^2 + v_2^2 ~{\approx (174~{\rm GeV})^2}$. The dominant two-loop
corrections are in turn proportional to $y_t^6\,v^2/(16\pi^2)^2$ and
to $y_t^4\,g_s^2\,v^2/(16\pi^2)^2$, where $g_s$ is the strong gauge
coupling.\footnote{With a slight abuse of notation, it has been common
  in the MSSM literature to denote the dominant one-loop corrections
  as ${\cal O}(\alpha_t)$ and the dominant two-loop corrections as
  ${\cal O}(\alpha_t^2)$ and ${\cal O}(\alpha_t\alpha_s)$, where
  $\alpha_t\equiv y_t^2/(4\pi)$ and $\alpha_s \equiv
  g_s^2/(4\pi)$. {We will follow t}his notation{, although it} leads
  to ambiguities when more couplings are involved in the corrections.}
We also note that the calculation of each physical scalar mass $M_i^2$
requires that \eqn{eq:det} be solved at the complex pole $p^2=M_i^2 -
i\Gamma_i M_i$, which turns it into an implicit equation.  This can be
solved either order by order or via a numerical iterative
solution. The latter approach leads to a mixing of orders in the
perturbative expansion, which can have undesirable consequences such
as, e.g., a violation of gauge invariance by terms that are of higher
order with respect to the accuracy of the calculation.

The complexity of a calculation of radiative corrections increases
with the number of loops and the number of scales (masses and external
momenta) on which the corresponding loop integrals depend. At the
one-loop level, a general solution in terms of analytic functions is
always possible, and the most complicated functions entering one- and
two-point diagrams such as tadpoles and self-energies are simple
logarithms. Hence, fully analytic calculations of the one-loop
corrections to the Higgs masses are by now available for most of the
SUSY models considered in the literature.  {Beyond the one-loop level,
  fully analytic results are currently available only for special
  cases, and} in general a numerical evaluation {of the loop
  integrals} is required.  {On the other hand,} much simpler results
for the two-loop and higher-order integrals can be obtained
analytically by adopting certain approximations; most notably, when
setting the external momentum in the self-energy to zero, the two-loop
integrals can be expressed in terms of at most dilogarithms. In
contrast, some three-loop integrals with vanishing external momentum
but arbitrary masses still need to be evaluated numerically. In the
presence of hierarchies between some of the masses, however,
analytical results for the three-loop integrals can be obtained via
asymptotic expansions. In order to obtain the most accurate
predictions available for the Higgs masses it is standard practice to
combine the full results for the one-loop corrections with approximate
results for the higher-loop corrections.

In the limit of vanishing external momentum, tadpole and self-energy
diagrams can be connected to the derivatives of the effective
potential $V_{\rm eff} = V_0 + \Delta V$, where $V_0$ is the
tree-level potential and $\Delta V$ is the sum of
one-particle-irreducible (1PI) vacuum diagrams, expressed in terms of
field-dependent particle masses and couplings. In particular, one has
\beq
\label{eq:effpot}
T_i ~=~-\left.\frac{\partial \Delta V}{\partial \phi_i}\right|_{\rm min}~,~~~~~~~
\Sigma_{ij}(0) ~=~
-\left.\frac{\partial^2 \Delta V}{\partial \phi_i\partial \phi_j}
\right|_{\rm min}~,
\eeq
where the derivatives are evaluated at the minimum of $V_{\rm eff}$.
As it requires only the calculation of vacuum diagrams, followed by a
straightforward application of the chain rule to obtain the
derivatives of $\Delta V$ with respect to the Higgs fields $\phi_i$,
the effective-potential approach typically allows for a simpler
calculation of tadpoles and zero-momentum self-energies when compared
to the direct calculation of Feynman diagrams with one or two external
legs.  The two approaches must of course give the same final result,
as long as the same approximations and the same renormalization
conditions are employed in each calculation.

Strictly speaking, the approximation of vanishing external momentum in
the two-loop corrections to a particle's mass is justified only if the
tree-level mass of that particle can itself be considered
vanishing. For the mass of the lighter CP-even Higgs boson of the
MSSM, in view of the tree-level upper bound $\Mh <\MZ|\cos2\beta|$,
that approximation can be consistently adopted in the so-called
``gaugeless limit'', in which the electroweak (EW) gauge couplings $g$
and $g^\prime$ are set to zero in the two-loop corrections (indeed,
this also implies $\MZ\rightarrow 0$). In the gaugeless limit the
two-loop corrections to the MSSM Higgs masses depend only on the
Yukawa couplings and on $g_s\,$; the numerically dominant corrections
are indeed those of ${\cal O}(\alpha_t\alpha_s)$ and ${\cal
  O}(\alpha_t^2)$, while the corrections involving the bottom and tau
Yukawa couplings become relevant only at large values of
$\tan\beta$. All complications of the non-Abelian $SU(2)\times
U(1)$~gauge group, such as, e.g., those related to gauge fixing and
ghost fields, are absent in this limit, and the number of diagrams
contributing to the Higgs-mass corrections is smaller than in the
general case.

For the heavier CP-even scalar, as well as for the CP-odd and charged
scalars, the approximation of vanishing tree-level mass is in general
not justified. However, in most of the phenomenologically relevant
MSSM scenarios the tree-level masses of the heavier Higgs bosons are
large enough that the impact of the radiative corrections is much
reduced with respect to the case of the lighter, SM-like Higgs
boson. In these scenarios, a precise calculation of the masses of the
heavier Higgs bosons is relevant only if their differences are
studied.  For example, in the MSSM with complex parameters a
CP-violating mixing between the two heavy neutral states can lead to
resonance-type effects that sensitively depend on the difference
between their masses.  In models beyond the MSSM, the approximation of
vanishing external momentum in the two-loop corrections to the mass of
the lighter Higgs boson can be consistently adopted only if all of the
couplings contributing to its tree-level mass are in turn
neglected. For example, as will be discussed in
section~\ref{sec:nmssmfixed}, in the NMSSM this approximation requires
that the doublet-singlet superpotential coupling $\lambda$ be set to
zero along with the EW gauge couplings, in which case the two-loop
corrections to the masses of the Higgs bosons residing in the two
$SU(2)$ doublets correspond to those computed in the MSSM.

\begin{figure}[t]
  \includegraphics[width=0.52\textwidth]
                  {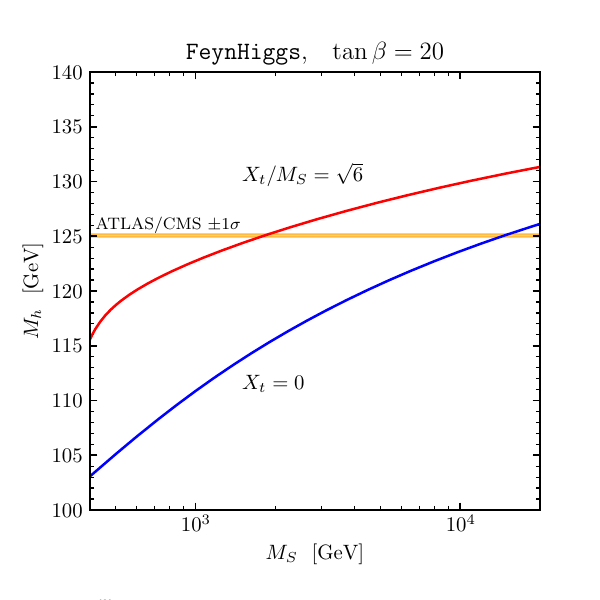}\hfill
                  \includegraphics[width=0.52\textwidth]
                                  {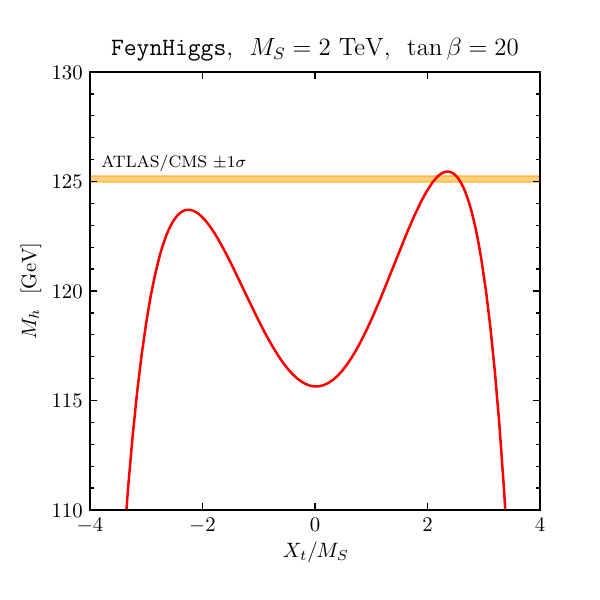}
\vspace*{-10mm}
\caption{{\em The lighter CP-even Higgs mass in the MSSM as a function
    of a common SUSY mass {parameter} $\MS$ and of the stop mixing
    parameter $X_t$ (normalized to $\MS$). Both parameters are defined
    in the $\drbar$ scheme at the scale $Q=\MS$.}}
\label{fig:first}
\end{figure}

To illustrate the relevance of the radiative corrections to the Higgs
masses in SUSY models, we show in figure~\ref{fig:first} the
predictions for the mass of the lighter CP-even Higgs boson in a
simplified MSSM scenario characterized by a degenerate mass
{parameter} $\MS$ for all SUSY particles as well as for the {heavier
  Higgs doublet}. We choose $\tan\beta = 20$ so that the tree-level
prediction for the lighter Higgs mass in eq.~(\ref{eq:treeMSSM})
essentially saturates its upper bound, i.e.~$M_h^{\rm tree}\approx
\MZ$. In the left plot of figure~\ref{fig:first} we show the
prediction for $M_h$ as a function of $\MS$ for two values of the
ratio $X_t/\MS$, where $X_t$ is the left-right stop mixing
parameter. In the right plot we set $\MS=2$~TeV and show instead the
prediction for $M_h$ as a function of $X_t/\MS$\,. The yellow band in
each plot corresponds to $M_h=125.10 \pm 0.14$~GeV, as results from a
recent combination of Run-1 and partial Run-2 data from ATLAS and
CMS~\cite{Zyla:2020zbs}. It appears from the plots in
figure~\ref{fig:first} that, in this simplified MSSM scenario, a
prediction for the Higgs mass compatible with the measured value can
be obtained with stop masses of about $2$~TeV when $|X_t/\MS| \approx
2$, whereas the stop masses need to exceed $10$~TeV when $X_t=0$.

The predictions for $M_h$ presented in figure~\ref{fig:first} were
obtained with the latest version ({\tt 2.17.0}) of the code \FH, and
they account for most of the advances that will be reviewed in
sections \ref{sec:fo}--\ref{sec:hybrid}. However, the bulk of the
dependence of $M_h$ on $\MS$ and $X_t$ can be traced to the
contributions of one-loop diagrams involving top and stops. In the
limit $\MS\gg M_t$, these so-called ${\cal O}(\alpha_t)$ contributions
can be approximated as
  \beq
  \left(\Delta M_h^2\right)_{{\cal O}(\alpha_t)} \approx~
  \frac{3\,M_t^4}{4\,\pi^2\,v^2}\left(\ln\frac{\MS^2}{M_t^2}
  \,+\, \frac{X_t^2}{\MS^2} \,-\,\frac{X_t^4}{12\,\MS^4}\right)~.
  \label{eq:1loop}
  \eeq
The logarithmic increase of $M_h$ as a function of $\MS$ visible in
the left plot of figure~\ref{fig:first} follows from the first term
within parentheses in eq.~(\ref{eq:1loop}), whereas the double-peaked
dependence of $M_h$ on $X_t$ visible in the right plot follows from
the second and third terms. We note that the one-loop correction in
eq.~(\ref{eq:1loop}) is symmetric with respect to the sign of $X_t$,
and it is maximized when $X_t/\MS=\pm\sqrt{6}$. The asymmetry between
positive and negative $X_t$ visible in the right plot of
figure~\ref{fig:first} is a two-loop effect, arising from terms linear
in $X_t$ in the one-loop correction to the top mass. Finally, we
stress that the quartic dependence of the dominant one-loop correction
on $M_t$ means that the prediction for the Higgs mass is particularly
sensitive to the measured value of the top mass, as well as to the
choices made for the renormalization of $M_t$ in the computation of
the Higgs-mass corrections beyond one loop.

\begin{figure}[t]
  \begin{center}
  \includegraphics[width=0.77\textwidth]
                  {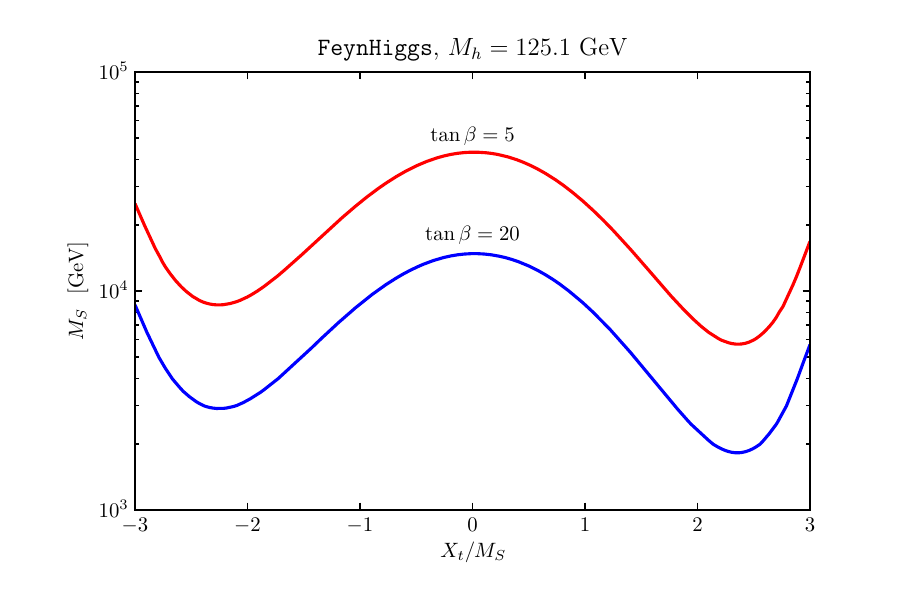}
\vspace*{-5mm}
\caption{{\em Values of the SUSY mass {parameter} $\MS$ and of the
    stop mixing parameter $X_t$ (normalized to $\MS$) that lead to the
    prediction $M_h=125.1$~GeV, in a simplified MSSM scenario with
    degenerate SUSY masses, for $\tan\beta=20$ (blue) or $\tan\beta=5$
    (red).}}
\vspace*{-5mm}
\label{fig:second}
\end{center}
\end{figure}

To further illustrate how the predictions for the SM-like Higgs mass
can constrain the parameter space of the MSSM, we plot in
figure~\ref{fig:second} the lines in the ($X_t/\MS\,,\,\MS$) plane
that, in our simplified scenario with degenerate SUSY masses, lead to
the prediction $M_h=125.1$~GeV. Note that neither theory nor
experimental uncertainties are taken into account in this example. The
lower (blue) line corresponds to $\tan\beta=20$, while the upper (red)
line corresponds to $\tan\beta=5$. The overall shape of the blue and
red lines in figure~\ref{fig:second} follows from the dependence on
$X_t/\MS$ of the Higgs-mass correction in eq.~(\ref{eq:1loop}). In
particular, the value of $\MS$ required to obtain an acceptable
prediction for $M_h$ is minimal for $|X_t/\MS| \approx 2$, and it has
a local maximum for $|X_t/\MS|\approx 0$ (for very large $|X_t/\MS|$,
on the other hand, the EW vacuum is unstable).  Since for
$\tan\beta=20$ the tree-level prediction for the lighter Higgs mass in
eq.~(\ref{eq:treeMSSM}) is essentially maximized, the blue line
implies a lower bound on $\MS$ of about $2$~TeV in this simplified
scenario.  On the other hand, the comparison between the blue and red
lines shows that, for lower values of $\tan\beta$, larger stop masses
are required to obtain $M_h\approx125$~GeV, reflecting the
$\tan\beta$-dependence of the tree-level prediction. It is also worth
pointing out that, in more-complicated MSSM scenarios where, e.g., the
gauginos are allowed to be lighter than the stops, an acceptable
prediction for $M_h$ can be obtained with somewhat smaller values for
the stop masses than those found here. In summary, the requirement
that the theory prediction for the SM-like Higgs mass agree with the
measured value establishes non-trivial correlations between the SUSY
parameters. However, even in the idealized situation of
figure~\ref{fig:second} where both experimental and theory
uncertainties are neglected, direct measurements of some of the SUSY
parameters will be necessary to obtain firm constraints on the
remaining ones.

\subsection{Input parameters and renormalization schemes}

When the theoretical prediction for an observable (e.g., the physical
mass of a particle) is computed beyond the leading order in
perturbation theory, it becomes necessary to specify a renormalization
scheme for the parameters entering the lower-order terms in the
calculation.  While the divergent parts of the counterterms are fixed
by the requirement that all divergences cancel out of the predictions
for physical observables up to the considered order in perturbation
theory, the finite parts of the counterterms define the
renormalization scheme. Which choice of renormalization condition is
the most ``sensible'' for a given parameter may depend on a
combination of factors, including practical convenience (e.g., some
choices can significantly simplify the calculations), explicit
gauge-independence, an improved convergence of the perturbative
expansion, and whether or not that parameter can be connected to an
already-measured physical quantity.

Technically the simplest renormalization scheme is ``minimal
subtraction'' (MS), which only absorbs poles in $\ep=(4-D)/2$ into the
counterterms, where $D$ is the number of space-time dimensions assumed
for the dimensionally regularized theory. In fact, the MS scheme is a
one-parameter class of schemes, distinguished by the renormalization
scale $Q$ on which the renormalized parameters depend.  Incomparably
more popular than the MS scheme itself is a variant, the
\msbar\ scheme, related to MS by the simple rescaling $Q^2\to
Q^2\,e^{\EulerGamma}/(4\pi)$, where $\EulerGamma=0.57721\ldots$ is the
Euler-Mascheroni constant.
In SUSY theories, dimensional regularization~(DREG) explicitly breaks
the balance between bosonic and fermionic degrees of freedom within a
superfield. Therefore, one usually works in a variant called
dimensional reduction~(DRED), where this balance is re-established by
supplementing each $D$-dimensional vector field~$A^{\mu}(x)$ with a
$2\ep$-dimensional ``$\ep$-scalar'' $\tilde A(x)$, where $x$ is the
$D$-dimensional space-time variable. Applying the modified minimal
subtraction to this theory results in the \drbar\ renormalization
scheme (we remark that, beyond one loop, the SUSY-preserving
properties of this scheme have been explicitly proven only for limited
classes of corrections).
In the prediction for a physical quantity the scale dependence of the
renormalized parameters cancels against an explicit logarithmic $Q$
dependence in the radiative corrections, up to the considered order in
the perturbative expansion. The residual scale dependence of the
prediction is formally a higher-order effect, and it is {therefore}
often {exploited as a (partial)} estimate of the theory uncertainty of
the calculation.

{Ideally, for a theory depending on a given number of free parameters,
  an equal number of physical observables would be chosen as input to
  determine those free parameters. In that case, predictions for
  further observables in terms of the input observables could be made
  within the theory. Such relations between physical observables are
  expected to be gauge-independent and free of ambiguities from, for
  instance, the recipe adopted for the treatment of tadpole
  contributions. This does not necessarily hold, however, for the
  relations between physical observables and parameters renormalized
  in minimal-subtraction schemes.}
It is also worth {noting} that, in {these} schemes, the contributions
of arbitrarily heavy particles do not necessarily decouple from the
predictions for low-energy observables.
{Particular care is therefore needed to avoid the occurrence of
  unphysical effects in calculations where some of the input
  parameters are defined directly in a minimal-subtraction scheme such
  as $\msbar$ or $\drbar$ rather than being connected to a physical
  observable.}
  
In order to directly relate the parameters entering the Higgs-mass
calculation to physical quantities, {non-minimal} renormalization
definitions can be chosen, which lead to non-vanishing finite parts of
the counterterms. Among the physical quantities that can be used to
define the parameters of a given SUSY model, an obvious distinction
can be made between those that have already been measured and those
that are still unknown. The former include the gauge-boson masses, the
third-generation\footnote{A common approximation in Higgs-mass
  calculations in SUSY models consists in neglecting the corrections
  that involve the Yukawa couplings of the first two generations, as
  they are usually negligible with respect to those that involve the
  third-generation couplings.} fermion masses, the Fermi constant
$G_F$ and the strong gauge coupling $\alpha_s(\MZ)$ (the latter
defined as a SM parameter in the $\msbar$ scheme). For example, the
$Z$-boson mass entering the tree-level mass matrix for the MSSM Higgs
masses can be naturally identified with the measured pole mass, in
which case the corresponding counterterm involves the $Z$-boson
self-energy. This kind of renormalization condition is usually denoted
as ``on-shell'' (OS). Even if the alternative choice is made to
renormalize the $Z$-boson mass in a minimal-subtraction scheme, the
corresponding $\msbar$ or $\drbar$ parameter still needs to be
computed starting from the {measured} pole mass at the required order
in the perturbative expansion.

In contrast, for parameters such as the masses and couplings of the
SUSY particles, the choice of renormalization conditions is a matter
of convenience, depending also on the kind of analysis that is being
performed on the model's parameter space. When the parameters of the
soft SUSY-breaking Lagrangian are obtained via renormalization-group
(RG) evolution from a set of high-energy boundary conditions, as in,
e.g., the gravity-mediation or gauge-mediation scenarios of SUSY
breaking, they are naturally expressed in the $\drbar$ scheme. It
might therefore be practical to perform the computation of the
radiative corrections to the Higgs masses directly in that scheme,
{taking care to avoid the unphysical effects discussed above}. If, on
the other hand, the SUSY parameters are taken as input directly at the
TeV scale, we may choose to express them in terms of
yet-to-be-measured physical observables.  While an OS definition for
the masses of the SUSY particles {can be formulated in an} unambiguous
{way} and connects the mass counterterms to the particles'
self-energies, for the parameters that determine their couplings
multiple options are available.
For example, particles that carry the same quantum numbers mix among
each other, and their couplings involve the rotation matrix that
diagonalizes their mass matrix. For $2\!\times\!2$ mixing, as in the
case of the ``left'' and ``right'' sfermions $(\tilde f_L\,, \tilde
f_R)$ that mix under EWSB, the rotation matrix can be parametrized by
a single mixing angle (plus a phase, in the case of complex
parameters). It is then possible to express the couplings of the
corresponding mass eigenstates in terms of this angle, and impose a
renormalization condition on the latter. A minimal-subtraction
condition would be straightforward, but it is known to be gauge
dependent. A proper OS definition, connecting the angle to a physical
process (such as a decay) that depends on it at tree level, brings
along a number of other disadvantages: the choice of a specific
process may destroy the symmetries between the particles that mix,
infra-red (IR) divergences may need to be dealt with, and seemingly
reasonable values for the input parameters may in fact correspond to
large couplings that undermine the convergence of the perturbative
expansion. An alternative non-minimal -- but process-independent --
definition requires that, in the renormalization of any interaction
vertex that involves external particles that mix, the counterterm of
the mixing angle cancel out the antisymmetric part in $ij$ of the
field-renormalization constants $\delta Z_{ij}$. This definition,
which is also commonly denoted as ``OS'', is often used for the
renormalization of the sfermion mixing angles in Higgs-mass
calculations, although it too becomes gauge dependent when EW
corrections are taken into account.

A further complication of non-minimal schemes is that, even if the
renormalization conditions are imposed on the masses and mixing angles
of the SUSY particles, what is often taken as input in
phenomenological studies are the underlying Lagrangian parameters. For
example, in the case of the third-generation squarks these parameters
include the soft SUSY-breaking masses $M_{{\tilde{Q}}}$,
$M_{{\tilde{t}}_R}$ and $M_{{\tilde{b}}_R}$, and the trilinear
couplings $A_t$ and $A_b$. The OS definition of the soft SUSY-breaking
parameters in the squark sector interprets them as the parameters
entering tree-level mass matrices for stops and sbottoms that are
diagonalized by the respective OS mixing angles (defined as described
above) and have the pole squark masses as eigenvalues. However, the
$SU(2)$ relation $M_{{\tilde{b}}_L}=M_{{\tilde{t}}_L}=M_{{\tilde{Q}}}$
only applies to bare or minimally-renormalized parameters. In this OS
scheme, as a result of EWSB, the soft SUSY-breaking mass entering the
{$LL$} element of the tree-level mass matrix for the sbottoms differs
from its stop counterpart by a finite shift. One possible approach is
to identify $M_{{\tilde{t}}_L}$ with the renormalized doublet-mass
parameter $M_{{\tilde{Q}}}$, which is taken as input, and compute
$M_{{\tilde{b}}_L}$ up to the required order in the perturbative
expansion by requiring that the bare doublet-mass parameter be the
same for stops and sbottoms. Alternatively, one can consider a
different OS scheme in which the relation
$M_{{\tilde{b}}_L}=M_{{\tilde{t}}_L}$ is imposed at the level of the
renormalized soft SUSY-breaking parameters. In that case only three of
the squark masses can be defined as pole masses, while the fourth is
treated as a dependent parameter, and is extracted from the $SU(2)$
relation that connects at tree level the masses and mixing angles of
stops and sbottoms.

Even when OS renormalization conditions are chosen for most of the
parameters entering a given calculation, it is quite common that some
parameters are still defined via minimal subtraction. For example, in
the calculation of the radiative corrections to the Higgs masses in
the MSSM, both $\tan\beta$ and the Higgs/higgsino mass parameter in
the superpotential, $\mu$, are usually renormalized in the $\drbar$
scheme. In addition, a $\drbar$ definition is commonly adopted for the
field-renormalization constants. We remark that the latter drop out of
the Higgs-mass calculation when \eqn{eq:det} is solved order by order
in the perturbative expansion, but they are nevertheless introduced to
ensure that the individual elements of $\Delta {\cal M}_{ij}^2(p^2)$
are free of UV divergences for all values of $p^2$. Finally, the
strong gauge coupling, whose definition becomes relevant in Higgs-mass
calculations beyond two loops, is usually renormalized by minimal
subtraction, irrespective of the choices made for the other
parameters.

In general, Higgs-mass calculations that are performed at the same
order in the perturbative expansion but employ different
renormalization schemes (or scales) will give numerically different
results. Since {this} difference is formally of higher order, it is
often included in the estimate of the theory uncertainty of the
prediction. We stress that a proper comparison between two
calculations must also take into account the different definitions of
the input parameters. In practice, the values of the renormalized
parameters must be given as input in the scheme employed by one
calculation, and then properly converted into the scheme employed by
the other one.  We postpone to section~\ref{sec:uncertainties} an
extensive discussion of the available estimates for the theory
uncertainty of the Higgs-mass calculation in SUSY models.

\subsection{Scenarios with large mass hierarchies}
\label{sec:hierarchies}

As shown in section~\ref{sec:general:radcor}, for values of $\MA$ and
$\tan\beta$ large enough to saturate the tree-level bound, a SM-like
Higgs boson with mass around $125$~GeV can be obtained in the MSSM
with an average stop mass $\MS$ of about $2$~TeV for $|X_t/\MS|\approx
2$, whereas for vanishing~$X_t$ the average stop mass needs to be
heavier than $10$~TeV. Lower values of $\MA$ and/or $\tan\beta$ imply
lower predictions for $\Mh$ at tree level, and thus require even
larger stop masses to ensure that the mass of the SM-like Higgs boson
is compatible with the observed value.  In contrast, SUSY models
beyond the MSSM -- such as, e.g., the NMSSM -- may allow for
additional contributions to the tree-level prediction, alleviating the
need for heavy SUSY particles.

In general, when the SUSY scale~$\MS$ is significantly larger than the
EW scale (which we can identify, e.g., with the top-quark mass $\Mt$),
any fixed-order computation of the Higgs boson masses may become
inadequate, because radiative corrections of order $n$ in the loop
expansion contain terms enhanced by as much as $\ln^n(\MS/\Mt)$. In
the presence of a significant hierarchy between the scales, the
computation of the Higgs masses needs to be reorganized in an EFT
approach: the heavy particles are integrated out at the scale $\MS$,
where they only affect the matching conditions for the couplings of
the EFT valid below $\MS$; the appropriate renormalization group
equations (RGEs) are then used to evolve these couplings between the
SUSY scale and the EW scale, where the running couplings are related
to physical observables such as the Higgs masses and the masses of
fermions and gauge bosons.  In this approach, the computation is free
of large logarithmic terms both at the SUSY scale and at the EW scale,
while the effect of those terms is accounted for to all orders in the
loop expansion by the evolution of the couplings between the two
scales. More precisely, large corrections can be resummed to the
(next-to)$^n$-leading-logarithmic (N$^n$LL) order by means of $n$-loop
calculations at the SUSY and EW scales combined with $(n\!+\!1)$-loop
RGEs.

In the simplest heavy-SUSY scenario, all of the superparticles as well
as all of the BSM Higgs bosons are clustered around a single scale
$\MS$, so that the EFT valid below this scale is just the SM. In this
case, the Higgs-mass calculation can rely in part on results already
available within the SM: full three-loop plus partial higher-loop RGEs
for all parameters of the SM Lagrangian, and full two-loop plus
partial higher-loop relations between the $\msbar$-renormalized
parameters and appropriate sets of physical observables at the EW
scale. These are combined with the matching conditions for the
Lagrangian parameters at the SUSY scale: in particular, the
calculation of the matching condition for the quartic Higgs coupling
at one, two and three loops is required for the resummation of the
large logarithmic corrections at NLL, NNLL and N$^3$LL, respectively.

Of particular interest from the phenomenological point of view are
SUSY scenarios in which an extended Higgs sector is within the reach
of the LHC. In the MSSM, direct searches for BSM Higgs bosons decaying
to pairs of down-type fermions already constrain significant regions
of the parameter space, favoring relatively low values of
$\tan\beta$. However, multi-TeV stop masses are required to obtain a
prediction for $\Mh$ of about $125$~GeV when $\tan\beta \lesssim 10$,
and an EFT approach is warranted. If all SUSY particles are heavy and
all Higgs bosons are relatively light, the EFT valid below the
matching scale is a two-Higgs-doublet model (\thdm), for which only
the NLL resummation of large logarithms (i.e., one-loop matching
conditions and two-loop RGEs) is currently available {in
  full}. More complicated hierarchical scenarios include the case in
which the BSM-Higgs masses sit at an intermediate scale between the
mass of the observed Higgs boson and the heavy SUSY masses, ``Split
SUSY'' scenarios in which gauginos and higgsinos are significantly
lighter than the sfermions, and, conversely, scenarios in which they
are significantly heavier. In each of these scenarios an appropriate
tower of EFTs needs to be constructed, which involves the computation
of threshold corrections at each of the scales where some heavy
particles are integrated out.

In the standard EFT approach to the Higgs-mass calculation, the
high-energy SUSY theory is matched to a renormalizable low-energy
theory (e.g., the SM or the \thdm) in the unbroken phase of the EW
symmetry, and the effect of non-zero vevs for the Higgs fields is
taken into account only at the EW scale. The resulting prediction for
the Higgs mass neglects terms suppressed by powers of $v^2/\MS^2\,$ --
where we denote by $v$ the vev of a SM-like scalar -- which can be
mapped to the effect of non-renormalizable, higher-dimensional
operators in the EFT, such as, e.g., dimension-six scalar interactions
$|\phi_i|^6$. These terms are clearly suppressed in the limit of large
$\MS$, where the resummation of logarithmic corrections provided by
the EFT approach is numerically important. In contrast, the FO
calculation of the Higgs masses is performed directly within the SUSY
theory and in the broken phase of the EW symmetry. Such a calculation
does not include the resummation of the large logarithms, but does
account for all $\MS$-suppressed effects up to the considered
perturbative order. In order to obtain accurate predictions for the
Higgs masses in SUSY scenarios with intermediate values of $\MS$, for
which neither the ${\cal O}(v^2/\MS^2)$ effects nor the higher-order
log-enhanced effects are obviously negligible, a number of ``hybrid''
approaches that combine existing FO and EFT calculations have been
proposed in recent years. To avoid double counting, these hybrid
approaches require a careful subtraction of the terms that are
accounted for by both the FO and the EFT parts of the calculation.
Indeed, a few successive adjustments -- some of which stemmed from
extensive discussions held during the KUTS meetings -- were necessary
to obtain predictions for the Higgs masses that, in the limit of very
large $\MS$, show the expected agreement with the pure EFT
calculation.

In the following three sections we will describe in detail the recent
advances of the Higgs-mass calculations in SUSY models in the FO, EFT
and hybrid approach, respectively.

\newpage
\section{Fixed-order calculations}
\label{sec:fo}


\subsection{Higgs-mass calculations in the MSSM}

The MSSM~\cite{Nilles:1983ge, Haber:1984rc} is one of the
best-motivated extensions of the SM, and probably the most
studied. The existence of a tree-level upper bound,
$\Mh<\MZ\,|\cos2\beta|$, on the mass of the lighter CP-even Higgs
boson of the MSSM was recognized early in the
1980s~\cite{Inoue:1982ej}. One-loop radiative corrections to this
bound were computed shortly thereafter~\cite{Li:1984tc}, but the
computation neglected the effects of the top Yukawa coupling, which at
the time was expected to be sub-dominant with respect to the EW gauge
couplings, resulting in an upper bound of at most $95$~GeV. In 1989,
two papers~\cite{Gunion:1989dp, Berger:1989hg} did consider the effect
of large Yukawa couplings, but focused on the corrections to mass sum
rules as opposed to individual masses. In 1991, as the searches for
the top quark and for the stops implied that they all had to be
heavier than the $Z$ boson, three seminal papers~\cite{Okada:1990vk,
  Ellis:1990nz, Haber:1990aw} pointed out that the one-loop
corrections controlled by the top Yukawa couplings had the potential
to increase the upper bound on $\Mh$ well above $\MZ$. The importance
of these corrections for Higgs phenomenology at the LEP was swiftly
recognized~\cite{Barbieri:1991tk, Ellis:1991zd, Brignole:1991pq}, and
by the mid-1990s full one-loop calculations of the Higgs masses had
become available~\cite{Chankowski:1991md, Brignole:1991wp,
  Brignole:1992uf, Chankowski:1992er, Dabelstein:1994hb,
  Pierce:1996zz} for a simplified {\em(``vanilla'')} version of the
MSSM Lagrangian that does not include CP-violating phases, flavor
mixing or $R$-parity violation. Between the late 1990s and the mid
2000s~\cite{Hempfling:1993qq, Heinemeyer:1998jw, Heinemeyer:1998kz,
  Zhang:1998bm, Heinemeyer:1998np, Espinosa:1999zm, Carena:2000dp,
  Espinosa:2000df, Degrassi:2001yf, Brignole:2001jy, Brignole:2002bz,
  Dedes:2002dy, Dedes:2003km, Allanach:2004rh, Heinemeyer:2004xw},
two-loop corrections to the masses of the neutral\,\footnote{The
  two-loop corrections to the mass of the charged Higgs boson in the
  MSSM were studied much later, under the same approximations, in
  refs.~\cite{Frank:2013hba, Hollik:2015ema}.} Higgs bosons were also
computed, under the approximations of vanishing external momenta in
the self-energies and of vanishing EW gauge couplings (i.e., adopting
the ``gaugeless limit'' described in
section~\ref{sec:general:radcor}). This combination of full one-loop
and gaugeless, zero-momentum two-loop corrections to the Higgs masses
was also implemented in widely-used codes for the determination of the
MSSM mass spectrum, such as \FH~\cite{Heinemeyer:1998yj, Hahn:2009zz,
  Bahl:2018qog}, \SSY~\cite{Allanach:2001kg, Allanach:2014nba}, {\tt
  SuSpect}~\cite{Djouadi:2002ze} and \SP~\cite{Porod:2003um,
  Porod:2011nf}.\footnote{Descriptions of all of the
    public codes mentioned here and thereafter, including extended
    citation guides, are collected in the appendix.}

In the past two decades, a substantial effort has been devoted to the
improvement of the fixed-order calculation of the Higgs masses in the
MSSM, along three general directions:

\begin{itemize}
  
\item Completing the two-loop calculation, including momentum
  dependence and corrections controlled by the EW gauge couplings;

\item Including the dominant three-loop corrections;

\item Extending the Higgs-mass calculation to the most general MSSM
  Lagrangian, including CP violation, $R$-parity violation and the
  effects of flavor mixing in the sfermion mass matrices.

\end{itemize}

In the following we summarize the developments along each of these
directions, highlighting in separate paragraphs the calculations that
were presented and discussed during the KUTS meetings.

\subsubsection{Completing the two-loop calculation in the vanilla MSSM}
\label{sec:vanilla}

A full calculation of the two-loop corrections to the neutral Higgs
masses in the effective potential approach -- i.e., including the
effects of the EW gauge couplings but neglecting external-momentum
effects -- became available already in 2002~\cite{Martin:2002wn}. The
two-loop self-energies of the scalar Higgs bosons, see
eq.~(\ref{eq:effpot}), were obtained from a numerical differentiation
of the two-loop effective potential of the MSSM, which had been
computed in the \drbar\ renormalization scheme and in the Landau gauge
in ref.~\cite{Martin:2002iu}.  It was found in
ref.~\cite{Martin:2002wn} that the two-loop EW corrections to the
Higgs masses suffer from singularities when the tree-level squared
masses of the would-be Goldstone bosons entering the loops pass
through zero, and it was argued that these singularities would be
cured by the inclusion of the full momentum dependence in the two-loop
self-energies.

Two years later, a calculation of the two-loop contributions involving
the strong gauge coupling or the third-family Yukawa couplings to the
self-energies of both neutral and charged Higgs bosons was presented
in ref.~\cite{Martin:2004kr}. That calculation, based on the methods
of refs.~\cite{Martin:2003qz, Martin:2005qm}, went beyond the two-loop
results implemented in the existing public codes in that it included
external-momentum effects, as well as contributions involving the
{$D$}-term-induced EW interactions between Higgs bosons and sfermions.
Combined with the effective-potential results of
ref.~\cite{Martin:2002wn}, the results of ref.~\cite{Martin:2004kr}
provided an almost-complete two-loop calculation of the Higgs masses
in the MSSM -- the only missing part being the external-momentum
dependence of diagrams that vanish when the EW gauge couplings are
turned off (namely, the part required to tame the above-mentioned
singularities). However, no code for the calculation of the MSSM mass
spectrum based on the results of refs.~\cite{Martin:2002wn,
  Martin:2004kr} was made available, and the way those results were
organized did not lend itself to a straightforward implementation in
the existing public codes.  On the one hand, the $\drbar$
renormalization scheme adopted in refs.~\cite{Martin:2002wn,
  Martin:2004kr} for the parameters of the MSSM Lagrangian does not
match the ``mixed OS--\drbar'' scheme adopted in \FH, where some of
the relevant parameters (e.g., the mass and mixing terms for the
squarks) are renormalized on-shell~\cite{Heinemeyer:2010mm,
  Fritzsche:2011nr, Fritzsche:2013fta}. On the other hand, the
implementation of the results of refs.~\cite{Martin:2002wn,
  Martin:2004kr} in \SSY, {\tt SuSpect} and \SP, which also adopt the
$\drbar$ scheme, is complicated by the different choice of gauge in
{these} codes, and by the singularities in the two-loop, zero-momentum
EW corrections. The latter restrict the applicability of the
calculation to the range of renormalization scales where none of the
tree-level Higgs masses is tachyonic, which may depend on the
considered scenario and should be determined by the codes on a
case-by-case basis.

\paragraph{Advances during KUTS:}

In 2014, at the early stages of the KUTS initiative, the two-loop
contributions to the Higgs self-energies that involve the strong gauge
coupling and the top Yukawa coupling, i.e.~those denoted as ${\cal
  O}(\alpha_t\alpha_s)$, were computed again with full momentum
dependence in refs.~\cite{Borowka:2014wla,Degrassi:2014pfa}. In
particular, ref.~\cite{Borowka:2014wla} relied on the
sector-decomposition algorithm of
refs.~\cite{Carter:2010hi,Borowka:2012yc} for the numerical
calculation of the two-loop integrals, and adopted the mixed
OS--\drbar\ renormalization scheme of
\FH. Ref.~\cite{Degrassi:2014pfa} relied instead on the methods of
refs.~\cite{Martin:2003qz, Martin:2005qm} for the loop integrals, and
obtained results both in the \drbar\ scheme and in a mixed
OS--\drbar\ scheme.  Ref.~\cite{Degrassi:2014pfa} also computed the
two-loop corrections that involve both the strong gauge coupling and
the EW couplings, under the approximation that the only non-vanishing
Yukawa coupling is the top one.\footnote{These mixed EW--QCD
  corrections are often called ${\cal O}(\alpha\alpha_s)$, which
  highlights the ambiguity of such notation: for example, they include
  both terms proportional to $g^4\,g_s^2$ and terms proportional to
  $g^2\,{y_t^2}\,g_s^2$.} The \drbar\ results of
ref.~\cite{Degrassi:2014pfa} were found to be in full agreement with
those of the earlier ref.~\cite{Martin:2004kr}. On the other hand, a
discrepancy between the OS--\drbar\ results of
ref.~\cite{Degrassi:2014pfa} and those of ref.~\cite{Borowka:2014wla}
was traced to different renormalization prescriptions for the
parameter $\tan\beta$ and for the field-renormalization constants. The
scheme dependence associated {with} the former is numerically small,
while the latter enter the prediction for the Higgs mass only at
higher orders, and their effect can be considered part of the theory
uncertainty~\cite{Borowka:2015ura}. As to the numerical impact of the
corrections, it was found that both the momentum-dependent part of the
${\cal O}(\alpha_t\alpha_s)$ corrections and the whole ${\cal
  O}(\alpha\alpha_s)$ corrections can shift the prediction for the
Higgs mass by a few hundred MeV in representative scenarios with stop
masses of about $1$~TeV, but there can be significant cancellations
between the two classes of corrections. {Finally, in 2018,
  ref.~\cite{Borowka:2018anu} computed all of the two-loop corrections
  to the Higgs mass that involve the strong gauge coupling, including
  also the mixed EW--QCD effects, with full dependence on the external
  momentum and allowing for complex parameters.  In the limit of real
  parameters, the calculation of ref.~\cite{Borowka:2018anu} improved
  on the one of ref.~\cite{Degrassi:2014pfa} in that it included also
  the effects of Yukawa couplings other than the top.}

\subsubsection{Dominant three-loop corrections}
\label{sec:mssm3loop}

Another obvious direction for the improvement of the fixed-order
calculation of the Higgs masses in the MSSM is the inclusion of
three-loop effects.  The logarithmic terms in the three-loop
corrections to the mass of the lighter, SM-like Higgs scalar can be
identified in the EFT approach by solving perturbatively the
appropriate system of boundary conditions and RGEs, without actually
computing any three-loop diagrams (see section~\ref{sec:eft} for
further details). In the approximation where only the top Yukawa
coupling and the strong gauge coupling are different from zero, the
three-loop logarithmic terms were computed at LL in
ref.~\cite{Degrassi:2002fi}, at NLL in ref.~\cite{Martin:2007pg} and
at NNLL in ref.~\cite{Draper:2013oza}.

The first genuinely three-loop computation of the corrections to the
lighter Higgs mass was presented in refs.~\cite{Harlander:2008ju,
  Kant:2010tf}. It was restricted to the terms of ${\cal
  O}(\alpha_t\alpha^2_s)$, i.e.~those involving the highest power of
the strong gauge coupling, {which can be consistently
  computed} in the limit of vanishing external momentum. Since some
three-loop integrals cannot {currently} be solved
analytically for arbitrary values of the masses, a number of possible
hierarchies among the SUSY masses was considered, for which analytical
results can be obtained via asymptotic expansions. The relevant
parameters were renormalized in the \drbar\ scheme, with the exception
of the stop masses for which a modified scheme, denoted as \mdrbar,
was introduced in scenarios with {a} heavy gluino. Indeed,
in the \drbar\ scheme potentially large contributions proportional to
powers of the gluino mass $\mgl$ affect the corrections to the Higgs
mass already at two loops~\cite{Degrassi:2001yf}.\footnote{These
  contributions cancel out if the stop masses and mixing are defined
  on-shell~\cite{Heinemeyer:1998np, Degrassi:2001yf}.} In the
\mdrbar\ scheme of refs.~\cite{Harlander:2008ju, Kant:2010tf} the
corrections proportional to $\mgl^2$ are instead absorbed in the
definition of the squared stop masses. It was found that the
three-loop ${\cal O}(\alpha_t\alpha^2_s)$ corrections to the Higgs
mass are typically of the order of a few hundred MeV in scenarios with
stop masses of about $1$~TeV. These corrections were made available in
the public code \Htm~\cite{Kant:2010tf}, which relies on \FH\ for the
one- and two-loop parts of the calculation. Since in \FH\ the
parameters in the top/stop sector are renormalized in the OS scheme,
an internal conversion of the relevant one- and two-loop corrections
to the \drbar\ (or \mdrbar) scheme is performed within \Htm.

\paragraph{Advances during KUTS:}

In 2014, ref.~\cite{Kunz:2014gya} re-examined the two-loop
determination of the \drbar-renormalized top mass of the MSSM, which
enters the corrections to the Higgs mass in the three-loop calculation
of refs.~\cite{Harlander:2008ju, Kant:2010tf}. It was shown that the
renormalization-scale dependence of the Higgs-mass prediction of
\Htm\ can be improved by performing the conversion from the
\msbar-renormalized top mass of the SM (in turn extracted from the
pole mass) to the \drbar-renormalized top mass of the MSSM at a fixed
scale of the order of the SUSY masses. In 2017, the three-loop
corrections of refs.~\cite{Harlander:2008ju, Kant:2010tf} were
implemented in the stand-alone module {\tt
  Himalaya}~\cite{Harlander:2017kuc}, which can be directly linked to
codes that perform the two-loop calculation of the MSSM Higgs masses
in the \drbar\ scheme. In 2018, ref.~\cite{Stockinger:2018oxe} studied
the compatibility of DRED with SUSY at three loops, extending the
earlier analyses of refs.~\cite{Hollik:2005nn, Harlander:2006xq,
  Harlander:2009mn}. It was shown that, in the gaugeless limit, the
Slavnov-Taylor relations expressing SUSY invariance are respected by
DRED, and no SUSY-restoring counterterms are required. Finally, in
2019 the ${\cal O}(\alpha_t\alpha^2_s)$ corrections to the lighter
Higgs mass in the limit of vanishing external momentum were computed
for arbitrary values of the SUSY masses in
refs.~\cite{R.:2019ply,R.:2019irs}. The three-loop integrals that do
not have analytical solutions were computed numerically with the
methods of ref.\cite{Bauberger:2017nct}. The effects of the ${\cal
  O}(\alpha_t\alpha^2_s)$ corrections on the prediction for the Higgs
mass were found to be in good agreement with those of the
corresponding corrections implemented in \Htm.

\subsubsection{Beyond the vanilla MSSM}
Going beyond the simplified MSSM Lagrangian considered in the previous
sections, the direction that has received the most attention so far is
the inclusion of the effects of complex parameters in the calculation
of the Higgs boson masses and mixing. At tree level, CP symmetry is
conserved in the MSSM Higgs sector. In the presence of complex
parameters in the MSSM Lagrangian, however, radiative corrections
induce a mixing among the CP-even bosons, $h$ and $H$, and the CP-odd
boson, $A$, such that beyond tree level they combine into three
neutral mass eigenstates usually denoted as $h_i$ (with
$i=1,\!2,\!3$). Since the CP-odd boson mass $\MA$ is no longer a
well-defined quantity beyond tree level, it is convenient to express
the tree-level mass matrix for the neutral Higgs bosons in terms of
the charged-Higgs mass $\MHp$. The dominant one-loop corrections to
the Higgs mass matrix in the presence of complex parameters, under
various approximations, were computed between the late 1990s and the
early 2000s in refs.~\cite{Pilaftsis:1998dd, Demir:1999hj,
  Pilaftsis:1999qt, Choi:2000wz, Carena:2000yi, Ibrahim:2000qj,
  Heinemeyer:2001qd, Carena:2001fw, Ibrahim:2002zk} (the calculations
of refs.~\cite{Pilaftsis:1999qt, Carena:2000yi, Carena:2001fw} also
included two-loop leading-logarithmic terms). {F}ull calculation{s} of
the one-loop corrections became available {in 2004~\cite{Ellis:2004fs}
  and} in 2006~\cite{Frank:2006yh}, and the two-loop ${\cal
  O}(\alpha_t\alpha_s)$ corrections were computed {in
  2007}~\cite{Heinemeyer:2007aq}. The results of
refs.~\cite{Choi:2000wz, Carena:2000yi, Carena:2001fw, Ellis:2004fs}
were implemented in the public code {\tt CPsuperH}~\cite{Lee:2003nta,
  Lee:2007gn, Lee:2012wa}, whereas the results of
refs.~\cite{Frank:2006yh, Heinemeyer:2007aq} {were} implemented in
\FH. For the two-loop corrections other than ${\cal
  O}(\alpha_t\alpha_s)$, which were only known for real MSSM
parameters at the time, the dependence on the relevant phases was
approximated in \FH\ through an interpolation between the corrections
obtained with positive and with negative values of the corresponding
parameters.

Another direction of development for the Higgs-mass calculation beyond
the vanilla MSSM was the inclusion of the effects of the mixing
between different generations of sfermions. In the most general MSSM
Lagrangian, the soft SUSY-breaking mass and trilinear-interaction
terms for the sfermions are $3\!\times\!3$ matrices in flavor space.
After EWSB, all sfermions with the same electric charge mix with each
other, via $6\!\times\!6$ mass matrices for the up- and down-type
squarks and the charged leptons, and a $3\!\times\!3$ mass matrix for
the sneutrinos.  A calculation of the one-loop corrections to the
Higgs masses allowing for generic mixing between the second and third
generations of squarks was first performed in
2004~\cite{Heinemeyer:2004by} (for further studies of the effects of
stop-scharm mixing see also refs.~\cite{Cao:2006xb,
  Brignole:2015kva}). A version of the calculation of
ref.~\cite{Heinemeyer:2004by} extended to full three-generation mixing
was implemented in \FH, and was later cross-checked (and amended) in
ref.~\cite{AranaCatania:2011ak}. It was found that corrections to the
mass of the lighter Higgs boson of up to several GeV can arise in the
presence of large mixing between the second and third generations in
the soft SUSY-breaking trilinear couplings, although for down-type
squarks the constraints from $B$ physics must be taken into
account. Finally, the effects of slepton-flavor mixing on the one-loop
corrections to the Higgs masses were studied in
ref.~\cite{Gomez:2014uha}, and found to be very small in the
considered scenarios.

\paragraph{Advances during KUTS:}

A fruitful line of activity in recent years has been the extension to
the case of complex MSSM parameters of all two-loop corrections to the
Higgs masses that were implemented in \FH\ beyond ${\cal
  O}(\alpha_t\alpha_s)$. In 2014, the two-loop corrections of ${\cal
  O}(\alpha_t^2)$, i.e.~those involving only the top Yukawa coupling,
were computed in the limit of vanishing external momentum in
refs.~\cite{Hollik:2014wea, Hollik:2014bua, Hahn:2015gaa}. In 2017,
the calculation of the two-loop Yukawa-induced corrections was
extended to ${\cal O}(\alpha_t^2,\alpha_t\alpha_b,\alpha_b^2)$ in
ref.~\cite{Passehr:2017ufr}, thus accounting for the terms controlled
by the bottom Yukawa coupling, which are relevant for large
$\tan\beta$, but still in the limit of vanishing momentum. Finally,
{as mentioned in section~\ref{sec:vanilla}, a complete calculation of
  the} two-loop corrections that involve the strong gauge coupling,
including {the} full dependence on the external momentum{, was
  presented in 2018} in ref.~\cite{Borowka:2018anu}.  All of these
calculations adopted the mixed OS--\drbar\ scheme of \FH, see
refs.~\cite{Heinemeyer:2010mm, Fritzsche:2011nr, Fritzsche:2013fta},
allowing for a seamless implementation in the code, and they
collectively removed the need for approximations in the dependence of
the two-loop corrections on the complex parameters. They also provided
useful cross-checks of the earlier calculations of
refs.~\cite{Brignole:2001jy, Dedes:2003km, Degrassi:2014pfa} for the
case of real MSSM parameters.

Another independent line of activity in recent years stemmed from the
inclusion of the two-loop corrections to the Higgs masses in
\SA~\cite{Staub:2008uz, Staub:2009bi, Staub:2010jh, Staub:2012pb,
  Staub:2013tta, Staub:2015kfa}, a package that automatically
generates versions of the ``spectrum generator'' \SP\ for generic,
user-specified BSM models. The calculation of the Higgs masses employs
the \drbar\ renormalization scheme, and in the two-loop part it is
restricted to the gaugeless limit and to the approximation of
vanishing external momentum. It relies on the earlier results of
ref.~\cite{Martin:2001vx} for the complete two-loop effective
potential in a general renormalizable theory, which is adapted within
\SA\ to the specific BSM model under consideration. In the original
2014 paper describing the two-loop extension of \SA,
ref.~\cite{Goodsell:2014bna}, the derivatives of the effective
potential were determined numerically, but an analytic computation of
the derivatives was provided soon thereafter in
ref.~\cite{Goodsell:2015ira}. This new version of \SA\ was quickly put
to work in a variety of SUSY (as well as non-SUSY) extensions of the
SM. For what concerns the vanilla MSSM, ref.~\cite{Goodsell:2015ira}
provided a useful cross-check of the \drbar\ formulas for the two-loop
corrections to the Higgs masses implemented in the standard version of
\SP, namely those of refs.~\cite{Degrassi:2001yf, Brignole:2001jy,
  Brignole:2002bz, Dedes:2002dy, Dedes:2003km,
  Allanach:2004rh}. Beyond the vanilla MSSM, in 2014
ref.~\cite{Dreiner:2014lqa} studied the effects of $R$-parity
violating couplings, showing that they can induce positive corrections
to the lighter Higgs mass of up to a few GeV.  In 2015,
ref.~\cite{Goodsell:2015yca} studied the effects of flavor mixing in
the soft SUSY-breaking terms, finding that a large stop-scharm-Higgs
coupling can induce shifts of a few GeV in the two-loop corrections to
the lighter Higgs mass. Finally, in 2016 ref.~\cite{Goodsell:2016udb}
studied the dependence of the Higgs masses on CP-violating phases for
the MSSM parameters. A direct comparison with the earlier results of
refs.~\cite{Frank:2006yh, Heinemeyer:2007aq, Hollik:2014wea,
  Hollik:2014bua, Hahn:2015gaa} was{,}
however{,} not feasible, due to the different (namely,
mixed OS--\drbar) renormalization scheme employed in that set of
calculations.

\subsection{Higgs-mass calculations in the NMSSM}
\label{sec:nmssmfixed}

In the NMSSM -- for reviews, see refs.~\cite{Maniatis:2009re,
  Ellwanger:2009dp} -- the Higgs sector is augmented with a
gauge-singlet superfield\,\footnote{Here and thereafter, a superfield
  is denoted by a ``hat'' over the symbol used for its scalar
  component. {We adopt the conventions of ref.~\cite{Ellwanger:2009dp}
    for the signs of the parameters in the superpotential and in the
    soft SUSY-breaking Lagrangian.}}  $\hat S$. The simplest and
most-studied version of the NMSSM involves a $Z_3$ symmetry that
forbids terms linear or quadratic in the Higgs superfields. The
superpotential mass term for the MSSM Higgs doublets is replaced by a
trilinear singlet-doublet interaction, plus a cubic interaction term
for the singlet
\beq
\label{eq:nmssm-supp}
\mu\,\hat H_1 \hat H_2 ~~~\longrightarrow~~~
\lambda\,\hat S\, \hat H_1 \hat H_2 \,-\, \frac\kappa3\,\hat S^3~,
\eeq
and the corresponding term in the soft SUSY-breaking scalar potential is
replaced as
\beq
\label{eq:nmssm-soft}
B_\mu\,H_1 H_2 ~~~\longrightarrow~~~ \lambda\,A_\lambda\, S\,H_1 H_2 \,-\, \frac\kappa3\,A_\kappa\,S^3~.
\eeq
In addition, the potential contains a mass term $m_S^2\,|S|^2$ for the
singlet. For appropriate values of the soft SUSY-breaking parameters
the singlet takes a vev, $v_s\equiv \langle S\rangle$, inducing
effective $\mu$ and $B_\mu$ {parameters} for the
doublets:\,\footnote{In the equations that illustrate the NMSSM Higgs
  sector we assume for simplicity that all parameters are real.}
\beq
\label{eq:muBmueff}
\mu_{\rm eff}~=~\lambda\,v_s~,~~~~~~~
B_{\mu\,{\rm eff}}~=~\lambda\,v_s(A_\lambda+\kappa\,v_s)~.
\eeq
This provides a solution to the so-called ``$\mu$ problem'' of the
MSSM, i.e.~the question of why the superpotential mass {parameter}
$\mu$ should be at the same scale as the soft SUSY-breaking
parameters.

In the absence of complex parameters in the NMSSM Lagrangian, the
CP-even component of the singlet mixes with the CP-even components of
the two doublets via a $3\!\times\!3$ mass matrix, while its CP-odd
component mixes with the CP-odd boson $A$ via a $2\!\times\!2$ mass
matrix (after the combination of CP-odd components of the doublets
that corresponds to the neutral would-be-Goldstone boson is rotated
out). In turn, the fermionic component of the singlet superfield --
the singlino -- mixes with the neutral components of higgsinos and EW
gauginos via a $5\!\times\!5$ mass matrix. Even in the CP-conserving
case, the presence of mixing in the CP-odd sector makes it unpractical
to express the tree-level mass matrix of the CP-even bosons in terms
of a pseudoscalar mass, as is done in the MSSM. The matrix is usually
expressed either directly in terms of $A_\lambda$ or in terms of
$\MHp$, which at tree level {is given by}:
\beq
\MHp^2 ~=~ \frac{B_{\mu\,{\rm eff}}}{\sin\beta\cos\beta}\,+\,\MW^2
{\,-\,\lambda^2\,v^2}~,
\eeq
where we define $\tan\beta \equiv v_2/v_1$ {and $v^2
  \equiv v_1^2+v_2^2\approx(174~{\rm
  GeV})^2$} as in the MSSM.  Beyond tree level, $A_\lambda$
is usually renormalized in the \drbar\ scheme, whereas $\MHp$ is
usually identified with the pole mass. For non-zero phases of the
parameters in the Higgs sector, CP violation can arise in the NMSSM
already at tree level.  In this case all of the neutral Higgs bosons
mix via a $5\!\times\!5$ mass matrix (again, after the neutral
would-be-Goldstone boson is rotated out). As in the case of the MSSM,
the radiative corrections can in turn induce CP violation in the Higgs
sector in the presence of non-zero phases for the Higgs-sfermion
trilinear couplings and for the gaugino and higgsino mass parameters.

In the NMSSM, the upper bound on the mass of the lightest CP-even
boson $h_1$ is weaker than the corresponding bound on $h$ in the MSSM,
thanks to an additional contribution to the quartic Higgs coupling
controlled by the singlet-doublet superpotential coupling:
\beq
\label{eq:nmssmbound}
M^2_{h_1}~<~ \MZ^2\,\cos^22\beta ~+~ \lambda^2 v^2\,\sin^22\beta~.
\eeq
In principle, this allows for a smaller contribution from radiative
corrections in order to reproduce the observed value of the SM-like
Higgs mass. Note that the second contribution to the upper bound in
eq.~(\ref{eq:nmssmbound}) is significant only for small or moderate
$\tan\beta$, which in turn suppresses the first contribution. Large
values of $\lambda$ are thus required for $M_{h_1}$ to be
significantly larger than $\MZ$ at tree-level. However, if $\lambda$
is larger than about $0.7\!-\!0.8$ at the weak scale it develops a
Landau pole below the GUT scale, in which case the NMSSM can only be
viewed as a low-energy effective theory to be embedded in a
further-extended SUSY model.
On the other hand, if $\lambda \rightarrow 0$, with $A_\lambda$ held
fixed, the singlet and the singlino decouple from the Higgs and
higgsino sectors, respectively. If in addition $v_s\rightarrow\infty$,
with both $\lambda\,v_s$ and $\kappa\,v_s$ held fixed, one recovers
the so-called ``MSSM limit'' of the NMSSM. In this limit{,}
the masses and couplings of the Higgs doublets are exactly the same as
in the MSSM, with the effective $\mu$ and $B_\mu$ parameters given in
eq.~(\ref{eq:muBmueff}).

The mechanism to dynamically generate a superpotential mass term for
the Higgs doublets via the coupling to a singlet was proposed already
in 1975~\cite{Fayet:1974pd}, and again by several groups in the early
1980s~\cite{Kaul:1981hi, Barbieri:1982eh, Nilles:1982dy, Frere:1983ag,
  Derendinger:1983bz}. The first detailed studies of the Higgs sector
of the NMSSM at tree level date to 1989~\cite{Ellis:1988er,
  Drees:1988fc}. However, in the course of the following two decades
the radiative corrections to the Higgs masses were computed only at
one loop, for vanishing external momentum and, with few exceptions, in
the gaugeless limit. In the CP-conserving NMSSM, analytic formulas for
the quark/squark contributions were obtained in the effective
potential approach in refs.~\cite{Ellwanger:1993hn, Pandita:1993hx,
  Pandita:1993tg, Elliott:1993uc, Elliott:1993bs, King:1995vk}; the
Higgs/higgsino contributions controlled by $\lambda$ and $\kappa$ were
obtained in ref.~\cite{Elliott:1993bs} from a numerical
differentiation of the corresponding contributions to the one-loop
effective potential; ref.~\cite{Ellwanger:2005fh} computed the
leading-logarithmic terms of the one-loop corrections induced by
Higgs, chargino and neutralino loops, including also the effects of
the EW gauge couplings. For the NMSSM with complex parameters, the
one-loop corrections to the Higgs masses (including the EW effects)
were computed in the effective potential approach in
refs.\cite{Ham:2001kf, Ham:2001wt, Ham:2003jf}.

In 2009, a one-loop calculation of the corrections to the neutral
Higgs masses in the NMSSM with real parameters was presented in
ref.~\cite{Degrassi:2009yq}, under the sole approximation of
neglecting the Yukawa couplings of the first two generations. One year
later that one-loop calculation was replicated (including also the
tiny effects from the first-two-generation Yukawa couplings) and
extended to the charged Higgs mass in ref.~\cite{Staub:2010ty},
relying on an early version of \SA~\cite{Staub:2008uz, Staub:2009bi,
  Staub:2010jh}. Both calculations assumed that the tree-level mass
matrices for the Higgs bosons are fully expressed in terms of
\drbar-renormalized parameters, and included the necessary one-loop
formulas to extract the EW gauge couplings and the parameter $v$ from
a set of physical observables (e.g., ref.~\cite{Degrassi:2009yq} used
$\MZ$, $\MW$ and the muon decay constant $G_\mu$). In 2011 the
one-loop calculation of the Higgs-mass corrections was performed again
in ref.~\cite{Ender:2011qh}, where several renormalization schemes
were considered: the pure \drbar\ scheme, in which the earlier results
of refs.~\cite{Degrassi:2009yq, Staub:2010ty} were reproduced; a mixed
OS--\drbar\ scheme in which the tree-level mass matrices are expressed
in terms of the physical masses $\MZ$, $\MW$ and $\MHp$, the electric
charge $e$ in the Thomson limit, plus the \drbar\ parameters
$\tan\beta$, $v_s$, $\lambda$, $\kappa$ and $A_\kappa$; a scheme,
denoted as OS, in which $\tan\beta$ is still \drbar, but $v_s$,
$\lambda$, $\kappa$ and $A_\kappa$ are traded for combinations of the
CP-odd Higgs masses and of the chargino and neutralino masses. In 2012
the one-loop calculation of the Higgs masses in the mixed
OS--\drbar\ scheme of ref.~\cite{Ender:2011qh} was extended to the
NMSSM with complex parameters in ref.~\cite{Graf:2012hh}.  Also,
\SA\ was used in ref.~\cite{Ross:2012nr} to extend the one-loop
calculation of ref.~\cite{Staub:2010ty} to the most general version of
the NMSSM, known as GNMSSM, in which there is no $Z_3$ symmetry that
forbids linear and quadratic terms in the superpotential and in the
soft SUSY-breaking potential.

Beyond one loop, an effective-potential calculation of the ${\cal
  O}(\alpha_t \alpha_s, \alpha_b\alpha_s)$ corrections\,\footnote{ We
  recall that by ${\cal O}(\alpha_t \alpha_s, \alpha_b\alpha_s)$ we
  denote the two-loop corrections to the Higgs masses that involve the
  strong gauge coupling and the third-family Yukawa couplings in the
  gaugeless limit. In the NMSSM, these corrections include terms that
  depend also on various powers of the singlet-doublet coupling
  $\lambda$\,.} to the masses of the neutral Higgs bosons in the NMSSM
with real parameters was also presented in
ref.~\cite{Degrassi:2009yq}. The results of {this} calculation are
restricted to the gaugeless and vanishing-momentum limits, and assume
that all of the relevant parameters in the tree-level mass matrices
and in the one-loop corrections are renormalized in the
\drbar\ scheme.  Note that, in contrast to the case of the MSSM, in
the NMSSM the parameter $v$ enters the tree-level Higgs mass matrix
even in the gaugeless limit, in combination with the coupling
$\lambda$, and it should therefore be extracted from physical
observables at two loops. However, the effective-potential approach of
ref.~\cite{Degrassi:2009yq} does not provide the necessary two-loop
contributions to the gauge-boson self-energies. Moreover, as already
mentioned in section~\ref{sec:general:radcor}, a vanishing-momentum
calculation of the lightest Higgs mass in which $\lambda$ itself is
not considered vanishing misses corrections that are of the same order
in the couplings as those that are being computed, because the mass
receives a tree-level contribution proportional to $\lambda$, see
eq.~(\ref{eq:nmssmbound}). In summary, it can be argued that an
effective-potential calculation such as the one of
ref.~\cite{Degrassi:2009yq} fully captures the ${\cal O}(\alpha_t
\alpha_s, \alpha_b\alpha_s)$ corrections to the mass of the lightest,
SM-like Higgs boson only in the limit $\lambda\rightarrow0$, where
they reduce to those already computed for the MSSM in
ref.~\cite{Degrassi:2001yf}.

The calculations described above were promptly implemented in public
codes for the determination of the NMSSM mass spectrum. In particular,
the full one-loop and ${\cal O}(\alpha_t \alpha_s, \alpha_b\alpha_s)$
two-loop corrections in the \drbar\ scheme from
ref.~\cite{Degrassi:2009yq} were implemented in
\NT~\cite{Ellwanger:2004xm, Ellwanger:2005dv} and in the
NMSSM-specific version of \SSY~\cite{Allanach:2013kza}. These codes
included also the MSSM results of ref.~\cite{Dedes:2003km} for the
remaining two-loop corrections controlled by the third-family Yukawa
couplings, in the gaugeless and vanishing-momentum limits. The
inclusion of these additional corrections, applied only to the
$2\!\times\!2$ sub-matrix that involves the Higgs doublets, allowed
the codes to better reproduce the MSSM predictions for the Higgs
masses in the ``MSSM limit'' of the NMSSM. The one-loop corrections in
the \drbar\ scheme from ref.~\cite{Staub:2010ty} were made available
in the NMSSM-specific version of \SP\ that is generated automatically
by \SA. These one-loop corrections, combined with the two-loop
corrections of ref.~\cite{Degrassi:2009yq} and, for the MSSM limit,
ref.~\cite{Dedes:2003km}, were also implemented in
\FS~\cite{Athron:2014yba, Athron:2017fvs}, a package that relies on
\SA\ to generate C++ spectrum generators for generic, user-specified
BSM models. Finally, the one-loop corrections of
refs.~\cite{Ender:2011qh, Graf:2012hh} were implemented in
\NC~\cite{Baglio:2013iia}, which computes masses and decay widths of
the Higgs bosons in the NMSSM with either real or complex
parameters. The code adopts a variant of the mixed OS--\drbar\ scheme
defined in ref.~\cite{Ender:2011qh}, slightly modified to comply with
the input format for complex parameters established in the ``SUSY Les
Houches Accord'' (SLHA)~\cite{Skands:2003cj, Allanach:2008qq}. {In
  addition, \NC\ provides the option to use the \drbar\ parameter
  $A_\lambda$ instead of the pole mass $\MHp$ as input for the mixed
  OS--\drbar\ scheme.}

\paragraph{Advances during KUTS:}
Broadly speaking, the recent developments in the Higgs-mass
calculations for the NMSSM followed four directions which we will
discuss separately below: the calculation of two-loop corrections
tailored for inclusion in \SA/\SP\ and in \NC, respectively; the
development of an NMSSM-specific version of \FH; detailed comparisons
between the predictions of the available codes.

In 2014, as soon as the automatic Higgs-mass calculation in \SA\ was
extended to two loops~\cite{Goodsell:2014bna}, the code was used to
reproduce the ${\cal O}(\alpha_t \alpha_s, \alpha_b\alpha_s)$
corrections of ref.~\cite{Degrassi:2009yq}. Shortly thereafter, in
ref.~\cite{Goodsell:2014pla}, it was used to compute the remaining
two-loop corrections to the full Higgs-mass matrices of the NMSSM with
real parameters, in the \drbar\ renormalization scheme and in the
gaugeless and vanishing-momentum limits. Compared to the earlier
practice of including the two-loop corrections beyond ${\cal
  O}(\alpha_t \alpha_s, \alpha_b\alpha_s)$ only in the MSSM limit, the
calculation of ref.~\cite{Goodsell:2014pla} allowed for the inclusion
of the two-loop corrections controlled by the NMSSM-specific
superpotential couplings $\lambda$ and $\kappa$. In 2016, the
calculation was extended to the NMSSM with complex parameters in
ref.~\cite{Goodsell:2016udb}. It should be noted that the two-loop
Higgs-mass calculations of refs.~\cite{Goodsell:2014bna,
  Goodsell:2014pla, Goodsell:2016udb} share the limitations described
earlier for the calculation of ref.~\cite{Degrassi:2009yq}: the limit
of vanishing momentum misses terms that are of the same order in the
couplings as those included in the two-loop result, and the extraction
of the \drbar-renormalized parameter $v$ from physical observables is
performed only at one-loop order.

A peculiarity of the Higgs-mass calculation in the NMSSM is the fact
that the singularities for vanishing tree-level masses of the
would-be-Goldstone bosons, first described in
ref.~\cite{Martin:2002wn} for the two-loop EW corrections in the MSSM,
affect the two-loop corrections even in the gaugeless limit, due to
the presence of Higgs self-couplings controlled by $\lambda$. The
origin of these singularities, known as ``Goldstone Boson
Catastrophe'' (GBC), was discussed in refs.~\cite{Martin:2014bca,
  Elias-Miro:2014pca, Pilaftsis:2015cka, Pilaftsis:2015bbs} for the SM
and in ref.~\cite{Kumar:2016ltb} for the MSSM. It was shown in these
papers that the singularities can be removed from the effective
potential and from its first derivatives by a resummation procedure
that effectively absorbs them in the mass terms of the
would-be-Goldstone bosons entering the one-loop corrections. However,
this resummation does not fully address the singularities in the
second derivatives of the effective potential, which enter the
zero-momentum calculation of the Higgs masses.  In 2016, the solution
to the GBC was extended to a general renormalizable theory in
ref.~\cite{Braathen:2016cqe}. It was shown that, at the two-loop
order, the resummation procedure of refs.~\cite{Martin:2014bca,
  Elias-Miro:2014pca, Pilaftsis:2015cka, Pilaftsis:2015bbs,
  Kumar:2016ltb} is equivalent to imposing an OS condition on the
masses of the would-be-Goldstone bosons. Moreover, the
momentum-dependent terms that are needed to compensate the
singularities of the second derivatives of the potential in the
gaugeless limit were obtained in ref.~\cite{Braathen:2016cqe} from an
expansion in $p^2$ of the full two-loop self-energies given earlier in
ref.~\cite{Martin:2003it}. In 2017, the implementation of these
results in \SA, described in ref.\cite{Braathen:2017izn}, eventually
allowed for a GBC-free calculation of the two-loop corrections to the
Higgs masses in the gaugeless limit of the NMSSM.

In 2014, the two-loop ${\cal O}(\alpha_t \alpha_s)$ corrections to the
Higgs masses in the NMSSM with complex parameters were computed in
ref.~\cite{Muhlleitner:2014vsa} and implemented in \NC. The
computation employed the mixed OS--\drbar\ scheme defined in
refs.~\cite{Ender:2011qh, Baglio:2013iia} for the parameters in the
Higgs sector, whereas the parameters in the top/stop sector were
renormalized either in the \drbar\ or in the OS scheme. The two-loop
results of ref.~\cite{Muhlleitner:2014vsa} were restricted to the
limit of vanishing external momentum, but, in contrast to the
effective-potential calculations of refs.~\cite{Degrassi:2009yq,
  Goodsell:2014bna, Goodsell:2014pla, Goodsell:2016udb}, they did
include the ${\cal O}(\alpha_t \alpha_s)$ contributions to the
gauge-boson self-energies that are involved in the renormalization of
$v$. It was shown that, in a representative scenario with stop masses
around $1$~TeV, these contributions shift the Higgs masses by
less than $50$~MeV. In 2019, the Higgs-mass calculation of \NC\ was
extended by the inclusion of the two-loop ${\cal O}(\alpha_t^2)$
corrections at vanishing external momentum and in the MSSM
limit. These corrections were computed in ref.~\cite{Dao:2019qaz}
starting from a CP-violating NMSSM setup, where the parameters of the
Higgs sector are renormalized in the mixed OS--\drbar\ scheme of
refs.~\cite{Ender:2011qh, Baglio:2013iia} but the limits
$\lambda,\kappa \to 0$ and $v_s \to \infty$ (with $\mu_{\rm eff} =
\lambda\, v_s$ fixed) are taken. The effects of different choices
(\drbar, OS or a mixture) for the renormalization of the parameters in
the top/stop sector were also investigated. Finally, the issue of a
residual gauge dependence affecting the Higgs-mass predictions at
higher orders when the one-loop self-energies are computed at the mass
pole was discussed in refs.~\cite{Dao:2019nxi,Domingo:2020wiy}.

A one-loop calculation of the Higgs masses in a renormalization scheme
suitable for implementation in \FH\ was performed for the NMSSM with
real parameters in 2016~\cite{Drechsel:2016jdg}. It was extended to
the NMSSM with complex parameters in 2017~\cite{Domingo:2017rhb}, and
to the GNMSSM with complex parameters in
2018~\cite{Hollik:2018yek}. The mixed OS--\drbar\ scheme employed for
the Higgs sector in refs.~\cite{Drechsel:2016jdg, Domingo:2017rhb,
  Hollik:2018yek} differs from the one of refs.~\cite{Ender:2011qh,
  Baglio:2013iia} in that the Lagrangian parameters $(g, g^\prime, v)$
are connected to the physical observables $(\MZ, \MW, G_\mu)$ instead
of $(\MZ, \MW, e)$.  In the GNMSSM calculation of
ref.~\cite{Hollik:2018yek} the additional $Z_3$-violating parameters
are all renormalized in the \drbar\ scheme.  The dominant two-loop
corrections, as well as a resummation of higher-order logarithmic
effects which will be discussed in section~\ref{sec:hybrid}, were
included in refs.~\cite{Drechsel:2016jdg, Domingo:2017rhb,
  Hollik:2018yek} only in the MSSM limit, exploiting the gaugeless,
zero-momentum results implemented in \FH\ at the time (namely, those
of refs.~\cite{Degrassi:2001yf, Brignole:2001jy, Brignole:2002bz,
  Dedes:2003km} in the real case, and refs.~\cite{Heinemeyer:2007aq,
  Hollik:2014wea, Hollik:2014bua, Hahn:2015gaa} in the complex case
and in the GNMSSM).  The effectiveness of this approximation for the
two-loop corrections was assessed in ref.~\cite{Drechsel:2016jdg} by
considering the impact of the same approximation on the one-loop
corrections from the top/stop sector.  However, the extensions of
\FH\ to the (G)NMSSM presented in refs.~\cite{Drechsel:2016jdg,
  Domingo:2017rhb, Hollik:2018yek} have not been made available to the
public so far.

A significant effort was also devoted during KUTS to the comparison
between the predictions of the available codes for Higgs-mass
calculations in the NMSSM. In 2015, ref.~\cite{Staub:2015aea} compared
the results of the \drbar\ calculations implemented in \FS, \NC, \NT,
\SSY\ and \SA/\SP, in six representative points of the NMSSM parameter
space. The bulk of the discrepancies between the predictions of the
five codes could be traced back to differences in the determination of
the running parameters (in particular, the top Yukawa coupling)
entering the calculation of the radiative corrections. Once these
differences were accounted for, the residual discrepancies were mainly
due to different approximations adopted by the codes in the two-loop
corrections. In points with large $\lambda$\,, the results of
\SA/\SP\ -- which {allow} for a more-complete determination of the
two-loop corrections controlled by that coupling -- differed from
those of the other codes by up to a few GeV. The predictions of
\NC\ were in general different from the others because, at the time,
the code included two-loop corrections only at ${\cal O}(\alpha_t
\alpha_s)$. Four years later, ref.~\cite{Dao:2019qaz} showed how the
inclusion in \NC\ of the ${\cal O}(\alpha_t^2)$ corrections in the
MSSM limit improves the agreement with the other \drbar\ codes in all
of the test points of ref.~\cite{Staub:2015aea}.

Detailed comparisons between the mixed OS--\drbar\ calculations
implemented in \NC\ and in \FH\ were presented in
refs.~\cite{Drechsel:2016jdg, Domingo:2017rhb, Drechsel:2016htw}. It
was shown that the effect of the different choices of renormalization
scheme for the EW parameters $(g, g^\prime, v)$ in the one-loop part
of the calculation is numerically small. At ${\cal O}(\alpha_t
\alpha_s)$, the two codes differ in that \NC\ implements the full
calculation of ref.~\cite{Muhlleitner:2014vsa}, whereas \FH\ includes
{these} corrections only in the MSSM limit. The effect of this
approximation on the Higgs masses is obviously more relevant at large
$\lambda$, {in which case, however, the corrections involving top and
  stop loops might not even be the dominant ones.}  In the four test
points considered in ref.~\cite{Drechsel:2016htw}{, all characterized
  by $\lambda < 0.7$, the effect of taking the ``MSSM limit'' in the
  ${\cal O}(\alpha_t \alpha_s)$ corrections was found to be below
  $1$~GeV}. The bulk of the differences found in
refs.~\cite{Drechsel:2016jdg, Domingo:2017rhb, Drechsel:2016htw}
between the predictions of the two codes at the two-loop level stemmed
instead from the fact that \FH\ did include the ${\cal O}(\alpha_t^2)$
corrections in the MSSM limit, while those corrections were not
implemented in \NC\ until later, see ref.~\cite{Dao:2019qaz}.

\subsection{Higgs-mass calculations in other SUSY models}
\label{sec:bnmmsmfixed}

Calculations of the radiative corrections to the Higgs boson masses,
of varying degrees of accuracy, have also been performed for a
plethora of non-minimal extensions of the (N)MSSM. In this section we
summarize a number of calculations that were presented and discussed
during the KUTS initiative. These include both automated calculations
obtained with the \SA\ package, and calculations performed directly
within specific models.

\paragraph{Models with Dirac gauginos:}

In this class of models, {first proposed in the late
  1970s~\cite{Fayet:1978qc},} a Dirac mass for each gaugino is
obtained via a superpotential term that couples the gauge-strength
superfield, whose fermionic component is the gaugino, to an additional
chiral superfield in the adjoint representation of the gauge group. In
the minimal Dirac-gaugino extension of the MSSM, or
MDGSSM~\cite{Belanger:2009wf}, the superfield content of the MSSM is
thus supplemented with a singlet, an $SU(2)$ triplet and an $SU(3)$
octet, and the superpotential and the soft SUSY-breaking Lagrangian
are supplemented with all gauge-invariant terms that involve the
adjoint (super)fields. In the scalar sector, the singlet and the
neutral component of the triplet mix with the neutral components of
the MSSM-like Higgs doublets, resulting in $4\!\times\!4$ mass
matrices when CP is conserved. Another well-studied model, known as
MRSSM~\cite{Kribs:2007ac}, involves an $R$-symmetry that forbids
Majorana mass terms for the gauginos, a $\mu$ term in the
superpotential, and MSSM-like trilinear interaction terms in the soft
SUSY-breaking Lagrangian. Adjoint superfields for each gauge group are
introduced as in the MDGSSM to allow for Dirac gaugino masses, and
additional chiral superfield doublets $\hat R_1$ and $\hat R_2$, which
couple to the MSSM-like Higgs doublets in the superpotential but do
not obtain vevs, are introduced to allow for higgsino mass terms.
 
In models with such intricate Higgs sectors, the \SA\ package proved
useful to compute automatically the radiative corrections to the Higgs
masses. Full one-loop results in the $\drbar$ scheme were obtained in
ref.~\cite{Benakli:2012cy} for the MDGSSM, in
ref.~\cite{Benakli:2014cia} for a variant of the MDGSSM with
additional fields allowing the unification of gauge couplings, and in
ref.~\cite{Diessner:2014ksa} for the MRSSM. For what concerns the
two-loop corrections, in Dirac-gaugino models those of ${\cal
  O}(\alpha_t\alpha_s)$ differ from their MSSM counterparts because
they contain a sum on two gluino mass eigenstates, as well as
additional contributions from diagrams involving the scalar component
of the octet superfield, the sgluon. In 2015, the ${\cal
  O}(\alpha_t\alpha_s)$ corrections were computed with \SA\ -- as
usual, in the \drbar\ scheme and in the gaugeless and
vanishing-momentum limits -- for the MDGSSM in
ref.~\cite{Goodsell:2015ura} and for the MRSSM in
ref.~\cite{Diessner:2015yna}. In the latter it was found that, for
multi-TeV values of the Dirac-gluino mass $M_3$, the contribution of
the two-loop diagrams involving sgluons can increase the prediction
for the SM-like Higgs mass by more than $10$~GeV. Subsequently, in
2016, ref.~\cite{Braathen:2016mmb} presented explicit analytic
formulae for the ${\cal O}(\alpha_t\alpha_s)$ corrections in both the
MDGSSM and the MRSSM, obtained with an effective-potential
calculation. It was pointed out in ref.~\cite{Braathen:2016mmb} that,
in the pure \drbar\ scheme adopted by \SA, the large two-loop
corrections stem from non-decoupling effects -- analogous to those
already discussed in section~{\ref{sec:mssm3loop}} -- that are
enhanced by $M_3^2/M_{{\tilde t}_i}^2\,$, where $M_{{\tilde t}_i}$
(with $i=1,2$) are the stop mass eigenstates. In contrast, the sgluon
contributions are much more {moderate} if an OS scheme is adopted for
the parameters in the stop sector. We stress that the two-loop results
of refs.~\cite{Goodsell:2015ura, Diessner:2015yna, Braathen:2016mmb}
share the limitations of the analogous results obtained in the NMSSM:
in the presence of non-vanishing Higgs self-couplings at tree-level, a
zero-momentum calculation does not fully capture the two-loop
{corrections} to the mass of the SM-like Higgs boson even if the
gaugeless limit is assumed.

\paragraph{Other automated calculations:}

The \SA\ package allowed for full one-loop and partial (i.e.,
gaugeless and zero-momentum) two-loop calculations of the Higgs masses
in a few other non-minimal extensions of the MSSM. In 2015,
ref.~\cite{Nickel:2015dna} considered a model in which the superfield
content of the MSSM is supplemented with a pair of vector-like top
superfields, identifying regions of the parameter space in which the
two-loop corrections involving the additional particles can be as
large as the MSSM-like corrections. Also in 2015,
ref.~\cite{Hirsch:2015fvq} studied a left-right model in which the
Higgs sector contains two bi-doublets of $SU(2)_L \times SU(2)_R$ and
two doublets of $SU(2)_R$, resulting in $6\!\times\!6$ mass matrices
for the neutral scalars when CP is conserved. The authors found that,
in this model, radiative corrections involving vector-like colored
states can significantly lower the mass of the scalar that is mostly
right-doublet, allowing for scenarios where a SM-like Higgs with mass
around $125$~GeV is accompanied by a light neutral scalar with mass of
${\cal O}(10)$~GeV.

In 2017, a SUSY model with an extended gauge group that originates
from the exceptional group $E_6$ was studied with \FS\ in
ref.~\cite{Athron:2017fvs}. This so-called E$_6$SSM features an
NMSSM-like Higgs sector, plus a large set of exotic particles. The
calculation of the Higgs masses in ref.~\cite{Athron:2017fvs} included
the full one-loop corrections generated by \SA. To facilitate the
 comparison between NMSSM and E$_6$SSM, a subset of two-loop
corrections that are in common between the two models was also
included, relying on the results of refs.~\cite{Dedes:2003km,
  Degrassi:2009yq}.

\paragraph{Models with right-handed neutrinos:}

The so-called ``MSSM see-saw model''~\cite{Borzumati:1986qx} is an
extension of the MSSM in which the neutrino masses are generated by a
supersymmetic version of the standard see-saw mechanism. The
superpotential includes a Yukawa interaction $Y^\nu_{ij}\,\hat H_2
\hat L_i\,\hat \nu^c_j$ and a Majorana mass term $\frac12 M_{ij} \hat
\nu^c_i\hat \nu^c_j$ for the right-handed neutrino superfields, so
that a suitable pattern of masses and mixing for the light neutrinos
can be obtained with Yukawa couplings of ${\cal O}(1)$ and Majorana
masses in the range of $10^{13}\!-\!10^{15}$~GeV. In 2010 (pre-KUTS),
ref.~\cite{Heinemeyer:2010eg} computed the one-loop corrections to the
Higgs masses arising from a single generation of right-handed
neutrinos and sneutrinos. These corrections turned out to be
negligible if the parameter $\tan\beta$ entering the tree-level Higgs
mass matrix is renormalized in an OS scheme, but they can amount to
several GeV if $\tan\beta$ is defined in the \drbar\ scheme (or in a
variant thereof denoted in the paper as m\drbar). In 2013,
ref.~\cite{Draper:2013ava} pointed out that the strong sensitivity of
the Higgs-mass predictions to the presence of additional SUSY
multiplets at arbitrarily high scales should be considered an artifact
of minimal-subtraction schemes such as $\drbar$, in which the
decoupling of heavy particles is not manifest. The authors of
ref.~\cite{Draper:2013ava} proposed additional ``OS-like'' definitions
for $\tan\beta$ that also lead to negligible contributions to the
Higgs masses from the heavy supermultiplets, and discussed how these
contributions match those that would be obtained in an EFT calculation
where the heavy particles are integrated out at a scale comparable to
their mass. Finally, in 2014 ref.~\cite{Heinemeyer:2014hka} extended
the calculation of ref.~\cite{Heinemeyer:2010eg} to the case of three
generations of right-handed neutrino superfields.

Another model for which radiative corrections to the Higgs masses were
computed in recent years is the so-called ``$\mu$-from-$\nu$
Supersymmetric Standard Model'', or
$\mu\nu$SSM~\cite{LopezFogliani:2005yw}, a variant of the NMSSM in
which the role of the singlet superfield is played by three
right-handed neutrino superfields $\hat \nu^c_i$. The superpotential
includes an additional Yukawa interaction $Y^\nu_{ij}\,\hat H_2 \hat
L_i\,\hat \nu^c_j$\,, but it does not include large Majorana mass
terms for the right-handed neutrinos. Therefore, a suitable pattern of
masses and mixing for the neutrinos is obtained through an ``EW
see-saw'' mechanism in which the role of the large mass scale is
played by the neutralino masses, and the new Yukawa couplings can be
comparable in size to the electron Yukawa coupling. In this model the
right-handed neutrino interactions that replace the singlet
interactions of eqs.~(\ref{eq:nmssm-supp}) and (\ref{eq:nmssm-soft})
break both $R$-parity and lepton number conservation. Consequently,
the Higgs doublets mix with the left- and right-handed sneutrinos,
resulting in $8\!\times\!8$ mass matrices for the neutral scalars when
CP is conserved.

In 2017, ref.~\cite{Biekotter:2017xmf} presented a full one-loop
calculation of the corrections to the Higgs masses in a simplified
version of the $\mu\nu$SSM with only one right-handed neutrino,
adopting a mixed OS-\drbar\ renormalization scheme that, for the
parameters that have a counterpart in the NMSSM, matches the one of
ref.~\cite{Drechsel:2016jdg}. The extension of this calculation to
three generations of right-handed neutrinos was presented two years
later in ref.~\cite{Biekotter:2019gtq}. In both papers, the dominant
two-loop corrections in the MSSM limit, as well as the resummation of
higher-order logarithmic effects, were also included following
ref.~\cite{Drechsel:2016jdg}. In this way, a comparison between the
$\mu\nu$SSM predictions of refs.~\cite{Biekotter:2017xmf,
  Biekotter:2019gtq} and the NMSSM predictions of
ref.~\cite{Drechsel:2016jdg} singles out the effects of the one-loop
corrections controlled by the neutrino Yukawa couplings. It was found
that, for values of the Yukawa couplings of ${\cal O}(10^{-7})$, which
in this model correspond to sub-eV neutrino masses, the corresponding
effects on the prediction for the mass of the SM-like Higgs are
negligible. On the other hand, in the full $\mu\nu$SSM the presence of
two additional singlets interacting with the Higgs doublets can lead
to predictions for the scalar sector that differ substantially from
those of the NMSSM. In 2020, the one-loop corrections computed in
refs.~\cite{Biekotter:2017xmf, Biekotter:2019gtq} were made available
in the public code {\tt munuSSM}~\cite{Biekotter:2020ehh}. Through an
automated link to \FH, the code includes in the Higgs-mass calculation
also the dominant two-loop and higher-order corrections that are in
common between the $\mu\nu$SSM and the MSSM.

\paragraph{A supersymmetric Goldstone-Higgs model:}

In this model, the idea of an elementary pseudo-Goldstone boson that
acquires mass through radiative corrections and plays the role of the
observed Higgs boson, see refs.~\cite{Alanne:2014kea, Gertov:2015xma},
is exploited in a supersymmetric setup to relate the SUSY-breaking
scale with the radiatively-generated EW scale. In 2016,
ref.~\cite{Alanne:2016uta} obtained an approximate picture of the mass
spectrum of the model, computing the leading contributions to the mass
of the pseudo-Goldstone boson via the one-loop effective
potential. For a more precise investigation of the viability of this
model, and of its ability to reproduce the observed value of the Higgs
mass, a calculation that goes beyond the one-loop effective-potential
approximation would be desirable.

\subsection{Prospects}
\label{sec:prospectsFO}

As discussed earlier in this section, a full two-loop calculation of
the corrections to the Higgs masses, including momentum dependence and
all of the effects controlled by the EW gauge couplings, is not yet
available in any SUSY extension of the SM. For what concerns the MSSM,
even in the calculation of refs.~\cite{Martin:2002wn, Martin:2004kr},
which did include an effective-potential calculation of the EW
effects, the momentum dependence of the two-loop corrections was
included only in the gaugeless limit. Beyond the MSSM, two-loop
corrections to the Higgs masses have so far been computed only under
the combined approximations of vanishing momentum and vanishing EW
gauge couplings. However, in models that feature additional Higgs
self-couplings at tree level, such as the NMSSM, a consistent
calculation of the corrections to the lighter Higgs mass requires the
inclusion of the momentum dependence even in the gaugeless
limit. Closing these gaps, and obtaining results for each of the
different renormalization schemes considered in the literature, will
certainly be a priority in the near future.

As an alternative to performing individual calculations of the missing
corrections in many different models, a sensible approach might
consist in computing the full two-loop corrections only once for a
general renormalizable theory, and then adapting the results to the
field-and-interaction content of the specific model under
consideration. This approach, pioneered already in the 2000s by
refs.~\cite{Martin:2001vx, Martin:2003it, Martin:2005eg}, was at the
origin of the MSSM-specific results of refs.~\cite{Martin:2002wn,
  Martin:2004kr}. It was also at the origin of the zero-momentum and
gaugeless calculation of the two-loop corrections in generic BSM
models implemented in \SA, see refs.~\cite{Goodsell:2014bna,
  Goodsell:2015ira, Braathen:2017izn}. However, a complete calculation
of the two-loop corrections controlled by the EW gauge couplings
remained elusive, because the momentum-dependent contributions from
diagrams that involve more than one massive-gauge-boson propagator
were not available until very recently. At last, in 2019 a complete
two-loop calculation of tadpoles and self-energies for the scalars of
a general renormalizable theory was presented in
ref.~\cite{Goodsell:2019zfs}. When adapted to specific SUSY models,
the results of ref.~\cite{Goodsell:2019zfs} will allow for complete
two-loop calculations of the Higgs masses, starting from a set of
\drbar-renormalized Lagrangian parameters. However, the choice of the
most convenient renormalization scheme depends on the kind of
phenomenological analysis that is aimed for, as well as on the
considered region of the model's parameter space. It thus remains
something to be determined on a case-by-case basis. Any change of
renormalization scheme for the parameters that enter only the
radiative corrections to the Higgs mass matrices -- e.g., the
parameters in the quark/squark sector -- amounts to a product of
one-loop effects, and should not, as such, present particular
difficulties. In contrast, in order to connect the \drbar\ parameters
$g$, $g^\prime$ and $v$ entering the tree-level mass matrices to a set
of physical observables -- usually chosen among $\MZ$, $\MW$, $G_\mu$
and $e$ -- it will still be necessary to obtain complete two-loop
results for the gauge-boson self-energies, and possibly a two-loop
(but zero-momentum) calculation of the muon-decay amplitude.

Beyond two loops, fixed-order calculations of the corrections to the
Higgs masses in SUSY models exist only for the MSSM. In particular,
the three-loop corrections to the lighter Higgs mass of ${\cal
  O}(\alpha_t\alpha_s^2)$, i.e.~those involving the top Yukawa
coupling and the highest power of the strong gauge coupling, have
already been computed by different groups in the vanishing-momentum
limit, see section~\ref{sec:mssm3loop}. A first direction of
improvement, still under the approximation of vanishing momentum and
vanishing EW gauge couplings, would be the calculation of the
three-loop corrections that involve lower powers of the strong gauge
coupling, among which the most relevant are expected to be those of
${\cal O}(\alpha_t^2\alpha_s)$ and ${\cal O}(\alpha_t^3)$. Indeed, it
was noticed already in refs.~\cite{Martin:2007pg, Hahn:2013ria,
  Draper:2013oza} that, at least for what concerns the logarithmic
terms, there can be significant cancellations between these
contributions and the ${\cal O}(\alpha_t\alpha_s^2)$ ones. As in the
case of the two-loop corrections, going beyond the gaugeless limit in
the MSSM, or extending the calculation to models with additional Higgs
self-couplings, would require for consistency also the inclusion of
external-momentum effects in the three-loop self-energies. This might
be achieved via numerical methods, see
e.g.~ref.~\cite{Bauberger:2019heh}.  However, at least for the MSSM,
it still has to be determined whether the effort would be justified by
the size of the resulting corrections.

To conclude this section we recall that,
in SUSY models, both the measured value of the SM-like Higgs mass and
the exclusion bounds from direct searches at the LHC are most easily
accommodated by scenarios with multi-TeV SUSY masses.  In such
scenarios,
any fixed-order calculation of the Higgs masses may become inadequate,
because the uncomputed higher-order corrections involve higher powers
of the logarithm of the ratio between the SUSY scale and the EW
scale. In order to obtain an accurate prediction for the Higgs masses,
such potentially large logarithmic corrections must be resummed to all
perturbative orders in an EFT approach. The current status of this
kind of calculations will be reviewed in the next section.

\newpage
\section{EFT calculations}
\label{sec:eft}
\subsection{Overview}

\label{sec:EFToverview}

As mentioned already in section~\ref{sec:hierarchies}, a generic
$n$-loop amplitude has a logarithmic dependence, up to the $n$th
power, on the masses of the particles circulating in the loops. In the
presence of hierarchies between the masses, terms enhanced by
logarithms of {large} mass ratios can counteract the suppression
arising from the loop factors, slowing down (or even endangering) the
convergence of the perturbative expansion. It is then necessary to
reorganize the calculation in an EFT approach: the heavy particles are
integrated out of the theory at a renormalization scale comparable to
their masses, leaving behind threshold corrections to the couplings of
the light particles. These couplings are then evolved via appropriate
RGEs to a scale of the order of the masses of the light particles,
where physical observables (e.g., on-shell masses for the Higgs
bosons) are computed including only the light particles in the
loops. In this approach, the calculations at both the heavy- and
light-particle mass scales are free of large logarithmic terms, while
the effect of those terms is accounted for by the RG evolution. If a
BSM theory involves multiple widely-split mass scales, a tower of EFTs
must be built, computing threshold corrections at each of the scales
where some heavy particles are integrated out.

In the simplest realization of the EFT approach for SUSY models, all
of the superparticles and all of the BSM Higgs bosons are integrated
out at a common scale $\MS$, so that the EFT valid below this scale is
just the SM. The calculation of the pole mass of the Higgs boson then
requires the determination of the matching condition at the scale
$\MS$ for the quartic Higgs coupling\,\footnote{We resort to this
  notation to distinguish the quartic Higgs coupling of the SM from
  the singlet-doublet superpotential coupling of the NMSSM, see
  eq.~(\ref{eq:nmssm-supp}). In our conventions the SM potential
  contains the quartic interaction term $\frac12 \lambdaSM |H|^4$.}
$\lambdaSM$\,, which we can decompose in a tree-level part and a loop
correction as $\lambdaSM = \lambdaSM^{\rm tree} + \Delta\lambda$. The
tree-level matching condition includes the original tree-level value
of the coupling in the SUSY model, plus possible contributions from
the decoupling of the heavy scalars that, in the limit of unbroken EW
symmetry, have a non-vanishing trilinear coupling to two SM-like Higgs
bosons. For example, in the MSSM there are no such couplings, and the
tree-level matching condition is just
\beq
\label{eq:lambdamssm}
\lambdaSM^{\rm tree}(\MS) ~=~ \frac{1}{4} \left(g^2 + g^{\prime\,2}\right)\,\cos^2 2\beta~.
\eeq
In contrast, in the NMSSM there is an additional contribution
proportional to $\lambda^2$ in the original quartic coupling, plus a
term arising from the decoupling of the singlet scalar:
\beq
\label{eq:lambdanmssm}
\lambdaSM^{\rm tree}(\MS) ~=~ \frac{1}{4} \left(g^2 + g^{\prime\,2}\right)\,\cos^2 2\beta~+~\frac{\lambda^2}{2}\,\sin^2 2\beta
~-~\frac{\left[2\,\lambda^2\,v_s
    - \lambda\left( A_\lambda + 2\,\kappa\,v_s\right)\sin2\beta
    \right]^2}{2\,\kappa\,v_s\,(A_\kappa + 4\,\kappa\,v_s )}~.
\eeq

In general, the correction $\Delta\lambda$ contains contributions of
three different kinds: $i)$ contributions of one-particle-irreducible
(1PI) diagrams with four external Higgs fields, which involve only
loop integrals with vanishing external momenta; $ii)$ contributions
involving the renormalization constant of the Higgs field, which
require a computation of the ${\cal O}(p^2)$ part of the Higgs-boson
self-energy; $iii)$ contributions that arise from changes in the
definition of the parameters entering the matching
condition. Concerning the third kind, a first contribution arises from
the fact that the SUSY model provides a prediction for the quartic
Higgs coupling in the $\drbar$ scheme, whereas $\lambdaSM$ is
generally interpreted as an $\msbar$-renormalized quantity. Moreover,
the prediction is expressed in terms of the $\drbar$-renormalized
parameters of the SUSY model, some of which need to be connected to
their $\msbar$-renormalized counterparts in the SM. For example, the
conversion of the EW gauge couplings in eqs.~(\ref{eq:lambdamssm}) and
(\ref{eq:lambdanmssm}) requires the computation of the ${\cal O}(p^2)$
part of the gauge-boson self-energies. Beyond one loop,
$\Delta\lambda$ contains also terms resulting from the product of
lower-order contributions of different kinds. As in the case of the
fixed-order calculation, the dominant contributions to $\Delta\lambda$
are generally those involving the top Yukawa coupling and, beyond one
loop, the strong gauge coupling.\footnote{At variance with the
notation adopted for the mass corrections, we denote the various
contributions to $\Delta\lambda$ via the full combinations of
couplings involved. For example, the dominant contributions involving
the top Yukawa coupling are those of ${\cal O}(y_t^4)$ at one loop,
and those of ${\cal O}(y_t^4 g_s^2)$ and ${\cal O}(y_t^6)$ at two
loops.}

Once the matching condition for $\lambdaSM$ is determined at the SUSY
scale $\MS$ from a full set of SUSY parameters, the quartic coupling
is evolved down to the EW scale. There $\lambdaSM$ can be used to
compute the pole Higgs mass, including only the contributions of SM
particles in the radiative corrections. Alternatively, $\lambdaSM$ can
be extracted at the EW scale from the measured value of the Higgs
mass, evolved up to the SUSY scale, and used there to constrain the
SUSY parameters. In this case, one requires that the coupling obtained
via RG evolution coincide with the prediction of the SUSY model. We
recall that a N$^n$LL resummation of the logarithms of the ratio
between the SUSY and EW scales requires $n$-loop calculations at each
scale, combined with $(n+1)$-loop RGEs. On the other hand, the
standard procedure of matching the full SUSY model to a renormalizable
EFT in the unbroken phase of the EW symmetry amounts to neglecting
corrections suppressed by powers of $v^2/\MS^2\,$, which can be mapped
to the effect of non-renormalizable, higher-dimensional operators in
the EFT Lagrangian.

The scenarios in which all of the BSM particles are integrated out at
the same scale have the advantage that existing SM calculations can be
exploited to extract the running parameters of the EFT Lagrangian from
a set of physical observables at the EW scale and evolve them up to
the SUSY scale. In particular, the full NNLL resummation of the large
logarithmic corrections can rely on the results of
refs.~\cite{Buttazzo:2013uya, Kniehl:2015nwa, Martin:2019lqd} for the
full two-loop relations between running SM parameters and physical
observables at the EW scale, and on the results of
refs.~\cite{Mihaila:2012fm, Chetyrkin:2012rz, Mihaila:2012pz,
  Bednyakov:2012rb, Bednyakov:2012en, Chetyrkin:2013wya,
  Bednyakov:2013eba} for the full three-loop RGEs of the SM. For a
partial N$^3$LL resummation that involves only the highest powers of
the strong gauge coupling, the three-loop relation between $\lambdaSM$
and the pole Higgs mass of refs.~\cite{Martin:2013gka, Martin:2014cxa,
  Martin:2016bgz, Martin:2017lqn} and the four-loop RGE for
$\lambdaSM$ of refs.~\cite{Martin:2015eia, Chetyrkin:2016ruf} can be
exploited.

In scenarios with more-complicated mass hierarchies, the EFT valid
below the SUSY scale may differ from the SM. For example, both Higgs
doublets might be significantly lighter than the superparticles, in
which case the considered SUSY model is matched at the scale $\MS$
with a \thdm, whose scalar potential reads 
\bea
V &=&m_{11}^2 \Phi_1^\dagger \Phi_1 ~+~ m_{22}^2 \Phi_2^\dagger \Phi_2
~-  (m_{12}^2 \,\Phi_1^\dagger\Phi_2 ~+~ {\rm h.c.})\nonumber\\
&+&\frac{\lambda_1}{2}\,(\Phi_1^\dagger\Phi_1)^2
+\,\frac{\lambda_2}{2}\,(\Phi_2^\dagger\Phi_2)^2
+\,\lambda_3\,(\Phi_1^\dagger\Phi_1)(\Phi_2^\dagger\Phi_2)
+\,\lambda_4\,(\Phi_1^\dagger\Phi_2)(\Phi_2^\dagger\Phi_1)\nonumber\\
&+& \left\{
\frac{\lambda_5}{2}\,(\Phi_1^\dagger\Phi_2)^2
+ \left[\lambda_6\,(\Phi_1^\dagger\Phi_1) \,+\, \lambda_7\,(\Phi_2^\dagger\Phi_2)
  \right]\Phi_1^\dagger\Phi_2 ~+~ {\rm h.c.} \right\}~,
\label{eq:VTHDM}
\eea
where $\Phi_1$ and $\Phi_2$ are $SU(2)$ doublets with the same
hypercharge, related to the Higgs doublets of the MSSM by $\Phi_1=
-i\sigma_2 H_1^*$ and $\Phi_2 = H_2$. We work in a basis where both of
the Higgs vevs are real {and non-negative}. While in the MSSM the
{tree-level} interactions of the Higgs doublets with quarks and
leptons are those of a ``Type-II'' \thdm~{\cite{Hall:1981bc}},
i.e.~$H_1$ couples only to down-type fermions and $H_2$ couples only
to up-type fermions, couplings of the Higgs doublets to the ``wrong''
fermion species are generated at loop level when the SUSY particles
are integrated out. As a result, the EFT valid below the SUSY scale is
in fact a ``Type-III'' \thdm, {which includes all possible
  dimension-four} Yukawa couplings that are allowed by gauge
invariance.
In the calculation of the Higgs masses, matching conditions are
computed for all of the quartic Higgs couplings {$\lambda_i$ (with
  $i=1\ldots7$),} and the loop corrections $\Delta\lambda_i$ include
contributions from diagrams involving the SUSY particles. The
couplings are then evolved either directly to the EW scale, where
masses and mixing are computed at once for the extended Higgs sector,
or to an intermediate scale $\MA$ where the heavier Higgs doublet is
integrated out, leaving again the SM as EFT. In this case the
tree-level matching condition for the quartic Higgs coupling of the SM
reads
\bea
\label{eq:lambdathdm}
\lambdaSM^{\rm tree}(\MA) &=&
\lambda_1\cos^4\beta+\lambda_2\sin^4\beta
+ 2\,(\lambda_3+\lambda_4+{\rm Re}\, \lambda_5)\sin^2\beta\cos^2\beta \nonumber
\\[2mm]
&+&4\,({\rm Re}\,\lambda_6\,\cos^2\beta +
  {\rm Re}\,\lambda_7\,\sin^2\beta)\sin\beta\cos\beta~.
\eea
The loop correction $\Delta\lambda$ includes contributions that arise
from diagrams involving the heavy doublet. The RG evolution of
$\lambdaSM$ then allows for the all-orders resummation of terms
enhanced by $\ln(\MA/M_t)$, where again we take the top mass as a
proxy for the EW scale.

Other examples of non-trivial mass {hierarchies} are given by ``Split
SUSY'' scenarios, in which the gauginos and the higgsinos are
significantly lighter than the sfermions. In this case the EFT valid
below the sfermion scale includes additional Higgs-higgsino-gaugino
couplings, which differ from the corresponding gauge couplings due to
the breaking of SUSY. These additional interactions contribute to both
the RGE(s) for the quartic Higgs coupling(s) and the corrections to
the Higgs mass(es) at the EW scale. Conversely, in scenarios where the
gluino is significantly heavier than the squarks it might be
convenient to decouple it from the full SUSY model at its own mass
scale, in order to avoid the occurrence of two-loop corrections to the
quartic Higgs couplings enhanced by gluino-squark mass ratios such as
$M_3^2/M_{\tilde Q}^2$ at the scale where the squarks are integrated
out. Scenarios in which one of the stops is much lighter than the
other sfermions have also been considered for their implications for
EW baryogenesis.

\subsection{Pre-KUTS developments}
\label{sec:EFTpreKUTS}

The EFT approach to the calculation of the Higgs mass in SUSY models
dates back to the early 1990s~\cite{Barbieri:1990ja, Espinosa:1991fc,
  Casas:1994us}. Over the years, it has also been exploited to
determine the coefficients of the logarithmic terms in the Higgs-mass
corrections at fixed order, by solving perturbatively the appropriate
systems of boundary conditions and RGEs. For example, in the case of
the MSSM the logarithmic corrections have been determined at
one~\cite{Haber:1993an}, two\,\footnote{The renormalization-scheme
  dependence of the two-loop logarithmic corrections was discussed in
  refs.~\cite{Espinosa:1999zm, Carena:2000dp}.}~\cite{Carena:1995bx,
  Carena:1995wu, Haber:1996fp}, three~\cite{Degrassi:2002fi,
  Martin:2007pg}, and even four loops and beyond~\cite{Hahn:2013ria,
  Draper:2013oza}.  However, as long as the focus was on ``natural''
scenarios with SUSY masses of a few hundred GeV, the omission of
$\ovms$ terms limited the accuracy of the EFT approach, and the effect
of the resummation of logarithmic corrections was not expected to be
important enough to justify abandoning the fixed-order calculations of
the Higgs mass in favor of a complicated EFT set-up with
higher-dimensional operators.\footnote{See, however,
  ref.~\cite{Espinosa:2001mm} for the effect of dimension-six
  operators in a scenario with only one light stop.}

Starting from the mid 2000s, however, an interest in ``unnatural''
scenarios with SUSY masses far above the TeV scale brought the EFT
approach to the calculation of the Higgs mass back into fashion. In
particular, in 2004 refs.~\cite{ArkaniHamed:2004fb, Giudice:2004tc}
pointed out that Split SUSY preserves some positive aspects of the
MSSM (such as gauge-coupling unification and a candidate for Dark
Matter) while getting rid of some negative ones (e.g., the flavor
problem). Early phenomenological studies of scenarios with light
gauginos and higgsinos involved a LL determination of the Higgs mass,
i.e., one-loop RGEs and tree-level boundary conditions. A Split-SUSY
scenario in which one of the stops is also light was studied at LL in
ref.~\cite{Carena:2008rt}. Beyond LL, the one-loop contributions of
gauginos and higgsinos to the radiative corrections to the Higgs mass
at the EW scale were computed already in 2004 in
ref.~\cite{Binger:2004nn}, and reproduced a few years later in
ref.~\cite{Bernal:2007uv}. The former paper also included the two-loop
RGE for the quartic Higgs coupling, obtained by adapting the results
valid for a general renormalizable theory from
refs.~\cite{Machacek:1983tz, Machacek:1983fi, Machacek:1984zw,
  Luo:2002ti}, while the latter included partial one-loop results for
the boundary conditions at the sfermion-mass scale. The remaining
ingredients for a NLL determination of the Higgs mass in Split SUSY
became available in 2011, when ref.~\cite{Giardino:2011aa} computed
the one-loop boundary conditions at the sfermion-mass scale for the
Higgs-higgsino-gaugino couplings, and ref.~\cite{Giudice:2011cg}
computed the one-loop boundary condition for the quartic Higgs
coupling (neglecting the effects of all Yukawa couplings except
$y_t$\,) as well as the two-loop RGEs for all of the parameters of the
Split-SUSY Lagrangian.\footnote{A number of errors and omissions in
  the one-loop boundary conditions of refs.~\cite{Giardino:2011aa,
    Giudice:2011cg} were later corrected in
  ref.~\cite{Bagnaschi:2014rsa}. Also, several errors in the two-loop
  RGEs of ref.~\cite{Giudice:2011cg} were pointed out in
  refs.~\cite{Tamarit:2012ie, Benakli:2013msa}.}  Finally, in 2013
ref.~\cite{Benakli:2013msa} obtained predictions for the Higgs mass in
a variant of Split SUSY inspired by Dirac gaugino models, in which the
Higgs-higgsino-gaugino couplings are suppressed. The paper also
highlighted the importance of decoupling the gluino at a separate
scale if its mass is in the multi-TeV range.

Abandoning naturalness as a criterion to fix the sfermion masses opens
up the scenario in which {\em all} of the BSM particles are
super-heavy, leaving the SM as an effective theory valid up to scales
well above the reach of the LHC. First evoked humorously in 2005 in an
April Fool's prank\,\footnote{The joke was apparently lost on the
  dozens of authors who cited that paper as if it had been a serious
  one.} on ``Supersplit Supersymmetry''~\cite{Fox:2005yp}, this
``high-scale SUSY'' scenario attracted renewed attention in 2009, when
ref.~\cite{Hall:2009nd} pointed out that the hypothesis of a
SUSY-breaking scale near the GUT scale singles out the relatively
narrow range of $128$\,--\,$141$~GeV for the mass of the SM-like Higgs
boson. In 2011 the predictions for the Higgs mass in the high-scale
SUSY scenario were further studied in ref.~\cite{Cabrera:2011bi},
which employed two-loop RGEs but only a partial one-loop calculation
of the boundary conditions, and in ref.~\cite{Giudice:2011cg}, which
employed a full NLL calculation. In 2012, after the {Higgs-boson}
discovery at the LHC, ref.~\cite{Degrassi:2012ry} updated the analysis
of ref.~\cite{Giudice:2011cg}, including also the dominant two-loop
corrections (in the gaugeless limit) to the relation between
$\lambdaSM$ and the Higgs mass at the EW scale.

The first phase of operations of the LHC also brought under the
spotlight scenarios where at least some of the SUSY particles have
masses of a few TeV. This was due to both the increasingly stringent
bounds from direct searches of colored SUSY particles, and the fact
that, at least in the MSSM, multi-TeV stop masses are needed to obtain
a prediction for the SM-like Higgs mass of about $125$~GeV. While this
kind of hierarchy seems too mild to endanger the convergence of the
perturbative expansion, it still implies that the uncertainty of a
fixed-order calculation of the Higgs mass arising from the uncomputed
two- and higher-loop corrections can be significantly larger than the
experimental precision of its measurement. In 2013, two papers
discussed the use of the EFT approach to improve the prediction for
the SM-like Higgs mass in the MSSM with multi-TeV stop
masses. Ref.~\cite{Hahn:2013ria}, whose main focus was the combination
of fixed-order and EFT techniques that will be discussed in detail in
section~\ref{sec:hybrid}, included a NLL resummation of the
logarithmic corrections controlled by the top Yukawa coupling in the
gaugeless limit. Ref.~\cite{Draper:2013oza} included additional NLL
effects (e.g., terms involving both the top-Yukawa and EW-gauge
couplings), plus a NNLL resummation of the top-induced corrections in
the gaugeless limit. To obtain the latter, simplified formulas for the
two-loop ${\cal O}(y_t^4 g_s^2)$ and ${\cal O}(y_t^6)$ contributions
to $\Delta\lambda$ were derived in the limit of degenerate masses of
the stops, the gluino and the heavy Higgs doublet, adapting the
results of the Higgs-mass calculation of
ref.~\cite{Espinosa:2000df}. Both analyses found that, in scenarios
with stop masses above $1$~TeV, the resummation of higher-order
logarithmic corrections leads to predictions for the SM-like Higgs
mass that could differ by as much as a few GeV from those of the
available fixed-order calculations, which included the two-loop
corrections in the gaugeless limit and only the ${\cal
  O}(\alpha_t\alpha_s^2)$ corrections at three loops. By highlighting
the impact of the resummation even in mildly hierarchical scenarios,
refs.~\cite{Hahn:2013ria, Draper:2013oza} made the case for a
systematic improvement, at the NLL level and beyond, of the EFT
calculation of the Higgs mass in SUSY models.

\subsection{Advances during KUTS}

\subsubsection{Matching the MSSM directly to the SM}
\label{sec:MSSMSMEFT}

As mentioned earlier, in scenarios where all of the SUSY particles as
well as the heavy Higgs bosons are clustered around the same high
scale the calculation of the Higgs mass can rely on existing results
for the RGEs of the parameters of the SM Lagrangian, and for the
relations between running parameters and physical observables. What is
left to compute in these scenarios is thus the matching condition at
the SUSY scale for the quartic Higgs coupling of the SM. We stress
that this requires also the calculation of the matching conditions for
the other SM couplings, at a perturbative order that depends on how
{these} couplings -- or their MSSM counterparts -- enter the matching
condition for $\lambdaSM$ (e.g., in a two-loop calculation, the
matching conditions for the couplings entering at tree level must be
computed at two loops, while those for the couplings entering only at
one loop can be computed at one loop).

During the years of the KUTS initiative, a substantial effort was
devoted to the calculation of $\Delta\lambda$ in the heavy-SUSY
scenario where the MSSM is matched directly to the SM. In 2014,
ref.~\cite{Bagnaschi:2014rsa} revised and corrected the one-loop
calculation of $\Delta\lambda$ of ref.~\cite{Giudice:2011cg}, and
computed in addition the two-loop ${\cal O}(y_t^4 g_s^2)$ contribution
for arbitrary values of all of the relevant MSSM parameters, thus
generalizing the result of ref.~\cite{Draper:2013oza} which was valid
in the limit of degenerate stop and gluino masses. It was found in
ref.~\cite{Bagnaschi:2014rsa} that the inclusion of the ${\cal
  O}(y_t^4 g_s^2)$ contribution to $\Delta\lambda$ can increase the
prediction for the Higgs mass by about $1$~GeV in scenarios with
$|X_t|\approx 2\,\MS$, where $M_tX_t$ is the off-diagonal entry in
the stop mass matrix and $\MS$ denotes an average stop mass.
In 2015, ref.~\cite{Vega:2015fna} included in $\Delta\lambda$ the
subset of one-loop contributions controlled by the bottom and tau
Yukawa couplings that are enhanced at large values of $\tan\beta$. It
also confirmed the result of ref.~\cite{Bagnaschi:2014rsa} for the
two-loop ${\cal O}(y_t^4 g_s^2)$ contribution, and corrected the
result of ref.~\cite{Draper:2013oza} for the two-loop ${\cal
  O}(y_t^6)$ contribution in the limit of degenerate stop and
heavy-Higgs masses.
In 2017, ref.~\cite{Bagnaschi:2017xid} provided the full one-loop
contributions to $\Delta\lambda$ involving the bottom and tau Yukawa
couplings, the full two-loop contributions of ${\cal O}(y_b^4 g_s^2)$,
and the full two-loop contributions that involve only the third-family
Yukawa couplings,\footnote{With an impossibly cumbersome notation,
  these contributions could be collectively denoted as ${\cal
    O}\left(\,(y_t^2+y_b^2+y_\tau^2)^3\,\right)$.} of which the
dominant ones are those of ${\cal O}(y_t^6)$. It also discussed how,
in order to avoid potentially large two-loop contributions enhanced by
$\tan\beta$, the one-loop contributions to $\Delta\lambda$ should be
expressed in terms of the bottom {Yukawa} coupling of the
MSSM. Combined with the earlier one- and two-loop results of
ref.~\cite{Bagnaschi:2014rsa}, the results of
ref.~\cite{Bagnaschi:2017xid} allow for a complete NNLL resummation of
the large logarithmic corrections to the Higgs mass in the gaugeless
limit, for arbitrary (but real) values of all of the relevant MSSM
parameters. A study of the numerical impact of the two-loop
contributions to $\Delta\lambda$ in scenarios where the stop masses
are not degenerate showed that the use of simplified formulas with an
average stop mass can lead to a rather poor approximation of the
results obtained with the exact formulas.

A first step beyond the NNLL resummation was taken in 2018, when
ref.~\cite{Harlander:2018yhj} provided the three-loop ${\cal O}(y_t^4
g_s^4)$ contribution to $\Delta\lambda$. Combined with the four-loop
${\cal O}(y_t^4 g_s^6)$ contribution to the RGE for $\lambdaSM$ and
the three-loop ${\cal O}(y_t^4 g_s^4 v^2)$ contribution to the
relation between $\lambdaSM$ and the Higgs mass, the results of
ref.~\cite{Harlander:2018yhj} allow for a N$^3$LL resummation of the
logarithmic corrections that involve the top Yukawa coupling and the
highest powers of the strong gauge coupling. The three-loop
contribution to $\Delta\lambda$ was extracted from the Higgs-mass
calculation of refs.~\cite{Harlander:2008ju, Kant:2010tf,
  Harlander:2017kuc}, which relied on a set of expansions around
various limiting cases for the SUSY masses. A study of the numerical
impact of the newly-computed contribution revealed a strong dependence
on the stop mixing term. For vanishing $X_t$ the inclusion of the
three-loop contribution shifts the Higgs mass by $20$~MeV or less, but
when the stop mixing term approaches the ``maximal'' value $|X_t| =
\sqrt6\,\MS$ the shift can reach up to $600$~MeV. However, the latter
figure should only be taken as an estimate of the possible size of the
mass shift, because in the scenario with degenerate squark and gluino
masses the three-loop calculation of ref.~\cite{Harlander:2018yhj}
involves an expansion in the stop mixing {parameter} that becomes
unreliable when $|X_t| \gtrsim \MS$.

In 2019, ref.~\cite{Bagnaschi:2019esc} took a step beyond the
gaugeless limit in the NNLL calculation of the Higgs mass, providing
the two-loop contributions to $\Delta\lambda$ that involve both the
strong and the EW gauge couplings,\footnote{Namely, the two-loop
  corrections of ${\cal O}(y_t^2\, g^2 g_s^2)$, ${\cal O}(y_t^2
  \,g^{\prime\,2} g_s^2)$, ${\cal O}(y_b^2\, g^2 g_s^2)$, ${\cal
    O}(y_b^2 \,g^{\prime\,2} g_s^2)$, ${\cal O}(g^4 g_s^2)$ and ${\cal
    O}(g^{\prime\,4} g_s^2)$.} for generic values of all the relevant
SUSY parameters. In contrast {to} the ``gaugeless'' two-loop
contributions of refs.~\cite{Draper:2013oza, Bagnaschi:2014rsa,
  Vega:2015fna, Bagnaschi:2017xid}, where $\lambdaSM(\MS)$ can be
considered vanishing at tree level -- see eq.~(\ref{eq:lambdamssm}) --
and all of the relevant two-loop diagrams can be computed at vanishing
external momenta, the calculation of the mixed QCD--EW contributions
requires also the ${\cal O}(p^2)$ parts of the two-loop self-energies
of the Higgs boson (for the field-renormalization contributions) and
of the gauge bosons (for the MSSM--SM conversion of the EW gauge
couplings). A study of the effects of the mixed QCD--EW contributions
to $\Delta\lambda$ in scenarios with multi-TeV stop masses showed that
they are largely sub-dominant with respect to the gaugeless two-loop
contributions, and their inclusion can shift the prediction for the
Higgs mass by ${\cal O}(100)$~MeV. Alternatively, it can shift the
values of the stop masses required to obtain the observed value of the
Higgs mass by ${\cal O}(100)$~GeV.

If the squark mass and mixing terms entering the one-loop
contributions to $\Delta\lambda$ are renormalized in the $\drbar$
scheme, the two-loop contributions controlled by the strong gauge
coupling contain terms that, in the limit of large gluino mass, depend
linearly or quadratically on the ratios of $M_3$ over the various
squark masses. Even for a mild hierarchy between the gluino and the
squarks, which would normally not warrant a separate decoupling scale
for the gluino, the dependence on mass ratios such as $M_3^2/M_{\tilde
  Q}^2$ may result in rather large two-loop effects. To circumvent
this problem in the ${\cal O}(y_t^4g_s^2)$ contribution to
$\Delta\lambda$, the authors of ref.~\cite{Vega:2015fna} had proposed
to renormalize the stop mass and mixing terms in an OS
scheme. However, it was subsequently pointed out in
ref.~\cite{Bahl:2019wzx} that the definition of
ref.~\cite{Vega:2015fna} for the stop mixing term leads to the
occurrence of terms enhanced by $\ln \MS/M_t$ in the ${\cal O}(y_t^6)$
contribution to $\Delta\lambda$, spoiling the underlying assumptions
of the EFT approach. As an alternative, the authors of
ref.~\cite{Bahl:2019wzx} proposed to adopt for the stop parameters the
$\mdrbar$ scheme of refs.~\cite{Harlander:2008ju, Kant:2010tf}, see
section~\ref{sec:mssm3loop}, which they extended with an appropriate
definition for $X_t$.

The calculations of the various contributions to $\Delta\lambda$
discussed so far were restricted to the case of real parameters in the
MSSM Lagrangian. In 2020, ref.~\cite{Bahl:2020tuq} extended the full
one-loop contributions, as well as the two-loop contributions in the
limit of vanishing EW-gauge and tau-Yukawa couplings, to the case of
complex parameters. The predictions for $M_h$ obtained with the full
dependence of $\Delta\lambda$ on the {CP-violating} phases were
compared with predictions in which the phase dependence is
approximated by an interpolation, observing deviations of up to $1$
GeV in scenarios with more than one non-zero phase.
Ref.~\cite{Bahl:2020tuq} also discussed the impact of including in the
determination of the Yukawa couplings corrections that are formally of
higher order with respect to the accuracy of the Higgs-mass
calculation. In particular, the inclusion of one-loop EW corrections
and two-loop ${\cal O}(\alpha_s^2)$ corrections -- the latter adapted
from {refs.~\cite{Noth:2008tw, Noth:2010jy,
    Mihaila:2010mp}} -- in the relation between the bottom Yukawa
coupling of the MSSM and its SM counterpart at the SUSY scale allows
for an improved treatment of effects that are enhanced at large
$\tan\beta$. The inclusion of three-loop ${\cal O}(\alpha_s^3)$
corrections in the relation between the top Yukawa coupling of the SM
and the top mass at the EW scale accounts for the bulk of the N$^3$LL
effects that involve the highest powers of the strong gauge coupling.
{This was found in ref.~\cite{Bahl:2020tuq} to be a
  sufficient approximation of those effects, in view of the
  uncertainty of the expansion in the stop mixing parameter that was
  employed in ref.~\cite{Harlander:2018yhj} to obtain the three-loop
  ${\cal O}(y_t^4g_s^4)$ contributions to $\Delta\lambda$.}

\subsubsection{Matching the MSSM to a \thdm}

Compared with the case in which the EFT valid at the EW scale is just
the SM, the heavy-SUSY scenario in which both Higgs doublets are
within reach of the LHC has an obvious appeal from the point of view
of phenomenology. However, the calculation of the Higgs masses in this
scenario cannot rely on the existing SM results, and the resummation
of large logarithmic corrections has so far been performed only at the
NLL order (i.e., involving one-loop corrections and two-loop RGEs).

For the MSSM with real parameters, the one-loop squark contributions
to $\Delta\lambda_i$ had already been obtained in the 1990s, see
ref.~\cite{Haber:1993an}. In 2015, ref.~\cite{Carena:2015uoe} extended
the results of ref.~\cite{Haber:1993an} to the case of complex
parameters, neglecting however some of the terms that involve the EW
gauge couplings. In 2019, the missing terms for the squark
contributions in the complex case were included in
ref.~\cite{Murphy:2019qpm}.
A full one-loop calculation of $\Delta\lambda_i$, including also the
higgsino-gaugino contributions and the contributions arising from the
$\drbar$--$\msbar$ translation of the quartic couplings, had become
available in 2009, see ref.~\cite{Gorbahn:2009pp}, and in 2018 it was
reproduced in ref.~\cite{Bahl:2018jom}. The latter paper also pointed
out that the parameter $\tan\beta$ of the \thdm\ differs from its MSSM
counterpart by a loop-induced shift.

For what concerns the two-loop contributions to $\Delta\lambda_i$, in
2015 ref.~\cite{Lee:2015uza} proposed a procedure to identify those of
${\cal O}(y_t^4g_s^2)$
from the $\tan\beta$-dependence of the various terms entering the
corresponding contribution to the quartic coupling of the SM. Shortly
thereafter, ref.~\cite{Carena:2015uoe} extended that calculation to
the case of complex parameters, and also resolved an ambiguity in the
results for $\Delta\lambda_3$, $\Delta\lambda_4$ and
$\Delta\lambda_5$, for which the procedure of ref.~\cite{Lee:2015uza}
determines only the sum.
These results were{,} however{,} restricted to
the case of degenerate soft SUSY-breaking masses for the stops. In
2020, ref.~\cite{Bahl:2020jaq} computed the ${\cal O}(y_t^4g_s^2)$
contributions to $\Delta\lambda_i$ for arbitrary complex values of all
of the relevant parameters, as well as the ${\cal O}(y_t^6)$
contributions in the limit of degenerate stop masses.

The two-loop RGEs for the \thdm\ can be extracted from the formulas of
refs.~\cite{Machacek:1983tz, Machacek:1983fi, Machacek:1984zw,
  Luo:2002ti} for a general renormalizable theory, which have been
implemented in public codes such as \SA~\cite{Staub:2013tta} and {\tt
  PyR@TE}~\cite{Lyonnet:2013dna, Lyonnet:2016xiz, Sartore:2020pkk,
  Sartore:2020gou}. In 2014, ref.~\cite{Dev:2014yca} used \SA\ to
obtain explicit results for the RGEs of the Type-II \thdm. In 2015,
{these} RGEs were revised and corrected in ref.~\cite{Lee:2015uza},
where they entered the EFT calculation of the Higgs masses under the
approximation of neglecting the RG evolution of the loop-induced
``wrong'' Yukawa couplings. In 2018, ref.~\cite{Bahl:2018jom} used
\SA\ to obtain RGEs for the Type-III \thdm, neglecting all Yukawa
couplings except those of the two doublets to top quarks. However,
later in 2018 refs.~\cite{Bednyakov:2018cmx, Schienbein:2018fsw} found
that the implementation in \SA\ and {\tt PyR@TE} of the general
results of refs.~\cite{Machacek:1983tz, Machacek:1983fi,
  Machacek:1984zw, Luo:2002ti} was not appropriate for models, such as
the \thdm, that feature mixing in the scalar sector.
{Ref.~\cite{Schienbein:2018fsw}} provided the {correct} two-loop RGEs
for the Type-III \thdm, and ref.~\cite{Bednyakov:2018cmx} computed the
three-loop contributions that involve only $\lambda_i$ to the RGEs for
the masses and quartic couplings of the Higgs doublets. {Also in late
  2018, the correct two-loop RGEs for the Type-III \thdm\ were
  independently derived in ref.~\cite{Oredsson:2018yho} and made
  available in the public code {\tt 2HDME}~\cite{Oredsson:2018vio}.}
Meanwhile, the three-loop RGEs for the gauge and Yukawa couplings of
the Type-III \thdm\ had been presented in 2017 in
ref.~\cite{Herren:2017uxn}.

After the quartic couplings are evolved down to the EW scale, they can
be used in conjuction with $\tan\beta$ and an input mass parameter --
usually taken as either $\MA$ or $\MHp$ -- to compute masses and
mixing angles in the Higgs sector.\footnote{Conversely, when the
  renormalization of the \thdm\ is studied independently of an
  underlying SUSY theory, the Higgs masses and mixing angles are
  usually treated as input parameters, see
  e.g.~refs.~\cite{Kanemura:2004mg, Krause:2016oke, Altenkamp:2017ldc,
    Kanemura:2017wtm}.}  In the approximation of neglecting the
``wrong'' Yukawa couplings, so that the relevant EFT is a type-II
\thdm, the one-loop contributions to the Higgs mass matrix from
fermions and gauge bosons are the same as in the MSSM and can be found
in the literature, see e.g.~refs.~\cite{Chankowski:1991md,
  Brignole:1991wp, Brignole:1992uf, Chankowski:1992er,
  Dabelstein:1994hb, Pierce:1996zz}. In contrast, the Higgs
contributions must be computed in terms of the quartic Higgs couplings
of the \thdm. In 2015, ref.~\cite{Bagnaschi:2015pwa} used \SA\ to
obtain the full one-loop corrections to the Higgs mass matrix for the
type-II \thdm. Two-loop corrections to the Higgs mass matrix at
vanishing external momentum are also generated by \SA\ in the
gaugeless limit, which in the \thdm\ includes also the contributions
of the quartic Higgs couplings. These corrections were discussed in
2017 in ref.~\cite{Braathen:2017izn}, but they have not yet been
applied to the case in which the \thdm\ is treated as the EFT of the
MSSM with heavy SUSY particles.

If there is a substantial gap between the masses of the heavy Higgs
bosons and the EW scale, the radiative corrections to the Higgs mass
matrix computed within the \thdm\ contain logarithmic terms involving
the ratio of the two scales, which might be large enough to require
resummation. To this effect, the heavier Higgs doublet is integrated
out at the scale $Q=\MA$, leaving the SM as EFT. The tree-level
matching condition for the quartic Higgs coupling $\lambdaSM$ is given
in eq.~(\ref{eq:lambdathdm}), and the threshold correction
$\Delta\lambda$ from one-loop diagrams involving the heavier Higgs
doublet was computed in ref.~\cite{Bahl:2018jom}.
The contribution to $\Delta\lambda$ from two-loop diagrams involving
the heavier Higgs doublet and top quarks was subsequently computed in
ref.~\cite{Bahl:2020jaq}.
After the RG evolution of the quartic coupling down to the EW scale,
typically at $Q=M_t$, the mass of the lighter Higgs boson can be
computed including the known SM results for the radiative corrections.

Refs.~\cite{Lee:2015uza, Bahl:2018jom, Murphy:2019qpm} also proposed a
procedure that resums the corrections enhanced by $\ln(\MA/M_t)$ while
still retaining information on the corrections to the masses of the
heavy Higgs bosons and on their mixing with the SM-like Higgs. While
the proposals of the three papers differ in minor details, their
central idea consists in computing the higher-order logarithmic
contributions to $\lambdaSM$ using the SM as EFT, and inserting them
in the full Higgs mass matrix of the \thdm, which is then diagonalized
to determine masses and mixing at once. It was shown in
ref.~\cite{Murphy:2019qpm} that this procedure provides a satisfactory
interpolation between the pure-\thdm\ calculation, which is
appropriate when both Higgs doublets are at the EW scale, and the
two-step calculation where the heavier Higgs doublet is integrated out
at an intermediate scale.

The availability of proper EFT calculations in the setup with heavy
SUSY particles and a light \thdm\ allowed for an assessment of the
benchmark scenarios used by ATLAS and CMS to interpret their Higgs
searches in the low-$\tan\beta$ region of the MSSM, where ultra-heavy
stops are required to obtain a prediction for the SM-like Higgs mass
around $125$~GeV. In the ``low-tb-high'' scenario proposed in
ref.~\cite{Bagnaschi:2015hka}, the Higgs masses had been computed with
an early version of \FH\ that performed the resummation of logarithmic
corrections by decoupling the SUSY particles and the heavy Higgs
doublet at the same high scale. It was shown in
refs.~\cite{Lee:2015uza, Bahl:2018jom} that, at low values of
$\tan\beta$ and $\MA$, this approximation can overestimate the
prediction for the SM-like Higgs mass by as much as $8$~GeV. In 2019,
a new benchmark scenario for the low-$\tan\beta$ region, based on the
EFT calculation of ref.~\cite{Bahl:2018jom}, was eventually proposed
in ref.~\cite{Bahl:2019ago}.

In some instances, rather than interpreting their Higgs searches in a
specific MSSM scenario, ATLAS and CMS relied on a simplifying
approach, the so-called ``hMSSM''~\cite{Djouadi:2013vqa,
  Maiani:2013hud, Djouadi:2013uqa, Djouadi:2015jea}. This
approximation assumes that the Higgs sector is CP conserving, that all
SUSY particles are too heavy to affect Higgs production and decays,
that any non-decoupling SUSY corrections to the Higgs couplings are
negligible, and that the radiative corrections to the elements other
than (2,2) in the mass matrix of the neutral CP-even components of
$H_1$ and $H_2$ are also negligible, i.e.~$\Delta {\cal M}_{1j}^2
\approx 0$ for $j=1,2$, see eq.~(\ref{eq:gamma}). In {this} case, the
remaining radiative correction $\Delta {\cal M}_{22}^2$ can be
expressed in terms of the parameters that determine the tree-level
mass matrix (i.e.~$\tan\beta$, $\MZ$ and $\MA$) plus the smaller
eigenvalue $M_h$, which is treated as an input and identified with the
mass of the observed Higgs boson. Consequently, the larger eigenvalue
$M_H$ and the angle $\alpha$ that diagonalizes the mass matrix can in
turn be expressed in terms of just those four input parameters, of
which only $\tan\beta$ and $\MA$ are unknown.
While the hMSSM approach does bring some benefits -- namely, the
limited number of input parameters, and the fact that the measured
value of the Higgs mass is one of them -- its predictions for the
Higgs properties can be mapped only to regions of the MSSM parameter
space in which the approximations of neglecting the $\Delta {\cal
  M}_{1j}^2$ corrections and the SUSY corrections to the Higgs
couplings are justified. Indeed, in refs.~\cite{Lee:2015uza,
  Bahl:2019ago} the comparison between the EFT calculations and the
hMSSM approach found regions of the MSSM parameter space where the
predictions for $\alpha$, which determines the couplings of the
CP-even Higgs bosons, can differ by more than $10\%$. Moreover, the
EFT calculations show that, for low values of $\tan\beta$ and $\MA$, a
prediction for the lighter Higgs mass of about $125$~GeV may require
stop masses as large as the GUT scale, putting into question the
validity of the MSSM as the underlying high-energy theory.

\subsubsection{Split-SUSY scenarios for the MSSM}

In the original Split-SUSY scenario of refs.~\cite{ArkaniHamed:2004fb,
  Giudice:2004tc}, where the heavy Higgs doublet is integrated out at
the same scale as the sfermions, the one-loop threshold corrections
and two-loop RGEs necessary to the NLL resummation of the large
logarithmic corrections had become available by the beginning of the
KUTS initiative, see section~\ref{sec:EFTpreKUTS}. In 2018,
ref.~\cite{Bahl:2018jom} included in the threshold corrections to the
Higgs-higgsino-gaugino couplings terms suppressed by $X_t^2/\MS^2$
that had been neglected\,\footnote{We recall that $X_t = A_t -
  \mu\cot\beta$, and that in Split SUSY the soft SUSY-breaking
  parameter $A_t$ is suppressed by the same symmetry that keeps $\mu$
  and the gaugino masses smaller than the scalar
  masses~\cite{ArkaniHamed:2004fb, Giudice:2004tc}.} in
refs.~\cite{Giardino:2011aa, Bagnaschi:2014rsa}.
Going beyond NLL, the two-loop threshold corrections to $\lambdaSM$
obtained in refs.~\cite{Bagnaschi:2014rsa, Bagnaschi:2017xid,
  Bagnaschi:2019esc} can be trivially adapted to this scenario by
taking the limits of vanishing gluino and higgsino masses (i.e.,
$M_3\rightarrow 0$ and $\mu \rightarrow 0$). However, a full NNLL
resummation of the logarithmic corrections in Split SUSY will require
not only the remaining two-loop corrections controlled by the EW
couplings, but also the three-loop part of the RGEs.

A Split-SUSY scenario that came under attention in the course of the
KUTS initiative is the one in which both Higgs doublets are
significantly lighter than the sfermions. In this case the EFT valid
below the sfermion mass scale is a \thdm\ augmented with the gauginos
and the higgsinos.  In 2014, one-loop RGEs for this EFT were presented
in ref.~\cite{Cheung:2014hya}, and in 2015
ref.~\cite{Bagnaschi:2015pwa} used \SA\ to obtain the 2-loop RGEs. In
2018, ref.~\cite{Bahl:2018jom} also used \SA\ to include in the RGEs
the effects of the ``wrong'' Yukawa couplings of the two Higgs
doublets with SM fermions and with higgsinos and gauginos. These
interactions, absent at tree level, are generated at one loop when the
sfermions are integrated out of the MSSM, but in Split SUSY they are
suppressed by ratios of the higgsino and gaugino masses over the
sfermion masses.  The EFT calculation of the Higgs masses in
ref.~\cite{Bahl:2018jom} employed independent decoupling scales for
the heavy Higgs doublet, for the EW gauginos and the higgsinos, and
for the gluino. One-loop threshold corrections to the effective
couplings were computed at each of these scales, under the
approximation of degenerate masses for the higgsinos and the EW
gauginos. This allowed for a NLL resummation of the logarithmic
corrections in all of the considered EFT towers.

It was shown in refs.~\cite{Lee:2015uza, Bagnaschi:2015hka} that, in
the Split-SUSY scenario with a light \thdm, an acceptable prediction
for the lighter Higgs mass at low $\tan\beta$ can be obtained for
lower values of the stop masses than in the scenario with a light
\thdm\ where all SUSY particles are decoupled at the high
scale. Finally, a benchmark scenario for Higgs searches in the MSSM
setup where the sfermions are very heavy while both Higgs doublets,
the higgsinos and the gauginos are at or below the TeV scale was
proposed in ref.~\cite{Bahl:2019ago}, relying on the NLL EFT
calculation of ref.~\cite{Bahl:2018jom}.

\subsubsection{Beyond the MSSM}

Compared with the case of the MSSM, the effort devoted so far to the
EFT calculation of the Higgs masses in non-minimal SUSY extensions of
the SM with hierarchical mass spectra has been relatively limited. In
view of the number of different models that could in principle be
studied, a sensible approach is the one already discussed in section
\ref{sec:prospectsFO} for the FO calculations: compute all of the
necessary corrections only once for a general theory, and then
specialize the results to the model under consideration. As mentioned
earlier, the two-loop RGEs for a general theory have been computed
long ago in refs.~\cite{Machacek:1983tz, Machacek:1983fi,
  Machacek:1984zw, Luo:2002ti}, and they are available in public codes
such as \SA\ and {\tt PyR@TE}. The calculation of the physical Higgs
mass(es) from the parameters of the EFT Lagrangian at the low scale
can rely on the existing SM results or, if the relevant EFT is an
extension of the SM, on the general Higgs-mass calculation implemented
in \SA. In addition, a NLL resummation of the large logarithmic
corrections requires the calculation of one-loop matching conditions
for the Higgs couplings between a general high-energy theory and a
general renormalizable EFT from which the heavy states have been
integrated out.

In 2018, the general one-loop matching conditions were computed
independently in refs.~\cite{Braathen:2018htl, Gabelmann:2018axh},
under the restriction that the high-energy theory does not contain
heavy gauge bosons. In particular, ref.~\cite{Braathen:2018htl}
discussed different choices that can be made in the renormalization of
the masses, couplings and mixing angles entering the {tree}-level part
of the matching conditions, as well as several subtleties concerning
the treatment of tadpoles, gauge dependence and infrared
divergences. As an application of the general formulas,
ref.~\cite{Braathen:2018htl} reproduced the MSSM results of
ref.~\cite{Bagnaschi:2014rsa}, and obtained novel results for the
one-loop matching condition for the quartic Higgs coupling in the
scenario where the high-energy theory is the MDGSSM and the EFT is the
SM plus higgsinos. Ref.~\cite{Gabelmann:2018axh} focused instead on
the implementation of the general one-loop matching conditions in the
package \SA. In addition to reproducing the results of
ref.~\cite{Bagnaschi:2014rsa} for the MSSM scenario with one light
Higgs doublet and those of ref.~\cite{Haber:1993an} for the MSSM
scenario with two light Higgs doublets, ref.~\cite{Gabelmann:2018axh}
provided a novel {NLL} calculation of the Higgs mass in the scenario
where the high-energy theory is the NMSSM and the EFT is the SM. In
2019, a follow-up paper~\cite{Gabelmann:2019jvz} employed \SA\ to
study the Split-SUSY scenario where the high-energy theory is the
GNMSSM and the EFT is the SM plus higgsinos, gauginos, and all of the
components of the singlet superfield.

Other EFT calculations of the Higgs masses performed in the years of
the KUTS initiative focused on SUSY models with Dirac gauginos. In
2018, ref.~\cite{Benakli:2018vqz} studied the conditions for ``Higgs
alignment'' -- i.e., one of the Higgs bosons being SM-like
independently of the masses of the others -- in Dirac-gaugino models
with an extended SUSY in the gauge sector. In the process,
ref.~\cite{Benakli:2018vqz} provided a novel EFT calculation of the
Higgs mass at the {NLL} level in the scenario where the high-energy
theory is the MDGSSM and the EFT is a type-II \thdm\ augmented with
Dirac bino and wino. It also considered the scenario where the
high-energy theory is the MRSSM and the EFT is just a type-II
\thdm. In both cases the two-loop RGEs were obtained with \SA. The
one-loop threshold corrections to the quartic Higgs couplings at the
matching scale were computed directly, although in the MRSSM case only
the contributions from loops involving the adjoint scalars were
included.

In 2019, ref.~\cite{Liu:2019hqt} computed the two-loop ${\cal
  O}(y_t^4g_s^2)$ corrections to the quartic coupling of the SM-like
Higgs boson that arise when a Dirac gluino and its associated octet
scalar (the sgluon) are integrated out of the theory at the respective
mass scales. In a rare departure from the gaugeless limit at two
loops, ref.~\cite{Liu:2019hqt} also obtained the threshold corrections
of ${\cal O}(y_t^4g^2)$ and ${\cal O}(g^6)$ that arise from diagrams
involving a Dirac wino and its adjoint scalar, in the ``Split Dirac
SUSY'' model\,\footnote{Note that in this model there is no ${\cal
    O}(g^2)$ contribution to the quartic Higgs coupling at tree level,
  therefore all of the relevant two-loop diagrams can be obtained from
  the effective potential.} of ref.~\cite{Fox:2014moa}.  A numerical
study showed that, in the scenarios with vanishing $X_t$ considered in
the paper, the shift induced by all of these two-loop corrections on
the prediction for the Higgs mass is small, typically below
$100$~MeV. Indeed, by explicitly decoupling the gluino from the EFT
one avoids the occurrence of corrections to the quartic Higgs coupling
enhanced by $M_3^2/M_{\tilde Q}^2$\,.

\subsubsection{Public codes for the EFT calculation of the Higgs masses in SUSY models}
\label{sec:EFTcodes}

In the course of the KUTS initiative, the EFT calculations of the
Higgs masses discussed in the previous sections have been implemented
in a number of public codes, which we list briefly here (detailed
descriptions and complete lists of references can be found in the
appendix).

\begin{itemize}
\item
\SH, based on ref.~\cite{Vega:2015fna}, provides a full NLL and
``gaugeless'' NNLL calculation of the Higgs mass in the MSSM scenario
where all SUSY particles and the heavy Higgs doublet are integrated
out at the same scale, as well as an NLL calculation in the original
Split-SUSY scenario with only one light Higgs doublet.
\item
\ME\ implements the calculation of ref.~\cite{Draper:2013oza} for the
MSSM scenario with heavy SUSY particles and only one light Higgs
doublet, and the calculation of ref.~\cite{Lee:2015uza} for the
scenario with two light Higgs doublets. In both scenarios, the code
also allows for light higgsinos and EW gauginos,\footnote{Since the
  gaugino masses feed into each other via two-loop corrections, this
  scenario involves additional fine tuning if the gap between the
  masses of gluino and EW gauginos is larger than a two-loop factor.}
under the approximation that the effects of the light SUSY particles
are included only in the one-loop RGEs, without distinguishing the
effective higgs-higgsino-gaugino couplings from the gauge couplings.
 \item
\FH\ provides, in addition to the ``hybrid'' calculation that will be
described in section~\ref{sec:hybrid}, the option of a pure EFT
calculation of the MSSM Higgs masses. For the heavy-SUSY scenario
where the EFT is the SM, it implements a full NLL and ``gaugeless''
NNLL calculation, relying on the one- and two-loop threshold
corrections
with full dependence on the {CP-violating} phases from
ref.~\cite{Bahl:2020tuq}.
For the scenario where the EFT is a \thdm,
it implements the NLL calculation of ref.~\cite{Bahl:2018jom}, which
covers eight different EFT towers depending on the relative position
of the thresholds for the heavy Higgs doublet, the higgsinos and EW
gauginos, and the gluino.
The matching conditions for the quartic Higgs couplings of the
\thdm\ also include the two-loop ${\cal O}(y_t^4g_s^2)$ and ${\cal
  O}(y_t^6)$ contributions from ref.~\cite{Bahl:2020jaq}.

\item
\FS\ contains several modules for the EFT calculation of the MSSM
Higgs masses. For the simplest heavy-SUSY scenario where the EFT valid
below the matching scale is the SM, the module \HS\ implements a full
NLL and ``gaugeless'' NNLL calculation that relies on the one- and
two-loop corrections from refs.~\cite{Bagnaschi:2014rsa,
  Bagnaschi:2017xid}, plus a partial N$^3$LL calculation that relies
on the three-loop corrections from ref.~\cite{Harlander:2018yhj},
provided by the code \HIM.  For the Split-SUSY scenario with only one
light Higgs doublet, the module {\tt SplitMSSM} implements one- and
two-loop corrections from refs.~\cite{Giudice:2011cg,
  Bagnaschi:2014rsa}. When the EFT valid below the matching scale is a
\thdm, the code contains separate modules for the scenarios with SUSY
particles all heavy, with light higgsinos, and with light higgsinos
and gauginos. They allow for the inclusion of either the dominant
one-loop corrections of ref.~\cite{Haber:1993an} or the full
corrections of ref.~\cite{Gorbahn:2009pp}, plus the two-loop ${\cal
  O}(y_t^4g_s^2)$ corrections in the approximation of
ref.~\cite{Lee:2015uza}.  {Note that \FS\ includes also a
  module, named \FE, that allows for the automated ``hybrid'' NLL
  calculation of the Higgs mass in any SUSY model matched directly to
  the SM.  This will be described in section~\ref{sec:hybrid}.}

\item
\SA\ allows for {automated} EFT calculations of the Higgs masses at
the NLL level, relying on the general one-loop matching conditions of
{refs.~\cite{Braathen:2018htl, Gabelmann:2018axh}}. The package comes
with model files for several heavy-SUSY scenarios. {In the case where
  the theory valid above the matching scale is the MSSM, these cover
  six different EFT towers depending on the relative position of the
  thresholds for the heavy Higgs doublet and for the SUSY fermions (in
  contrast to \FH, the gluino is always decoupled at the same scale as
  the EW gauginos). There are also model files for the NMSSM matched
  either directly to the SM, or to the SM plus higgsinos, gauginos,
  singlet and singlino.  Finally, \SA\ allows for automated ``hybrid''
  Higgs-mass calculations similar to those in \FE, but the accuracy of
  the resummation of the large logarithmic effects is only LL in this
  case.}

\end{itemize}

While the EFT calculations implemented in the codes listed above
differ from each other in several aspects -- e.g., in the classes of
threshold corrections that they include at the SUSY scale, and in the
renormalization scheme adopted for some of the SUSY parameters --
their predictions for the Higgs masses are generally in good agreement
with each other in the appropriate limits. For the heavy-SUSY scenario
where the high-energy theory is the MSSM and the EFT is the SM, a
comparison between \SH\ and \HS\ was presented in
ref.~\cite{Athron:2016fuq}, a comparison between \SH\ and \FH\ was
presented in ref.~\cite{Bahl:2017aev}, and a comparison between
\SH\ and the relevant module of \SA\ was presented in
ref.~\cite{Gabelmann:2018axh}. For the scenario where the high-energy
theory is the MSSM and the EFT is a \thdm, a comparison between
\ME\ and the relevant module of \FS\ was presented in
ref.~\cite{Athron:2017fvs}, a comparison between \ME\ and \FH\ was
presented in ref.~\cite{Bahl:2018jom}, and a comparison between
\ME\ and the relevant module of \SA\ was presented in
ref.~\cite{Gabelmann:2018axh}.

\subsection{Prospects}

As discussed earlier in this section, full EFT calculations of the
Higgs masses at the NLL level -- i.e., involving one-loop threshold
corrections and two-loop RGEs -- are already available for a variety
of SUSY models and of mass hierarchies within {these} models. For any
other model (or mass hierarchy, e.g.~one light stop) that should come
under attention in the future, the necessary ingredients for the {NLL}
calculation of the Higgs masses can in principle be obtained
``automatically'' from general formulas, with the current limitation
that the high-energy theory must not involve heavy gauge bosons.
In contrast, calculations beyond NLL have so far been
performed only for the simplest heavy-SUSY scenario where the MSSM is
matched directly to the SM, and they are restricted to subsets of
contributions: at NNLL they neglect most of the effects that involve
the EW gauge couplings, while at N$^3$LL they account only for the
effects that involve the top Yukawa coupling combined with the highest
powers of the strong gauge coupling.

In heavy-SUSY scenarios where the high-energy theory is matched
directly to the SM, a full NNLL calculation of the Higgs mass should
be well within reach. Indeed, in these scenarios one can rely on the
existing SM results for the three-loop RGEs and for the two-loop
relations between Lagrangian parameters and physical masses at the EW
scale, and all is left to compute is the full two-loop matching
condition for the quartic Higgs coupling at the SUSY scale. In
contrast {to} the case of the FO calculations described in
section~\ref{sec:fo}, all of the relevant two-loop diagrams can be
computed in the limit of unbroken EW symmetry and through an expansion
in the external momentum, and should not present particular
difficulties. The most economic approach could again be the one of
computing the two-loop matching condition only once for a general
high-energy theory, and then adapting the result to the particular
SUSY model under consideration. However, some additional work will
still be required, on a case-by-case basis, to establish the most
convenient renormalization scheme for the Lagrangian parameters, also
in order to avoid the occurrence of spuriously large corrections such
as, e.g., those enhanced by powers of $\tan\beta$ or by
$M_3^2/M_{\tilde Q}^2$\,.

In scenarios where the EFT valid below the SUSY scale is an extension
of the SM, an NNLL calculation of the Higgs mass(es) requires the
three-loop RGEs for the parameters of the EFT. Lacking those, the
inclusion of two-loop matching conditions for the couplings of the EFT
can be considered an improvement of the calculation only if the
hierarchy between the SUSY and EW scales is not so large that the
resummation of higher-order logarithmic effects is really
mandatory. The computation of the two-loop matching conditions for the
quartic Higgs coupling(s) should not involve additional conceptual
difficulties with respect to the case in which the EFT is just the
SM. However, if the EFT contains singlets or triplets of $SU(2)$,
there are also cubic interactions for which the computation of the
two-loop matching conditions is required.

For what concerns the N$^3$LL calculation of the Higgs mass in the
scenario where the MSSM is matched to the SM, a generalization of the
three-loop ${\cal O}(y_t^4 g_s^4)$ matching condition for the quartic
Higgs coupling of ref.~\cite{Harlander:2018yhj} to arbitrary values of
$X_t/\MS$ could be envisaged. In view of the modest impact of this
presumably dominant correction, however, it is doubtful that the
effort necessary to compute additional three-loop corrections and
four-loop RGEs -- in this scenario or even in more complicated ones --
will be considered justified in the short term.  We stress here that
the smallness of the gain that results from going to higher
perturbative orders in the calculation is in fact a desirable feature
of the EFT approach, in which the dominant effects are accounted for
by the evolution of the parameters between the SUSY scale and the EW
scale. For example, the large cancellations that had been noticed
between the ${\cal O}(\alpha_t\alpha_s^2)$, ${\cal
  O}(\alpha_t^2\alpha_s)$ and ${\cal O}(\alpha_t^3)$ corrections in
the FO calculation of the Higgs mass are already incorporated in the
RGEs. Consequently, one can speculate that the omission of the
three-loop ${\cal O}(y_t^6 g_s^2)$ and ${\cal O}(y_t^8)$ contributions
to the matching condition in the EFT calculation has a far less
dramatic impact than the omission of the corresponding terms in the FO
calculation.

Another possible direction of improvement, aimed at increasing the
accuracy of the EFT calculation of the Higgs masses in scenarios where
the hierarchy between the SUSY scale and the EW scale is mild, could
be the inclusion of terms suppressed by $v^2/\MS^2$\,. As mentioned in
section~\ref{sec:hierarchies}, these terms can be mapped to the effect
of dimension-six operators in the EFT Lagrangian, and they are
neglected when the high-energy theory is matched to a renormalizable
EFT in the unbroken phase of the EW symmetry. For example, in the EFT
approach the one-loop corrections to the Higgs mass proportional to
$y_t^2\,M_t^4/\MS^2$ arise from the inclusion in the scalar potential
of the term $c_6\,|H|^6$, where $c_6$ is a Wilson coefficient that
scales like $\MS^{-2}$, induced when the stops are integrated out of
the high-energy theory. In general, multiple dimension-six operators
contribute to the Higgs mass. In recent years, several
papers~\cite{Henning:2014gca, Huo:2015nka, Drozd:2015rsp,
  Wells:2017vla, Kramer:2019fwz} provided the one-loop contributions
to the Wilson coefficients of the relevant dimension-six operators
that arise when the squarks are integrated out of the MSSM, employing
a technique known as ``covariant derivative
expansion''~\cite{Gaillard:1985uh, Chan:1986jq, Cheyette:1987qz,
  Henning:2014wua}.\footnote{Ref.~\cite{Wells:2017vla} also showed how
  this technique can be used to obtain the one-loop contributions to
  the coefficients of the renormalizable operators of the EFT.}  In
addition, ref.~\cite{Bagnaschi:2017xid} presented a direct computation
of the two-loop ${\cal O}(y_t^6 g_s^2)$ contribution to $c_6$ in the
MSSM, relying on the effective-potential approach. However,
ref.~\cite{Bagnaschi:2017xid} also found that, for the values of the
stop masses that lead to a Higgs-mass prediction in the vicinity of
$125$~GeV, the dominant one- and two-loop ${\cal O}(v^2/\MS^2)$
effects arising from dimension-six operators are already largely
suppressed. Moreover, in the context of the Higgs-mass calculation,
the usefulness of a full inclusion of the dimension-six operators in
the EFT setup -- with matching conditions computed at the SUSY scale,
and subsequent RG evolution to the EW scale -- can be questioned on
general grounds, as the logarithmic enhancement of the higher-order
corrections that are thus resummed is always trumped by their
power-like suppression.

An alternative approach to the inclusion of the ${\cal O}(v^2/\MS^2)$
effects stems from the consideration that these effects are
automatically accounted for in the FO calculation of the Higgs masses,
which is usually performed in the broken phase of the EW theory and
does not necessarily involve any expansion in $v^2$. In order to cover
the whole spectrum of scenarios -- from those with a mild hierarchy
between the SUSY and EW scales, where the ${\cal O}(v^2/\MS^2)$
effects can be relevant, to those with a strong hierarchy, where the
resummation of large logarithmic effects is required -- it is
conceivable to combine a FO calculation of the former effects with an
EFT calculation of the latter. Indeed, a number of such ``hybrid''
approaches to the Higgs-mass calculation in SUSY models have been
proposed in the course of the KUTS initiative, as will be reviewed in
the next section.

\newpage
\section{Hybrid calculations}
\label{sec:hybrid}
\subsection{Motivation}

As discussed in the previous sections, EFT calculations of the Higgs
masses account, to all orders in the perturbative expansion, for the
logarithmic corrections that involve the ratio between different mass
scales (e.g., the SUSY scale $\MS$ and the EW scale $v$), and are
therefore suited to scenarios with {large} hierarchies between
scales. However, they neglect contributions to the Higgs masses
suppressed by powers of the ratio of scales, e.g.~$v^2/\MS^2\,$,
unless higher-dimensional operators are included in the EFT, at the
price of a significant increase in the complexity of the
calculation. In contrast, FO calculations of the Higgs masses do not
necessarily involve any expansion in ratios of scales, hence they can
be applied without loss of accuracy to scenarios with new physics near
the EW scale. However, they are unsuited to scenarios with {large}
hierarchies between scales, because the uncomputed higher-order
corrections involve higher powers of the logarithm of their ratio.

A novel approach to the determination of the Higgs masses consists in
combining the resummation of the logarithmic effects from the EFT
calculations with the complete treatment of the contributions
suppressed by powers of $v^2/\MS^2$ from the FO calculations. The aim
of this ``hybrid'' approach is to obtain a single calculation that can
be applied to the whole spectrum of SUSY scenarios, from those with
light SUSY particles to those with a {large} hierarchy between the
SUSY and EW scales, covering also the intermediary region with SUSY
masses of $0.5 \!-\!2$~TeV. Indeed, while the latter region is of
particular interest in view of LHC phenomenology, it sits at the
border of the domains of applicability of the FO and EFT calculations,
where it is not immediately obvious which, if either, of the two
approaches can be considered sufficiently accurate. It should be
stressed that there is no unique way to realize such combination of
the FO and EFT calculations, and that the proverbial Devil resides in
the detail: the contributions to the Higgs masses that are included in
both calculations must be subtracted to avoid double counting, and
possible differences in the definition of the parameters entering the
two calculations must be accounted for, all in a way that does not
spoil the resummation of higher-order logarithmic effects.

Since late 2013, three distinct methods for combining the FO and EFT
calculations in a hybrid approach have been proposed, and they
have been thoroughly discussed during the KUTS meetings. In the
following we summarize their main features.

\subsection{The hybrid approach of \FH}
\label{sec:hybridFH}
  
A version of \FH\ combining a two-loop FO calculation of the MSSM
Higgs masses with a resummation of higher-order logarithmic
corrections was first presented in 2013 in
ref.~\cite{Hahn:2013ria}. The hybrid approach of \FH\ was subsequently
refined in refs.~\cite{Bahl:2016brp, Bahl:2017aev, Bahl:2018jom,
  Bahl:2018ykj, Bahl:2019hmm, Bahl:2020tuq, Bahl:2020mjy}, and in
ref.~\cite{Bahl:2018zmf} it was used in the production of benchmark
scenarios for MSSM Higgs searches at the LHC. The main idea consists
in supplementing the FO corrections to the Higgs mass matrix, see
eq.~(\ref{eq:gamma}), with higher-order logarithmic terms computed
numerically in the EFT approach. In MSSM scenarios where the mass of
the heavier Higgs doublet is comparable to the SUSY masses, this
amounts to the replacement:
\beq
\label{eq:hybridFH}
\Delta {\cal M}_{hh}(p^2) ~~\longrightarrow~~
\Delta {\cal M}_{hh}(p^2) ~+~ 2\,\lambdaSM(M_t)\,v^2 ~-\,
\left[\Delta {\cal M}_{hh}(p^2)\right]_{{\rm d.c.}}~,
\eeq
where $\Delta {\cal M}_{hh}(p^2)$ is the FO correction to the $hh$
element of the mass matrix, in the basis of tree-level mass
eigenstates $(h,H)$, which \FH\ computes in full at one loop and in
the gaugeless limit at two loops; $\lambdaSM(M_t)$ is a SM-like quartic
Higgs coupling obtained in the EFT approach through a numerical
solution of the appropriate RGEs, starting from boundary conditions at
the SUSY scale; the subtraction term $\left[\Delta {\cal
    M}_{hh}(p^2)\right]_{{\rm d.c.}}$ is meant to avoid double
counting, removing the contributions that are present in both the FO
result and the EFT result. In the latest implementation of the hybrid
approach of \FH, see ref.~\cite{Bahl:2019hmm}, the subtraction term
contains the tree-level Higgs mass plus the ${\cal O}(v^2)$ terms of
an expansion of the SUSY contributions to $\Delta {\cal M}_{hh}(p^2)$
in powers of $v^2$ (i.e., the terms that do not vanish in the limit
$v^2/\MS^2 \rightarrow 0$). If the one-loop stop contributions to
$\Delta {\cal M}_{hh}(p^2)$ are expressed in terms of OS-renormalized
stop masses and mixing, the two-loop contributions of the
corresponding counterterms are not included in the subtraction term
(this will be further discussed below). Once the mass matrix has been
improved with the inclusion of the higher-order logarithmic terms, the
pole masses of the MSSM Higgs bosons are numerically determined from
the zeroes of the inverse-propagator matrix as in the regular FO
calculation, see eq.~(\ref{eq:det}).

In the original implementation of the hybrid approach in \FH,
ref.~\cite{Hahn:2013ria}, the resummation of higher-order logarithmic
effects was performed only in the scenario where the EFT valid below
the SUSY scale is the SM, and it included only the LL and NLL
contributions controlled by the top Yukawa coupling and the strong
gauge coupling. In 2016, ref.\cite{Bahl:2016brp} extended the hybrid
approach by including also the LL and NLL contributions controlled by
the EW gauge couplings, as well as the NNLL contributions controlled by
the top Yukawa coupling and the strong gauge
coupling. Ref.\cite{Bahl:2016brp} also adapted the hybrid approach to
split-SUSY scenarios in which the EW gauginos and the higgsinos, and
possibly also the gluino, are integrated out at intermediate scales
between the SUSY and EW scales. In 2017, ref.~\cite{Bahl:2017aev}
identified some spurious higher-order logarithmic contributions that
are included in the hybrid result for the lighter Higgs mass when the
poles of the inverse-propagator matrix, eq.~(\ref{eq:det}), are
determined numerically. In principle, these spurious contributions
would cancel out order by order in a complete FO calculation, and they
can be removed by truncating the determination of the propagator poles
at the perturbative order covered by the available FO calculation (in
\FH, this means full one-loop and gaugeless two-loop). It was found in
ref.~\cite{Bahl:2017aev} that this modification can shift the
prediction for the lighter Higgs mass by about $1.5$~GeV when $\MS$ is
of ${\cal O}(10~{\rm TeV})$.

In 2018, refs.~\cite{Bahl:2018jom, Bahl:2018ykj} extended the hybrid
approach of \FH\ to MSSM scenarios in which both Higgs doublets are
much lighter than the SUSY scale. In this case the EFT valid below the
SUSY scale is a \thdm, and the resummation of the logarithmic effects
is performed at NLL, with independent decoupling scales for the gluino
and for higgsinos and EW gauginos. It was pointed out in
ref.~\cite{Bahl:2018ykj} that, in the presence of scalar mixing, the
perturbative determination of the propagator poles proposed in
ref.~\cite{Bahl:2017aev} can lead to discontinuities in the Higgs-mass
predictions near the crossing points where the masses of the scalars
that mix with each other are degenerate.\footnote{This issue had first
  surfaced during the preparation of the benchmark scenarios of
  ref.~\cite{Bahl:2018zmf}.}  Ref.~\cite{Bahl:2018ykj} proposed an
alternative procedure in which the spurious logarithmic terms that
would cancel out only in a complete FO calculation are removed from
the Higgs self-energies via a redefinition of the Higgs fields, after
which the poles of the propagator can be determined numerically.

In 2020, ref.~\cite{Bahl:2020tuq} extended the hybrid approach of
\FH\ -- in scenarios with only one light Higgs doublet -- to include
the full NLL and gaugeless NNLL resummation of the corrections
controlled by the bottom Yukawa coupling, which were previously
computed only at fixed (i.e., two-loop) order.  To facilitate the
combination with the EFT component of the calculation, the
renormalization scheme for the bottom Yukawa coupling and for the soft
SUSY-breaking term $A_b$ in the FO component of the calculation was
changed from the OS scheme of refs.~\cite{Brignole:2002bz,
  Dedes:2003km} to the $\drbar$ scheme. It was found in
ref.~\cite{Bahl:2020tuq} that, in scenarios where the SUSY
contributions enhance the bottom Yukawa coupling, the differences in
its treatment between the pure FO calculation and the hybrid
calculation can lead to significant variations in the predictions for
$\Mh$ at large $\tan\beta$.
Also in 2020, ref.~\cite{Bahl:2020mjy} extended the hybrid approach of
\FH\ -- in scenarios where both Higgs doublets are light -- to the
case of complex parameters in the MSSM Lagrangian, largely relying on
ref.~\cite{Murphy:2019qpm} for the EFT component of the Higgs-mass
calculation.

An open issue in the hybrid approach of \FH\ is the possible mismatch
between the renormalization schemes employed in the FO and EFT
calculations. In the original implementation, the FO calculation
adopted an OS definition for the input parameters that determine the
stop masses and mixing, whereas the EFT calculation required
$\drbar$-renormalized parameters. It was therefore necessary to either
convert the input parameters from OS to $\drbar$ before passing them
to the EFT calculation, or modify the EFT calculation in such a way
that the boundary conditions at the SUSY scale are expressed in terms
of OS parameters. However, with the usual OS definition of the
stop-mixing parameter -- in which $M_t\,X_t$ is the off-diagonal
element of a $2\!\times\!2$ matrix whose eigenvalues are the pole stop
masses, and $M_t$ is the pole top mass -- the one-loop conversion of
$X_t$ between the OS and $\drbar$ schemes involves potentially large
logarithmic terms:
\beq
\label{eq:convXt}
X_t^{\drbar}(\MS) ~=~ X_t^{\rm OS} \,\left[
  ~1\,+\,\left(\frac{\alpha_s}{\pi} ~-~ \frac{3\,\alpha_t}{16\pi}\,(1-X_t^2/\MS^2)
  \right)\,\ln\frac{\MS^2}{M_t^2} ~+~ (...)\,\right]~,
\eeq
where the ellipses denote additional, non-logarithmic terms of ${\cal
  O}(\alpha_s)$ or ${\cal O}(\alpha_t)$, as well as terms depending on
other couplings. The alternative OS definition proposed in
ref.~\cite{Vega:2015fna}, in which the pole top mass in the
off-diagonal element of the stop mass matrix is replaced by the
running parameter $m_t(\MS)$, removes some of the logarithmic terms in
eq.~(\ref{eq:convXt}), but it does not affect the term proportional to
$X_t^2/\MS^2$. The latter stems from a threshold effect in the loop
integrals, and is specific to the case of degenerate stop masses (see
ref.~\cite{Sobolev:2020cjh} for a detailed discussion).
In the case of a strong hierarchy between the SUSY and EW scales, the
presence of large logarithmic terms either in the conversion of the
input parameters or in the boundary conditions at the SUSY scale
spoils the resummation of the logarithmic corrections. To circumvent
this problem, ref.~\cite{Bahl:2017aev} modified the hybrid calculation
of \FH, adding the option to use directly a $\drbar$ definition for
the stop parameters entering the FO part of the calculation. In that
case no conversion is needed, and the logarithmic corrections are
fully resummed at the desired order (i.e.~NLL or beyond, depending on
the scenario). If however the input parameters in the stop sector are
defined in the OS scheme, \FH\ includes only the logarithmic terms of
eq.~(\ref{eq:convXt}) in their conversion to the $\drbar$ scheme. The
presence of counterterm contributions in the FO part of the
calculation -- as mentioned above, those are not subtracted in
$\left[\Delta {\cal M}_{hh}(p^2)\right]_{{\rm d.c.}}$ -- ensures that
the prediction for the Higgs mass is correct up to the two-loop order,
but the resummation of the higher-order logarithmic corrections is
incomplete. As will be discussed in section~\ref{sec:uncertainties},
this is duly accounted for in the estimate of the theoretical
uncertainty of the Higgs-mass prediction of \FH.

\subsection{The hybrid approach of \FE}
\label{sec:hybridFE}

An alternative method to combine the EFT resummation of large
logarithmic corrections with the FO calculation of corrections
suppressed by powers of $v^2/\MS^2$ was proposed in 2016 in
ref.~\cite{Athron:2016fuq}, and it was implemented in the \FS\ module
\FE. In 2017 a similar approach was implemented in
\SA~\cite{Staub:2017jnp}.
The main idea of this approach consists in incorporating the
corrections to the Higgs mass suppressed by powers of $v^2/\MS^2$ into
the boundary condition for the quartic Higgs coupling at the SUSY
scale, $\lambdaSM(\MS)$, then proceeding as in a regular EFT calculation
(i.e., evolving $\lambdaSM$ down to the EW scale and computing there the
pole Higgs mass $M_h$). The boundary condition is determined by the
requirement that the FO result for the pole Higgs mass computed at the
SUSY scale be the same in the low-energy EFT (which is assumed to be
the SM) and in the high-energy SUSY model. Decomposing the FO result
for the Higgs mass computed in the SM as $(M_h^2)_{\smallSM} =
2\,\lambdaSM(\MS)\,v^2(\MS) + (\Delta M_h^2)_{\smallSM}$, one obtains
\beq
\label{eq:FEFTH}
\lambdaSM(\MS) ~=~ \frac{1}{2\,v^2(\MS)}\left[(M_h^2)_{\smallHET} - (\Delta
  M_h^2)_{\smallSM}\right]~,
\eeq
where $(M_h^2)_{\smallHET}$ is the FO result for the Higgs mass
computed in the high-energy theory (HET). Since the FO calculation
does not involve any expansion in $v^2/\MS^2$, the $\MS$-suppressed
terms are included in $\lambdaSM(\MS)$, and after the RG evolution of
$\lambdaSM$ they enter the result for $M_h$ computed at the EW
scale. We remark that, in this approach, the resummation of
higher-order logarithmic effects is correct only for the terms that
{are not suppressed by powers of} $v^2/\MS^2$, because the RG
evolution of the higher-dimensional operators that, in a pure EFT
approach, would account for the $\MS$-suppressed terms differs from
the RG evolution of the quartic coupling. On the other hand, the
$\MS$-suppressed terms are fully included in the Higgs-mass prediction
up to the loop order covered by the FO calculation.

An advantage of the hybrid approach of \FE\ is that the matching
procedure is largely independent of the considered high-energy theory,
and is therefore well-suited to be implemented in ``automated'' codes
such as \FS\ and \SA, which can compute the Higgs masses in generic
SUSY (and non-SUSY) extensions of the SM. Indeed, in
ref.~\cite{Athron:2016fuq} \FE\ was employed to obtain predictions for
the Higgs mass in several SUSY models beyond the MSSM, namely the
NMSSM, the E${_6}$SSM and the MRSSM.  On the other hand, the ``pole
matching'' condition of eq.~(\ref{eq:FEFTH}) can only be applied to
scenarios in which only one Higgs doublet is light, although it could
in principle be extended to cases in which the EFT includes additional
light particles that do not mix with the Higgs boson (e.g., to the
original Split-SUSY scenario).

In the early implementations of \FE, see refs.~\cite{Athron:2016fuq,
  Athron:2017fvs}, the FO calculation of the Higgs mass entering the
boundary condition in eq.~(\ref{eq:FEFTH}) contained only the one-loop
corrections computed ``automatically'' by \FS, allowing for the NLL
resummation of the higher-order logarithmic terms in a generic SUSY
model matched to the SM. In early 2020, ref.~\cite{Kwasnitza:2020wli}
improved the accuracy of the boundary condition for the MSSM case by
including the two-loop corrections in the gaugeless limit, as well as
the dominant three-loop corrections of refs.~\cite{Harlander:2008ju,
  Kant:2010tf, Harlander:2017kuc} which are obtained from
\HIM. Combined with full three-loop and partial four-loop RGEs for the
SM, this allows for the resummation of the NNLL corrections in the
gaugeless limit, and also for the resummation of the N$^3$LL
corrections that involve only the top Yukawa coupling and the highest
powers of the strong gauge coupling.  The $\MS$-suppressed effects are
in turn included fully up to one loop and in the gaugeless limit at
two loops. As the calculation of the three-loop corrections to the
Higgs mass in refs.~\cite{Harlander:2008ju, Kant:2010tf,
  Harlander:2017kuc} relied on an expansion to the first order in
$v^2$, no $\MS$-suppressed effects are actually included at three
loops.

A crucial aspect of the \FE\ approach, discussed in
refs.~\cite{Athron:2017fvs, Kwasnitza:2020wli}, is that each of the
terms on the right-hand side of eq.~(\ref{eq:FEFTH}) involves
potentially large logarithms of the ratio between the SUSY scale and
the EW scale, but these logarithms must cancel out in the
combination. If the parameters entering the various terms are defined
differently -- e.g., HET couplings for $(M_h^2)_{\smallHET}$ and SM
couplings for $(\Delta M_h^2)_{\smallSM}$ -- the cancellation of the
large logarithms holds only up to the loop order covered by the FO
calculation. However, the residues of the cancellation include
spurious, higher-loop {\em logarithmic} terms of the same order as
those that are being resummed by the RG evolution, thus spoiling the
resummation. To circumvent this problem,\footnote{For the time being
  the problem has not been addressed in \SA, hence the accuracy of the
  resummation in that code's hybrid mode is only LL.} the FO
calculation of \FE\ is reorganized in such a way that $(\Delta
M_h^2)_{\smallSM}$ and $v^2(\MS)$ are expressed in terms of HET
parameters. An external-momentum expansion of the self-energies up to
the considered loop order is also necessary to ensure the full
cancellation of the large logarithms.

Finally, it was shown in ref.~\cite{Kwasnitza:2020wli} that the choice
of expressing all loop corrections in terms of HET parameters ensures
that the two- and higher-loop ``leading-QCD'' contributions to
$\lambdaSM(\MS)$, i.e. those that are controlled by the highest powers
of the strong gauge coupling, do not involve powers of the ratio
$X_t/\MS$ higher than the fourth. This should result in a better
convergence of the perturbative expansion in scenarios where that
ratio is greater than~$1$.

\subsection{A third hybrid approach}
\label{sec:hybridHIM}

In late 2019, ref.~\cite{Harlander:2019dge} presented yet another
hybrid approach to the calculation of the Higgs mass in the MSSM. A
prediction for $M_h$ that includes both the resummation of
higher-order logarithmic corrections and the effects suppressed by
powers of $v^2/\MS^2$ is obtained from:
\beq
\label{eq:hybHIM}
(M_h^2)_{\rm hyb} ~=~ (M_h^2)_{\rm {\scriptscriptstyle EFT}} \,+\, \Delta_v^{0\ell+1\ell}
\,+\, \Delta_v^{2\ell}~,
\eeq
where $(M_h^2)_{\rm {\scriptscriptstyle EFT}}$ is the result of the
pure EFT calculation of ref.~\cite{Harlander:2018yhj}, see
section~\ref{sec:MSSMSMEFT}, which includes the full NLL resummation
of large logarithmic effects, plus a NNLL resummation in the gaugeless
limit and a N$^3$LL resummation of the effects that involve only the
top Yukawa coupling and the highest powers of the strong gauge
coupling. The remaining terms on the right-hand side of
eq.~(\ref{eq:hybHIM}) account for the $\MS$-suppressed effects at
different loop orders. In particular, the term
$\Delta_v^{0\ell+1\ell}$ accounts for the tree-level and one-loop
effects, and is obtained by subtracting the result of the pure EFT
calculation of $M_h^2$ provided by \HS\ from the result of the hybrid
calculation of $M_h^2$ provided by \FE, including only the NLL
resummation of logarithmic effects (i.e., one-loop threshold
corrections and two-loop RGEs) in each of the calculations. The term
$\Delta_v^{2\ell}$ contains instead the two-loop, $\MS$-suppressed
effects controlled by $y_t^4 g_s^2$ or by $y_t^6$, and is computed
from the difference between the known analytic formulas for the
two-loop corrections to the lighter Higgs mass in the gaugeless limit
and the same formulas expanded to the first order in $v^2$.

We remark that the proposal of ref.~\cite{Harlander:2019dge} remains
at the level of ``proof of concept'', as the script that combines the
various ingredients entering eq.~(\ref{eq:hybHIM}) has not been
released to the public so far. However, this hybrid calculation
accounts for both the logarithmic and the $\MS$-suppressed effects at
the same order in the relevant couplings as the calculation
implemented in the latest version of \FE, see
ref.~\cite{Kwasnitza:2020wli}. Indeed, despite being organized quite
differently from each other, the two hybrid calculations lead to very
similar predictions for the Higgs mass in the MSSM scenarios
considered in refs.~\cite{Kwasnitza:2020wli} and
\cite{Harlander:2019dge}.

\subsection{Comparing the FO, EFT and hybrid calculations}
\label{sec:hybridplot}

\begin{figure}[t]
  \includegraphics[width=0.52\textwidth]
                  {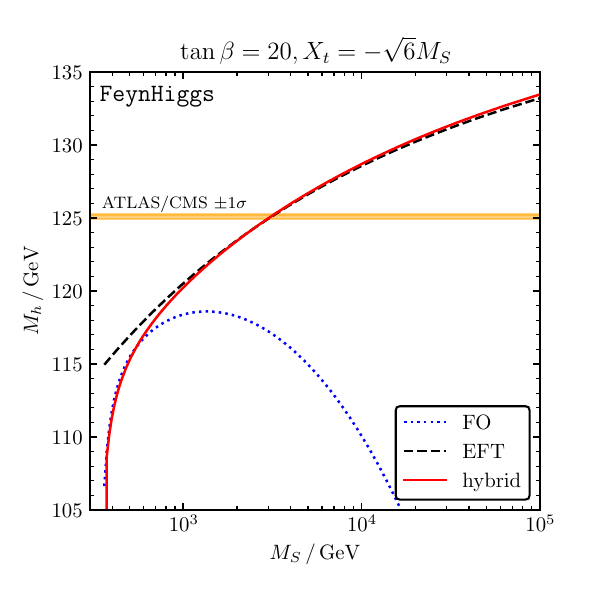}\hfill
                  \includegraphics[width=0.52\textwidth]
                                  {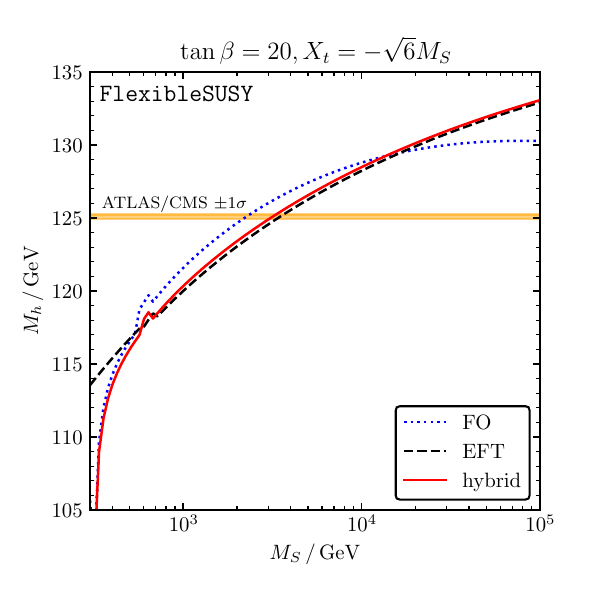}
  \vspace*{-6mm}                                
  \caption{\em Comparison between the pure FO, pure EFT and hybrid
    calculations of the mass {of the SM-like Higgs boson} in an MSSM
    scenario with degenerate SUSY masses. The sfermion mass and mixing
    parameters are defined in the $\drbar$ scheme at the scale
    $Q=\MS$. The left plot is produced with {\rm \FH}, the right one
    with different modules of {\rm \FS.}}
\label{fig:hybrid}
\end{figure}

The hybrid calculations of the Higgs mass described in this section
are meant to provide a combination of the results of pure FO
calculations, which are expected to be more reliable when the SUSY
masses are near the EW scale, with those of pure EFT calculations,
which are expected to be more reliable in heavy-SUSY scenarios. To
illustrate this point, we compare in figure~\ref{fig:hybrid} the
predictions for the mass {of the SM-like Higgs boson} obtained with
the three approaches, in a simplified MSSM scenario defined as
follows: all of the SUSY-breaking masses for sfermions and gauginos,
as well as the CP-odd Higgs-boson mass $\MA$ and the higgsino mass
$\mu$, are set equal to a common scale $\MS$, which is varied between
$300$~GeV and $100$~TeV; the stop mixing parameter is taken as $X_t =
-\sqrt{6}\,\MS$, and $\tan\beta=20$ (this fixes the value of the
trilinear Higgs-stop coupling $A_t$); the trilinear Higgs couplings to
all other sfermions are set to zero. The sfermion masses and $X_t$ are
interpreted as $\drbar$-renormalized parameters at the scale
$Q=\MS$. The left plot in figure~\ref{fig:hybrid} is obtained with
\FH, while the right plot is obtained with different modules of the
\FS\ package -- namely, \FS\ proper, \HS\ and \FE. In each plot the
blue dotted line is the result of the pure FO calculation, the black
dashed line is the result of the pure EFT calculation, and the red
solid line is the result of the hybrid calculation. The yellow band
corresponds to the value of the Higgs mass, as measured by ATLAS and
CMS~\cite{Aad:2015zhl} within one standard deviation of the
experimental accuracy.

The comparison between the different curves in figure~\ref{fig:hybrid}
shows that, both in \FH\ and in \FS, the hybrid calculation does
indeed agree with the FO calculation at small $\MS$ and with the EFT
calculation at large $\MS$. The small residual deviation between the
hybrid and EFT curves for \FH\ at large $\MS$ is due to two-loop
corrections involving the EW gauge couplings that are included in the
pure EFT calculation but not in the EFT component of the hybrid
calculation.  The kinks visible in the \FS\ curves around $\MS\approx
750$~GeV originate from a switch of the mass hierarchy used to
approximate the three-loop corrections in the calculation of
refs.~\cite{Harlander:2008ju, Kant:2010tf, Harlander:2017kuc}.
Moreover, for $\MS \lesssim 600$~GeV none of the mass hierarchies
implemented in \HIM\ reproduces this scenario, and the three-loop
corrections are switched off.  At small $\MS$, the comparison between
the three curves of each plot also shows that, in this scenario, the
$\MS$-suppressed effects that are not accounted for by the EFT
calculation can be relevant only for values of $\MS$ that result in a
very low prediction for the Higgs mass, which would be incompatible
with the measured value even if one assumed a theoretical uncertainty
of several GeV.

It is also worth noting that, while the EFT and hybrid predictions for
the Higgs mass show a good agreement between \FH\ and \FS\ at large
$\MS$, the predictions of the FO components of the two hybrid
calculations differ strikingly in this scenario: the prediction of
\FH\ decreases steeply when $\MS$ reaches a few TeV, whereas the
prediction of \FS\ has a much milder behavior at large $\MS$, and
starts differing significantly from the EFT result only for $\MS$
above $20\!-\!30$~TeV. The origin of this difference resides in the
treatment of the running top mass and the strong gauge coupling
entering the loop corrections, which in \FH\ are defined as the SM
parameters at the EW scale, whereas in \FS\ they are defined as the
MSSM parameters at the SUSY scale. These choices are compensated for
by appropriate counterterm contributions, so that the two FO
calculations are both correct at the considered perturbative
order. However, it appears that the choice implemented in the FO
calculation of \FS\ provides a better approximation of the
higher-order logarithmic corrections in this scenario.
As discussed in ref.~\cite{Bahl:2019hmm}, the FO prediction of
\FH\ would indeed show a much milder dependence on $\MS$, similar to
the one in \FS, if the top mass entering the loop corrections was
defined in the same way.
{It was also shown in ref.~\cite{Bahl:2019hmm} that the FO prediction
  of \FH\ has a milder dependence on $\MS$ in scenarios defined in
  terms of OS parameters for the stop sector {(the latter is the
    recommended choice for FO predictions in \FH)}. In general,}
the fact that effects that are formally of higher order can induce
such a strong variation in the results of the FO calculation
highlights the importance of resumming the large logarithmic
corrections in scenarios with a {large} hierarchy of
scales (indeed, \FH\ returns by default the results of its hybrid
calculation).


Finally we stress that, in all of the considered hybrid approaches,
the FO component of the calculation retains the $n\!\times\!n$
structure of the Higgs mass matrix. As a result, the hybrid
calculation accounts for the mixing effects in the Higgs sector -- in
the MSSM, these are contributions to the Higgs mass suppressed by
powers of $v^2/\MA^2$ -- up to the loop level covered by the FO
calculation, independently of the EFT used for the resummation of the
higher-order logarithmic effects. Indeed, it was found in
ref.~\cite{Bahl:2018jom} that, in an MSSM scenario with squark masses
of $100$~TeV and $\MA$ as low as $200$~GeV, the predictions for the
Higgs mass of the hybrid calculation in which the EFT includes only
one light Higgs doublet, see eq.~(\ref{eq:hybridFH}), differ from
those of the proper hybrid calculation in which the EFT is a \thdm\ by
at most $2\!-\!3$~GeV. In contrast, in a pure EFT calculation the use
of a theory with only one light Higgs doublet to describe a scenario
with heavy SUSY and low $\MA$ can lead to much larger deviations from
the correct predictions for the Higgs mass.

\subsection{Prospects}

It is natural to expect that any future improvement in the accuracy of
the FO and EFT calculations of the Higgs masses in SUSY models will
eventually trickle down to the hybrid calculations. When a full
two-loop calculation is finally completed, including all corrections
controlled by the EW gauge couplings, a hybrid version that also
allows for a full NNLL resummation of the logarithmic corrections will
certainly follow. A generalization of the three-loop calculation of
refs.~\cite{Harlander:2008ju, Kant:2010tf, Harlander:2017kuc} that
avoids the expansion in $X_t/\MS$ and possibly also the recourse to
different mass hierarchies would in turn improve the hybrid results of
\FS\ and \FH.

As discussed in section~\ref{sec:hybridFH}, another outstanding issue
in the hybrid calculation of the Higgs masses is the treatment of the
terms proportional to $\ln(\MS^2/M_t^2)$ that spoil the resummation of
the large logarithms when the stop mixing parameter is renormalized in
the OS scheme. If the stops are so heavy that the un-resummed
logarithmic terms significantly degrade the accuracy of the
calculation, it is probably not worth adopting an OS scheme for the
stop parameters in the first place, and they {can} be fixed directly
in the $\drbar$ (or, for heavy gluino, $\mdrbar$) scheme. In SUSY
scenarios with stop masses of a few TeV, however, the use of an OS
scheme might still be preferable in order to directly connect the
Higgs-mass predictions to the hoped-for future measurements of the
stop properties at the (HL{-})LHC. In {this} case, a possible path
forward would be to devise {a} definition for the stop mixing
{parameter} that {connects it to some measurable quantity but} does
not induce the unwanted corrections. As long as the standard OS
definition for $X_t$ is adopted, the effect of the non-resummed
logarithmic terms needs to be accounted for in the theoretical
uncertainty of the Higgs-mass prediction. How such {uncertainties}
should be estimated in the FO, EFT and hybrid calculations of the
Higgs masses has been the subject of intensive study in recent years,
as will be reviewed in the next section.

\vfill
\newpage

\section{Estimating the theory uncertainty of the Higgs-mass calculations}
\label{sec:uncertainties}
\subsection{Generalities}
\vspace*{-1mm}

\label{sec:unc_gen}

As is virtually always the case in theoretical particle physics, the
calculation of the Higgs masses in realistic SUSY models is too
complex to allow for exact solutions. Instead, it involves the
truncation of some perturbative expansion, where the expansion
parameter can be a loop factor, a whole tower of logarithmic
corrections, or the ratio between two mass scales.
Therefore, the result of any calculation of the Higgs masses should be
accompanied by an estimate of its ``theory uncertainty'', obtained by
simulating the effect of the terms in the expansion that are of higher
order with respect to the level of the truncation.\footnote{An
independent source of uncertainty for the Higgs-mass predictions in
SUSY models are the experimental uncertainties in the measurements of
the SM input parameters. This will be briefly discussed in section
\ref{sec:parunc}.}
If the calculation is organized properly, {it should be} sufficient to
simulate the effect of the first uncomputed term, which should be the
one that gives the largest contribution to the final result.

The{re is a wide range of methods} available to simulate the
uncomputed terms of an expansion: {for example, one can figure} out
their dependence on the relevant parameters and multiply by arbitrary
factors of order one, {or one can vary} the renormalization scheme and
scale of the parameters that enter the known terms of the
expansion. It is usually advisable to compare different estimates, to
make sure that the chosen method is not an outlier. On the other hand,
some methods might be more or less appropriate to the specific
calculation under consideration: e.g., scale variation may provide a
poor simulation of the higher-order effects if most of the parameters
are renormalized in an OS scheme {and are thus scale-independent; even
  in minimal-subtraction schemes, scale variation may not simulate
  classes of terms that do not include an UV divergence}. Moreover,
most of these methods have in common a degree of subjectiveness: the
choices of the arbitrary ${\cal O}(1)$ factors, or of the range for
the scale variation, depend to some extent on how aggressive (or
conservative) one wants the uncertainty estimate to be.

It is also worth noting that a given calculation is usually affected
by different sources of theory uncertainty at the same time. For
example, a FO calculation of the Higgs mass involves by definition a
truncation in the loop expansion, but it might also neglect the effect
of a subset of couplings within the considered loop order (e.g., when
the gaugeless limit is adopted) or rely on the approximation of
vanishing external momentum. Some of these sources of uncertainty may
be correlated, as in the case of the MSSM where the gaugeless limit
and the vanishing-momentum approximation neglect terms that involve
the same couplings. However, since the theory uncertainty does not
lend itself to a statistical interpretation, there is no definite
prescription to combine different sources, and again more-conservative
or less-conservative choices are possible.

Finally, we stress that the theory uncertainty of the Higgs-mass
calculation in a given SUSY model depends inevitably on the considered
point of the parameter space. In general, the uncertainties from
uncomputed higher-order effects tend to be larger in points where
there is a larger radiative correction to the tree-level
prediction. However, even in points where the correction is comparable
in size, the estimated uncertainties might differ, depending on
whether the correction is dominated by the effects of large couplings
(e.g., in scenarios with large $X_t$ {or, in the NMSSM,
  large $\lambda$}) or by the effects of large scales (e.g., in
scenarios with large $\MS$).
{In EFT calculations, the accuracy with which a given EFT
  describes the mass spectrum of the underlying SUSY model also
  depends on the considered point of the parameter space, and this
  should be reflected in the estimated uncertainty.}
 In summary, any ``one size fits all'' estimate of the theory
 uncertainty should be treated with care, and any code that computes
 the Higgs masses in SUSY models should also provide a point-by-point
 estimate of the uncertainty of its prediction.

\bigskip
 
After a summary of the state of the uncertainty estimates up to the
mid 2010s, in this section we will describe the considerable progress
achieved in this context in the years of the KUTS initiative.

\subsection{Pre-KUTS uncertainty estimates}

\bigskip

\paragraph{OS calculation:}

The first systematic attempt to estimate the theory uncertainty of the
Higgs-mass prediction in the MSSM dates to 2002, when
ref.~\cite{Degrassi:2002fi} discussed the status of the FO calculation
that was then implemented in \FH. This calculation included the full
one-loop corrections, the dominant two-loop corrections in the
gaugeless and vanishing-momentum limits, namely those of ${\cal
  O}(\alpha_t\alpha_s)$ and ${\cal O}(\alpha_t^2)$, and also the
two-loop ${\cal O}(\alpha_b\alpha_s)$ corrections relevant at large
$\tan\beta$. An OS scheme was {adopted} for the renormalization of the
parameters in the quark/squark sector, while $\tan\beta$, $\mu$ and
the Higgs field-renormalization constants were defined as $\drbar$
parameters.

The missing two-loop effects that were taken into account in the
uncertainty estimate of ref.~\cite{Degrassi:2002fi} were the
corrections controlled by the EW gauge couplings and the
external-momentum dependence of the gaugeless corrections. The impact
of {these} corrections was guessed by assuming that their size
relative to the dominant corrections (i.e., the gaugeless and
momentum-less ones) was the same as in the case of the fully-known
one-loop corrections. For the mass of the lighter Higgs scalar, this
resulted in an estimate of $1\!-\!2$~GeV for the corrections arising
from diagrams with {$D$}-term-induced Higgs-squark interactions,
$1$~GeV for the purely EW corrections (e.g.~ those from Higgs, gauge
boson and chargino or neutralino loops), and $1$~GeV for the momentum
effects.
An alternative estimate of the two-loop EW corrections was obtained by
varying the scale associated {with} the $\drbar$-renormalized
parameters by a factor of two above and below the central value, which
was chosen as $Q=M_t$. This yielded a shift in the Higgs mass of about
$\pm 1.5$~GeV.
It was assumed in ref.~\cite{Degrassi:2002fi} that there
would be at least some partial compensation between the different
missing two-loop corrections, and that their combined effect would
induce a shift of less than $3$~GeV in the prediction for the Higgs
mass.

The other source of uncertainty taken into account in
ref.~\cite{Degrassi:2002fi} were the three-loop effects. Their impact
on the prediction for the Higgs mass was estimated by changing the
definition of the top mass entering the two-loop corrections, which
was taken as either the pole mass or the $\msbar$-renormalized running
mass of the SM. An alternative estimate consisted in computing
explicitly the coefficient of $\ln^3(\MS/M_t)$ -- i.e., the
leading-logarithmic term in the three-loop corrections -- in the
gaugeless limit.  In an MSSM scenario with $\MS=1$~TeV, both
approaches yielded an estimate of $1\!-\!2$~GeV for the effects of the
missing three-loop corrections on the lighter Higgs mass. Again, the
combination of this uncertainty with the ones arising from the missing
two-loop effects required some amount of guesswork. In view of
possible cancellations between different missing corrections, a {\em
  ``realistic''} estimate of the theory uncertainty of the Higgs-mass
prediction was considered to be $\pm3$~GeV. At the end of 2004, a
point-by-point uncertainty estimate based on
ref.~\cite{Degrassi:2002fi} was implemented in \FH~\cite{Hahn:2005cu}.

\paragraph{\drbar\ calculations:}

The next systematic study of the theory uncertainty of the Higgs-mass
prediction dates to 2004, when ref.~\cite{Allanach:2004rh} discussed
the FO calculations implemented in the public codes \SSY, \Su\ and
\SP. These calculations included the full one-loop corrections, plus
all of the two-loop corrections that involve the third-family Yukawa
couplings in the gaugeless and vanishing-momentum limits. The $\drbar$
renormalization scheme was employed for all of the parameters entering
the corrections.

Since the three codes allowed for the RG evolution of the MSSM
Lagrangian parameters between different scales, the renormalization
scale at which the pole Higgs masses are computed could be varied at
will, and the shift in the Higgs mass resulting from this scale
variation was used to estimate the impact of the missing two-loop and
higher-order corrections. To fully capture the potentially large
logarithmic effects, the scale of the Higgs-mass computation was
varied between a value comparable to the EW scale (either $\MZ$ or
$150$~GeV, depending on the scenario) and twice the average of the
stop masses. In a number of scenarios where the prediction for the
lighter Higgs mass was below $119$~GeV -- an entirely realistic value
back then -- the resulting estimate of the theory uncertainty was of
$2\!-\!3$~GeV. However, in a scenario with large stop mixing and a
Higgs-mass prediction\,\footnote{Such a high prediction for $M_h$ in a
scenario with $\MS=1$~TeV stemmed from the fact that in
ref.~\cite{Allanach:2004rh} the pole top mass had been set equal to
$178$~GeV, following ref.~\cite{Azzi:2004rc}.} around $129$~GeV the
estimated uncertainty reached $4\!-\!5$~GeV.

As an alternative estimate of the theory uncertainty,
ref.~\cite{Allanach:2004rh} compared the results of the
\drbar\ calculation of \Su\ with those of the OS calculation of \FH,
with an appropriate conversion of the MSSM input parameters between
the two schemes. As \FH\ had in the meantime been upgraded to include
all of the two-loop corrections controlled by the bottom Yukawa
coupling in the gaugeless and vanishing-momentum limits, the
predictions of the two codes differed only by two-loop effects
controlled by the tau Yukawa coupling, which are all but negligible
unless $\tan\beta$ is very high, and by two-loop-EW and three-loop
effects resulting from the difference in the renormalization schemes,
which could be considered representative of the uncomputed
corrections. The estimate of the theory uncertainty obtained in this
way was in agreement with the one obtained from the scale variation:
$2\!-\!3$~GeV in most scenarios, rising to $4\!-\!5$~GeV in the
scenario with large stop mixing.

Finally, ref.~\cite{Allanach:2004rh} estimated the impact of the
missing momentum dependence of the two-loop corrections, using the
method introduced in ref.~\cite{Degrassi:2002fi} and finding shifts in
the light Higgs mass of half a GeV or even less in the considered
scenarios.
Overall, ref.~\cite{Allanach:2004rh} quoted a range of $3\!-\!5$~GeV,
depending on the considered point of the MSSM parameter space, for a
{\em ``reasonably conservative''} estimate of the global theory
uncertainty of the Higgs-mass calculation.

\paragraph{Discussion:}

Although refs.~\cite{Degrassi:2002fi, Allanach:2004rh} had been quite
explicit on the fact that the estimate of the theory uncertainty
should be treated as a function of the considered point of the
parameter space, in the following years it became customary for
phenomenological analyses of the MSSM to associate a fixed uncertainty
of $\pm 3$~GeV to the Higgs-mass prediction. Explicit computations of
some of the missing corrections -- namely, the two-loop EW effects at
vanishing momentum~\cite{Martin:2002wn}, the two-loop
momentum-dependent effects in the gaugeless limit~\cite{Martin:2004kr,
  Borowka:2014wla, Degrassi:2014pfa}, and the dominant three-loop
effects~\cite{Harlander:2008ju, Kant:2010tf} -- appeared to confirm
the estimates of refs.~\cite{Degrassi:2002fi,
  Allanach:2004rh}. However, the discovery in 2012 of a SM-like Higgs
boson with mass around $125$~GeV singled out precisely the regions of
the MSSM parameter space in which an estimate of $\pm 3$~GeV for the
theory uncertainty of the then-available FO calculations might have
been considered too optimistic. Indeed, in order to obtain through
radiative corrections a squared mass for the light Higgs that is at
least twice the tree-level value, it is necessary to have multi-TeV
stop masses, in which case the higher-order logarithmic effects can
become problematic, or large stop mixing, which also entails larger
uncertainties, or both.

The considerable effort devoted over the years of the KUTS initiative
to the improvement of the Higgs-mass calculations in SUSY models,
going beyond FO and including the resummation of large logarithmic
effects, was described in the previous sections.
For example, the predictions of \FH\ for the light Higgs mass in
scenarios with SUSY masses around $1$~TeV and large stop mixing have
shifted by more than $4$~GeV, and are now in better agreement with
those of the \drbar\ codes.
A parallel effort was devoted to point-by-point estimates of the
theory uncertainty of the improved calculations, as will be discussed
in the rest of this section. The adoption in phenomenological analyses
of these new uncertainty estimates has lagged behind, and as late as
2020 the benchmark scenarios proposed in ref.~\cite{Bahl:2020kwe} for
MSSM Higgs searches at the HL-LHC and the ILC still assumed a fixed
$\pm 3$~GeV uncertainty in the prediction for the light Higgs
mass. However, in view of the numerous improvements that the
Higgs-mass calculations have undergone since the times of
refs.~\cite{Degrassi:2002fi, Allanach:2004rh}, it should now be
legitimate to consider $\pm 3$~GeV a rather conservative estimate of
the theory uncertainty.

\subsection{Advances during KUTS}

For the sake of clarity, in this section we discuss separately the
recent developments in the uncertainty estimates of the EFT, FO and
hybrid approaches to the calculation of the Higgs mass. However, some
of the studies we refer to addressed more than one approach (e.g.,
this was obviously the case for all of the papers devoted to the
hybrid calculations).

\vspace*{-1mm}
\subsubsection{Uncertainty of the EFT calculations}
\label{sec:EFTunc}

Since the beginning of the KUTS initiative, the renewed focus on the
EFT calculations of the Higgs masses in SUSY models brought along the
need for an estimate of the associated theory uncertainty. In
scenarios with a strong hierarchy between mass scales, this
uncertainty is expected to be smaller than the one of the FO
calculations. Indeed, in the EFT approach the loop corrections
computed at the various matching scales tend to be smaller than those
encountered in the FO approach, since they are free from the
logarithmically enhanced terms that are accounted for to all
perturbative orders by the RG evolution. In 2014, {in the context of}
discussing the uncertainty estimate of the EFT calculation in the
simplest scenario where all SUSY masses are clustered around the same
high scale and the EFT valid below that scale is just the SM,
ref.~\cite{Bagnaschi:2014rsa} identified three distinct sources of
uncertainty:

\begin{itemize}
\renewcommand\labelitemi{--}

\item {\em SM uncertainties:}~ arising from uncomputed higher-order
  terms in the relations between physical observables and running
  parameters of the SM Lagrangian at the EW scale, and in the evolution
  of the running parameters up to the SUSY scale;

\item {\em SUSY uncertainties:}~ arising from uncomputed higher-order
  terms in the boundary condition for the quartic Higgs coupling at
  the SUSY scale;

\item {\em EFT uncertainties:}~ arising from the restriction to a
  renormalizable EFT in the unbroken phase of the EW symmetry, which
  amounts to neglecting effects suppressed by powers of $v^2/\MS^2$\,,
  where $v$ represents the EW scale and $\MS$ represents the SUSY
  scale.

\end{itemize}

This distinction and nomenclature\,\footnote{The nomenclature was
introduced in ref.~\cite{Vega:2015fna}. In scenarios with a
more-general EFT, the first source of uncertainty should obviously be
renamed. It is also conceivable to split it into ``low-scale'' and
``RGE'' components, see ref.~\cite{Bahl:2019hmm}.}  have been adopted
in a number of studies over the years. In the following we describe
how the individual sources of uncertainty are estimated in the NNLL
calculations of the Higgs mass implemented in the codes
\SH~\cite{Vega:2015fna}, \HS~\cite{Allanach:2018fif} and
\FH~\cite{Bahl:2019hmm}, as well as in the N$^3$LL calculation that
combines \HS\ and \HIM~\cite{Harlander:2018yhj,Harlander:2019dge}.

\paragraph{SM uncertainties:}

The estimates of the theory uncertainty associated {with} the
low-energy part of the EFT calculation and {with} the RG evolution of
the parameters take into account two contributions, which are expected
to be the dominant ones: the missing higher-order terms in the
relation between the pole Higgs mass and the parameters of the SM
Lagrangian at the EW scale, and the effect of higher-order terms in
the extraction of the top Yukawa coupling from the pole top mass. For
an NNLL resummation of the large logarithms both calculations need to
be performed at two loops, thus the estimate of the associated
uncertainty requires a simulation of the corresponding three-loop
effects.\footnote{For the two-loop calculations that are performed in
  the gaugeless limit the uncertainty estimate should also simulate
  the missing two-loop effects controlled by the EW gauge couplings.}

Concerning the determination of the pole Higgs mass, \SH\ assumes a
fixed uncertainty of $\pm 150$~MeV, as estimated in
ref.~\cite{Buttazzo:2013uya} for the full two-loop calculation in the
SM; \HS\ estimates the uncertainty as the largest of the shifts
induced by a variation of the renormalization scale in the calculation
of the Higgs mass by a factor $2$ or $1/2$ with respect to the central
value $Q=M_t$\,; \FH\ estimates it as the shift induced by changing
the definition the top mass entering the Higgs-mass corrections from
the $\msbar$ parameter evaluated at $Q=M_t$ to the pole mass.

Concerning the extraction of the top Yukawa coupling from the top
mass, all codes simulate the higher-order effects by including the
known three-loop QCD corrections of ${\cal O}(\alpha_s^3)$ from
refs.~\cite{Chetyrkin:1999ys, Chetyrkin:1999qi, Melnikov:2000qh}. In a
FO calculation of the Higgs mass, the resulting shift in $y_t$ would
only correspond to a four-loop effect. However, in the EFT calculation
a three-loop shift in $y_t$ affects the RG evolution of the quartic
Higgs coupling between the EW and SUSY scales at the N$^3$LL level,
providing an estimate of the uncertainty associated {with} the NNLL
resummation.

Finally, in the N$^3$LL calculation that combines \HS\ and \HIM, both
the determination of the pole Higgs mass and the extraction of the top
Yukawa couplings are performed at three loops, including only the
three-loop corrections that involve the highest powers of the strong
gauge coupling. The associated uncertainties are estimated as in the
NNLL calculation of \HS, but the additional corrections included in
the extraction of the top Yukawa coupling are the four-loop ${\cal
  O}(\alpha_s^4)$ ones from ref.~\cite{Marquard:2015qpa}, allowing for
an estimate of the missing N$^4$LL effects. It is{,}
however{,} worth noting that the impact of the
still-uncomputed N$^3$LL effects, i.e.~those that do not involve the
highest powers of $\alpha_s$, is not estimated by this procedure.

\vspace*{-1mm}
\paragraph{SUSY uncertainties:}

In the simplest heavy-SUSY scenario where the EFT is just the SM, as
well as in split MSSM scenarios with only one light Higgs doublet, the
missing higher-order terms in the boundary condition for the quartic
Higgs coupling at the SUSY scale are the two-loop contributions that
involve the EW gauge couplings
(with the exception of the mixed QCD-EW corrections from
ref.~\cite{Bagnaschi:2019esc}, which{,}
however{,} are not yet implemented in public codes),
and all contributions from three loops on, with the exception of the
three-loop ${\cal O}(y_t^4g_s^4)$ ones implemented in \HIM.

To estimate the impact of these missing terms, \SH, \HS\ and
\FH\ consider the dependence of the Higgs-mass prediction on the
matching scale where the boundary condition for $\lambdaSM$ is computed
(indeed, in a full calculation this dependence would be compensated
for by the explicit scale dependence of the higher-order terms). We
remark that this procedure requires that the running MSSM parameters
entering the known part of the boundary conditions, which are usually
given as input directly at the matching scale, be in turn evolved to
the new matching scale. The uncertainty is identified with the largest
of the shifts induced in the Higgs-mass prediction by a variation of
the matching scale by a factor $2$ or $1/2$ with respect to the
central value $\MS$\,.

An alternative estimate of the SUSY uncertainty, implemented only in
\HS\ and \FH, consists in changing the definition of the top Yukawa
coupling entering $\Delta\lambda$, adapting accordingly the formulas
for the two-loop contributions. In this case, the uncertainty is
identified with the shift induced in the Higgs-mass prediction when
the top Yukawa coupling entering the known contributions to
$\Delta\lambda$ is defined as the $\drbar$-renormalized parameter of
the MSSM instead of the $\msbar$-renormalized parameter of the
SM. Since this scheme variation induces also higher-order shifts that
do not depend explicitly on the renormalization scale, it is treated
as an independent source of uncertainty with respect to the scale
variation, and the absolute values of the two estimates are added
linearly.

In the N$^3$LL calculation that combines \HS\ and \HIM, additional
sources of uncertainty are the combined expansions in ratios of
particle masses and in the ratio $x_t = X_t/\MS$ that are used to
approximate the three-loop integrals in the ${\cal O}(y_t^4g_s^4)$
contributions to $\Delta\lambda$. As detailed in
ref.~\cite{Harlander:2018yhj}, the uncertainty associated {with} the
terms that are omitted in the expansion in mass ratios can be
estimated by comparing the approximate result for the logarithmic part
of the ${\cal O}(y_t^4g_s^4)$ contributions with the exact result that
can be extracted from the three-loop RGE for $\lambdaSM$.  The
uncertainty associated {with} the terms that are omitted in the second
expansion, i.e. those containing the highest powers of $x_t$, can be
estimated from the effect of the analogous terms entering a part of
the calculation where the expansion is not needed.  It was later
pointed out in ref.~\cite{Kwasnitza:2020wli} that the variation of the
matching scale provides an estimate of both of these sources of
uncertainty.
{Ref.~\cite{Harlander:2018yhj} and, later, ref.\cite{Bahl:2020tuq}
  also showed how} for large $x_t$ -- e.g., for the often-considered
value $|x_t| = \sqrt 6$, which maximizes the one-loop stop
contribution to $\Delta\lambda$ -- the uncertainty associated {with}
missing powers of $x_t$ can become larger than the effect of the whole
${\cal O}(y_t^4g_s^4)$ contributions {to $\Delta\lambda$}, in which
case the inclusion of the three-loop terms does not actually improve
the accuracy of the Higgs-mass calculation.

In phenomenological analyses of SUSY models, it sometimes happens that
-- either for the sake of simplicity or for lack of a better option --
an EFT is used that does not fully reflect the considered hierarchy
between mass scales. For example, the SM might be used as EFT below
the SUSY scale even in scenarios where higgsinos and gauginos are much
lighter than the sfermions. This would induce large logarithms in the
boundary conditions for the couplings of the EFT at the SUSY scale,
contrary to the spirit of the EFT approach. However, the presence of
large terms in the threshold corrections would also be reflected in an
enlarged estimate of the SUSY uncertainty. 

\vspace*{-1mm}
\paragraph{EFT uncertainties:}

As already discussed in sections~\ref{sec:THoverview} and
\ref{sec:eft}, the standard approach of matching the high-energy SUSY
theory to a renormalizable EFT in the unbroken phase of the EW
symmetry leads to neglecting terms suppressed by powers of
$v^2/\MS^2$\,, which can be mapped to the effect of non-renormalizable
operators with dimension six and higher. 

To estimate the impact of the neglected terms, \SH, \HS\ and
\FH\ consider the shift induced in the Higgs-mass prediction when the
boundary condition for the quartic Higgs coupling is shifted by an
arbitrary term that scales like $v^2/\MS^2$ times a loop factor. More
specifically, \SH\ estimates an upper and a lower uncertainty by
rescaling the one-loop contribution of each particle to
$\Delta\lambda$ by a factor $(1\pm 2\, v^2/M_i^2)$, where $M_i$ is
{the} mass {of the particle}; \FH\ does the same with a common
rescaling factor $(1\pm2\, v^2/\MS^2)$; \HS\ estimates symmetric upper
and lower uncertainties from the single rescaling factor $(1+2\,
v^2/\MS^2)$. Incidentally, we remark that the presence of a $2$ in the
rescaling factors stems {from} the fact that in
ref.~\cite{Vega:2015fna}, where this procedure was first introduced,
the Higgs vev was normalized as $v\approx 246$~GeV. Had the authors of
ref.~\cite{Vega:2015fna} adopted the same normalization for the Higgs
vev as in this report, where $v\approx 174$~GeV, their estimate of the
uncertainty associated {with} the missing ${\cal O}(v^2/\MS^2)$ terms
might well have been smaller by a factor of $2$. This should be taken
as a reminder of the degree of subjectiveness that inevitably affects
any estimate of the effect of uncomputed corrections.

In 2017, ref.~\cite{Bagnaschi:2017xid} compared the estimate of the
EFT uncertainty implemented in \SH\ with a direct calculation of the
one- and two-loop effects proportional to $y_t^2\,M_t^4/\MS^2$ and
$y_t^2\,g_s^2\,M_t^4/\MS^2$, respectively, which arise from the
introduction in the EFT Lagrangian of the dimension-six operators
$|H|^6$ and $|H|^2\,\overline{t_{\rm {\scriptscriptstyle R}}}\, H^{\rm
  {\scriptscriptstyle T}} \epsilon \,q_{\rm {\scriptscriptstyle
    L}}$\,. It was found in ref.~\cite{Bagnaschi:2017xid} that the
uncertainty estimate of \SH\ is at its most conservative -- namely,
larger than the directly-computed effects by about a factor of three
in the considered scenarios -- when {$|X_t|/\MS \approx \sqrt6$},
i.e.~near the value {for which a Higgs-mass prediction in the vicinity
  of $125$~GeV can be obtained with a stop-mass parameter of about
  $2$~TeV}. In contrast, for $X_t \ll \MS$ the estimate falls short of
the computed effects. However, in that case stop masses of more than
$10$~TeV are necessary to obtain {a phenomenologically acceptable
  prediction for the Higgs mass}, rendering the ${\cal O}(v^2/\MS^2)$
effects essentially irrelevant.

\vspace*{-1mm}
\paragraph{Numerical example:}

\begin{figure}[t]
  \includegraphics[width=0.52\textwidth]
                  {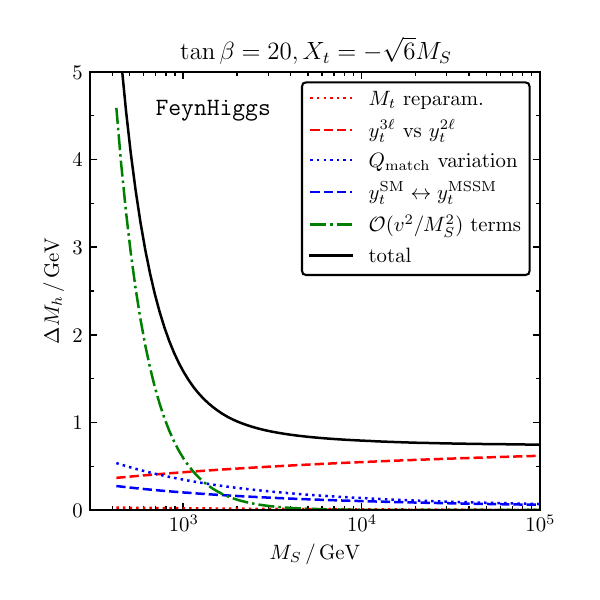}\hfill
                  \includegraphics[width=0.52\textwidth]
                                  {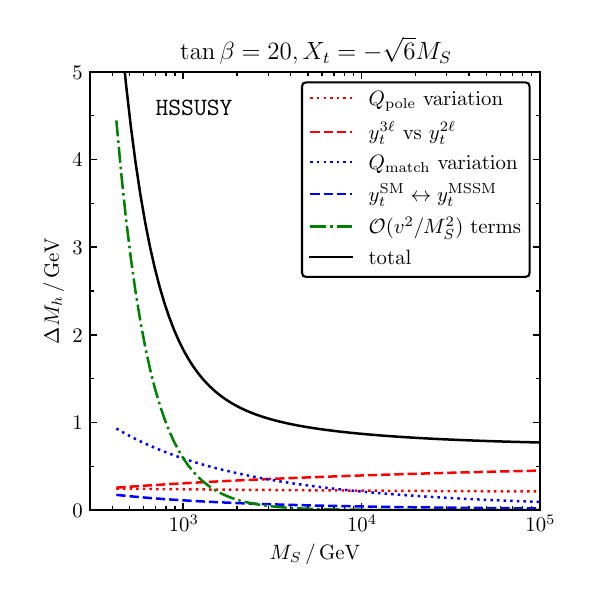}
  \vspace*{-6mm}
  \caption{\em Estimates of the different sources of theory
    uncertainty for the EFT calculation, as a function of the SUSY
    scale $\MS$. The left plot is produced with {\rm \FH}, the right
    one with {\rm \HS}.}
\label{fig:EFTunc}
\end{figure}

To illustrate the relative importance of the different sources of
uncertainty described so far, we show in figure~\ref{fig:EFTunc} the
corresponding estimates as a function of a common SUSY scale $\MS$, in
the same MSSM scenario {as} considered in
figure~\ref{fig:hybrid}. The left plot is produced with {\rm \FH}, the
right one with {\rm \HS}.\footnote{Results qualitatively similar to
  the ones discussed in this section can be obtained also with \SH,
  see ref.~\cite{Vega:2015fna}.}  We recall that, for the value of the
stop mixing parameter adopted in this scenario, the expansion in
$X_t/\MS$ employed in the three-loop contribution to the threshold
correction $\Delta\lambda$ is unreliable, hence in the case of {\rm
  \HS} we restrict our discussion to the uncertainty of the NNLL
calculation.

In each plot, the dotted red line represents the uncertainty of the
Higgs-mass determination at the EW scale, which \FH\ estimates by
scheme variation of the top mass and \HS\ estimates by scale
variation; the dashed red line represents the uncertainty associated
{with} the extraction of the top Yukawa coupling from the top mass;
the dotted blue line represents the uncertainty associated {with} the
variation of the matching scale around the central value $\MS$; the
dashed blue line represents the uncertainty associated {with} the
scheme change in the top Yukawa coupling entering the boundary
condition for $\lambdaSM$; the dot-dashed green line represents the
estimate of the uncertainty associated {with} the missing ${\cal
  O}(v^2/\MS^2)$ terms; finally, the solid black line corresponds to
the total estimate of the theory uncertainty, obtained by summing
linearly the absolute values of the individual estimates.

The comparison between the dotted and the dashed red lines in each
of the plots of figure~\ref{fig:EFTunc} shows that the largest
contribution to the ``SM uncertainty'' comes from the determination of
the top Yukawa coupling. The estimated uncertainty grows with $\MS$,
reflecting the effect of the top Yukawa coupling on the RG evolution
of the quartic Higgs coupling. The comparison between the dotted red
lines in the left and right plots shows that, in this scenario, the
scale variation implemented in \HS\ yields a significantly larger
estimate for the uncertainty of the Higgs-mass determination at the EW
scale than the scheme variation implemented in \FH.

The comparison between the dotted and the dashed blue lines in the
plots of figure~\ref{fig:EFTunc} shows that the estimate of the ``SUSY
uncertainty'' arising from a variation of the matching scale is
somewhat larger than the one arising from a change of scheme in the
top Yukawa coupling. In addition, both of the estimates of the SUSY
uncertainty tend to decrease with increasing $\MS$. The latter is a
consequence of the renormalization-scale dependence of $y_t$ and
$g_s$, which in the SM become both smaller at higher scales. As a
result, the overall impact of the threshold correction $\Delta\lambda$
on the Higgs-mass prediction is suppressed, and the associated
uncertainty follows suit.

The dot-dashed green lines in the plots of figure~\ref{fig:EFTunc}
show that the estimate of the ``EFT uncertainty'' associated {with}
the missing ${\cal O}(v^2/\MS^2)$ terms can reach about $4.5$~GeV at
the lowest considered values of $\MS$, but it gets quickly suppressed
as $\MS$ increases. In particular, in the range $\MS \approx
2\!-\!4$~{TeV}, which corresponds to a Higgs-mass prediction in the
vicinity of $125$~GeV (see figure~\ref{fig:hybrid}), the EFT
uncertainty is already sub-dominant with respect to the other sources.

Finally, the solid black lines in the plots of figure~\ref{fig:EFTunc}
show that at small $\MS$ the total uncertainty estimate of the EFT
calculation is dominated by the missing ${\cal O}(v^2/\MS^2)$ effects,
whereas for $\MS \gtrsim 1\!-\!2$~{TeV} the largest source of
uncertainty is the extraction of the top Yukawa coupling from the top
mass. At large $\MS$ the total uncertainty is estimated to be less
than $\pm 1$~GeV in this scenario, with a rather mild dependence on
$\MS$ which results from a partial cancellation between the opposite
scale dependences of the ``SM'' and ``SUSY'' components. We also
mention that the choice of adding up the absolute values of the
different sources of uncertainty can be considered conservative, as it
seems unlikely that the missing higher-order terms would all enter the
Higgs-mass prediction with the same sign.

\subsubsection{Uncertainty of the FO calculations}
\vspace*{-0.5mm}
\label{sec:FOunc}

The intense activity aimed at improving the calculation of the Higgs
masses in scenarios with heavy SUSY particles also highlighted the
need for a reassessment of the existing uncertainty estimates for the
FO calculations. In 2014, ref.~\cite{Bagnaschi:2014rsa} stressed that
a lingering spread of about $5$~GeV between the predictions for $M_h$
of the OS calculation of \FH\ and those of the $\drbar$ calculations
of \SSY, \Su\ and \SP\ in scenarios with stop masses around $1$~TeV
and large stop mixing pointed to a large theory uncertainty,
{possibly} exceeding the $\pm 3$~GeV that were commonly
assumed since the early 2000s.
In 2016, while discussing the two-loop ${\cal O}(\alpha_t\alpha_s)$
corrections to the Higgs masses in a SUSY model with Dirac gluinos,
ref.~\cite{Braathen:2016mmb} noted that changes in the definition
and/or scale of the strong gauge coupling $\alpha_s$ -- which had not
been considered in the original uncertainty estimate of
ref.~\cite{Degrassi:2002fi} -- can induce a shift of up to $7$~GeV in
the prediction for $M_h$. This is significantly larger than the shift
induced by a change in the renormalization scheme of the parameters in
the top/stop sector, despite the fact that both changes amount to
three-loop ${\cal O}(\alpha_t\alpha_s^2)$ effects in the Higgs
mass. Later in 2016, the importance of the definition and scale choice
for $\alpha_s$ was also stressed in the context of the NMSSM in
ref.~\cite{Drechsel:2016htw}, which found that differences of up to
$6$~GeV in the Higgs-mass predictions of \NC\ and \FH\ could be
greatly reduced once the strong gauge coupling was computed at the
same scale in the two codes. Again, the discrepancies induced by the
scale choice for $\alpha_s$ in the ``out-of-the-box'' predictions of
the two codes amount to three-loop ${\cal O}(\alpha_t\alpha_s^2)$
effects in the Higgs mass, and could in principle have been considered
as part of the uncertainty estimate.

We now summarize how, over the years of the KUTS initiative, new
estimates of the theory uncertainty were discussed for the FO
calculations of the Higgs masses. Since the inadequacy of such
calculations in scenarios with even moderately heavy SUSY particles had
by then become apparent, the main practical purpose of these estimates
was the comparison with the corresponding estimates for the EFT and
hybrid calculations. Starting from the case of the MSSM, we discuss
separately the uncertainties of the two-loop, fully $\drbar$
calculation implemented in codes such as \FS\ and \SSY\ (with the
possible addition of the dominant three-loop effects from \HIM), and
those of the two-loop, mixed OS--$\drbar$ calculation of \FH.

\vspace{-1mm}
\paragraph{\FS/\SSY/\HIM:}
In 2016, ref.~\cite{Athron:2016fuq} proposed a method to estimate the
uncertainty of the two-loop calculation of the MSSM Higgs masses
implemented in \FS, with the purpose of a comparison with the hybrid
calculation of \FE. A first estimate of the uncertainty relied on the
extraction of the running top Yukawa coupling $y_t$ from the pole top
mass. The uncertainty was defined as the maximum difference between
the predictions for the Higgs masses obtained with four definitions of
$y_t$ that are all equivalent at one loop, but differ by higher-order
terms. Since two of these definitions differ from the others by
two-loop ${\cal O}(\alpha_s^2)$ terms enhanced by $\ln^2(\MS/M_t)$,
the variation of $y_t$ in the one-loop part of the Higgs-mass
calculation allows for a simulation of the three-loop LL terms --
i.e., those of ${\cal O}(\alpha_t\alpha_s^2)$ enhanced by
$\ln^3(\MS/M_t)$ -- that are expected to be dominant among the missing
higher-order effects (at least until $\MS$ becomes so large that the
loop expansion breaks down). An alternative estimate of the
uncertainty, added in quadrature to the first one, relied on a
variation of the renormalization scale at which the calculation of the
pole Higgs mass is performed. In particular, the uncertainty was
defined as the maximal variation in the Higgs mass when the scale is
varied by a factor $2$ above and below its default value, which is
taken as the average of the stop masses. It was pointed out in
ref.~\cite{Athron:2016fuq} that, as the considered scale variation
does not cover the full range between the EW scale and the SUSY scale,
this second method cannot simulate the missing LL effects but only the
NLL ones, and it leads to a smaller uncertainty estimate than the
first method.

In 2018, ref.~\cite{Allanach:2018fif} estimated the theory uncertainty
of the three-loop calculation of the MSSM Higgs masses obtained from
the combination of \SSY\ and \HIM, with the purpose of a comparison
with the EFT calculation of \HS. The uncertainty of the FO calculation
was defined as the linear sum of five estimates, namely: the two
already proposed in ref.~\cite{Athron:2016fuq}, i.e., a variation in
the extraction of $y_t$ and a variation in the scale at which the
Higgs masses are computed; a variation -- again, by a factor $2$ above
and below the default value -- of the scale at which the running gauge
and Yukawa couplings are extracted from physical observables; two
additional estimates of higher-order effects in the determination of
the strong and electromagnetic gauge couplings, respectively. In
contrast {to} ref.~\cite{Athron:2016fuq}, the ${\cal
  O}(\alpha_t\alpha_s^2)$ corrections were included in the calculation
of the Higgs masses through \HIM. Thus, the first of the estimates
mentioned above compared two definitions of $y_t$ that differ by
two-loop LL terms of ${\cal O}(\alpha_t\alpha_s)$, allowing for a
simulation of the uncomputed three-loop LL terms of ${\cal
  O}(\alpha_t^2\alpha_s)$ in the Higgs-mass prediction. In simplified
scenarios with a common SUSY scale $\MS$, ref.~\cite{Allanach:2018fif}
found that the total uncertainty estimate of the FO calculation, which
is more precise for a low SUSY scale, and the one of the EFT
calculation, more precise for a high SUSY scale, coincide for values
of $\MS$ between $1$ and $1.3$~TeV, depending on the considered
scenario.

In 2019, ref.~\cite{Harlander:2019dge} estimated the theory
uncertainty of the three-loop calculation of the MSSM Higgs masses
obtained from the combination of \FS\ and \HIM, again with the purpose
of a comparison with the EFT calculation of \HS. In contrast {to}
ref.~\cite{Allanach:2018fif}, the uncomputed three- and four-loop LL
terms were simulated by varying the renormalization scale at which the
Higgs masses are computed in the whole range between $M_t$ and
$\MS$. An additional estimate of the uncomputed four-loop LL terms,
added linearly to the first estimate, consisted in including the
two-loop SUSY-QCD corrections from refs.~\cite{Harlander:2005wm,
  Bauer:2008bj, Bednyakov:2010ni} in the determination of the strong
gauge coupling. The procedure proposed in
ref.~\cite{Harlander:2019dge} leads to a larger estimate for the
theory uncertainty of the FO calculation compared to the one proposed
in ref.~\cite{Allanach:2018fif}, and the value of $\MS$ for which the
FO and EFT calculations have a similar estimated uncertainty is
lowered below $1$~TeV.

\vspace{-1mm}
\paragraph{\FH:}
In 2019, ref.~\cite{Bahl:2019hmm} presented a new estimate for the
theory uncertainty of the FO (namely, full one-loop and gaugeless
two-loop) calculation of the Higgs masses implemented in \FH, updating
the estimate based on ref.~\cite{Degrassi:2002fi} that had been
available in the code since 2004. Even in this case, the main purpose
was the comparison with the uncertainties of the EFT and hybrid
calculations of the Higgs masses implemented in the same code.

The first method proposed in ref.~\cite{Bahl:2019hmm} for estimating
the uncertainty of the Higgs-mass calculation consists in changing the
definition of the top mass entering the one- and two-loop corrections,
switching between the $\msbar$-renormalized mass parameter of the SM
evaluated at the scale $Q=M_t$, which is used by default in the code,
and the pole mass. This method simulates the uncomputed two- and
three-loop corrections that involve the top Yukawa coupling. However,
since the considered definitions of the top mass do not differ by
large logarithmic terms, the higher-order corrections are simulated
only at the NLL level.
A second estimate is therefore introduced to capture the three-loop LL
terms that are expected to give the largest contribution to the
uncertainty at large $\MS$, i.e.~those involving the highest powers of
the strong gauge coupling. In this case, the uncertainty is defined as
the shift induced in the Higgs mass when the ${\cal
  O}(\alpha_t\alpha_s)$ part of the two-loop corrections is multiplied
by a factor $1\pm\alpha_s/(4\pi)\ln(\MS^2/M_t^2)$, thus simulating the
effect of ${\cal O}(\alpha_t\alpha_s^2)$ corrections enhanced by
$\ln^3(\MS^2/M_t^2)$. Once again, we note how the choice of the
numerical coefficient for the factor that simulates the higher-order
terms introduces an element of subjectiveness in this kind of
estimates.
Finally, a third uncertainty estimate targets the corrections
controlled by the bottom Yukawa coupling, which can be numerically
relevant only at large values of $\tan\beta$. In particular, the
uncertainty is defined as the shift induced in the Higgs mass when the
so-called $\Delta_b$ terms, i.e.~a class of $\tan\beta$-enhanced terms
that are by default absorbed in the one-loop corrections via a
redefinition of the Higgs-sbottom coupling, see
refs.~\cite{Brignole:2002bz, Dedes:2003km, Heinemeyer:2004xw}, are
made to appear explicitly in the two-loop part of the corrections. We
remark that this uncertainty estimate should be considered
conservative, because the higher-order $\tan\beta$-enhanced effects
that it simulates -- e.g., three-loop terms enhanced by $\tan^2\beta$
-- are in fact already accounted for (``resummed'') in the default
determination of the Higgs mass.  The absolute values of the three
uncertainty estimates are added linearly.

In MSSM scenarios with $\MS=1$~TeV and large stop mixing, the total
uncertainty of the FO calculation of the lighter Higgs mass in
\FH\ was estimated in ref.~\cite{Bahl:2019hmm} to be about
$2\!-\!3$~GeV, depending on the renormalization scheme employed for
the stop parameters. As a result of the logarithmic enhancement of the
uncomputed higher-order corrections, the uncertainty estimate of the
FO calculation grows quickly for increasing $\MS$, reaching as much as
$10\!-\!15$~GeV for $\MS=10$~TeV. Once again, this highlights the need
for a resummation of the large logarithmic corrections in scenarios
where the stops have masses of even a few TeV.

\vspace{-1mm}
\paragraph{Beyond the MSSM:}
Over the years of the KUTS initiative, estimates of the theory
uncertainty of the FO calculation of the Higgs masses have also been
developed for a number of non-minimal SUSY models, see e.g.~the
studies in ref.~\cite{Braathen:2016mmb}, already mentioned at the
beginning of this section~\ref{sec:FOunc}, concerning SUSY models with
Dirac gluinos.

For the NMSSM, refs.~\cite{Ender:2011qh,Graf:2012hh} (pre-KUTS)
estimated the theory uncertainty of the one-loop calculation of the
Higgs masses by varying between OS and $\drbar$ the renormalization
scheme of the parameters entering the tree-level Higgs mass matrix, as
well as the scheme of the quark masses entering the one-loop
corrections. In addition, the scale at which the Higgs masses are
computed in the \drbar\ calculation was varied by a factor $2$ above
and below the default value, which was taken as the average stop
mass. This yielded estimates of about $10\%$ for the uncertainty of
the one-loop prediction for the Higgs masses in scenarios with stop
masses around $500$~GeV or less. In 2014,
ref.~\cite{Muhlleitner:2014vsa} discussed the improvement in the
theory uncertainty of the NMSSM Higgs-mass prediction that comes from
the inclusion of the two-loop ${\cal O}(\alpha_t\alpha_s)$ corrections
in the limit of vanishing external momentum. The uncertainty was
estimated only by varying the renormalization scheme of the parameters
in the top/stop sector between OS and $\drbar$. In scenarios with stop
masses around $1$~TeV, it was found that the inclusion of the ${\cal
  O}(\alpha_t\alpha_s)$ corrections drastically reduces the estimated
uncertainty from $15\%\!-\!25\%$ (depending on the stop mixing) to
about $1.5\%$. {In 2016, ref.~\cite{Drechsel:2016htw} compared the
  Higgs-mass predictions obtained by \NC\ adopting either the OS or
  the \drbar\ scheme for the top/stop sector, and found discrepancies
  of less than $2$~GeV in four representative NMSSM scenarios.}  It
should{,} however{,} be {recalled} that{, in the FO calculation of the
  Higgs masses,} the uncertainty associated {with} the definition of
$\alpha_s$ -- see the discussions in refs.~\cite{Braathen:2016mmb,
  Drechsel:2016htw} mentioned at the beginning of this section --
{significantly exceeds the uncertainty that can be estimated from a
  change of scheme in the top/stop sector}.
{Ref.~\cite{Drechsel:2016htw} also investigated the effect
  of switching between the different prescriptions of \FH\ and
  \NC\ for the OS renormalization of the EW parameters $(g, g^\prime,
  v)$ in the one-loop part of the calculation. This implied an
  estimate of at most $\pm 1$~GeV for the uncertainty of the
  Higgs-mass prediction associated with the uncomputed two-loop
  corrections that involve the EW gauge couplings.}
The findings of
ref{s}.~\cite{Muhlleitner:2014vsa, Drechsel:2016htw} were {also}
reassessed in 2019, when ref.~\cite{Dao:2019qaz} discussed the
inclusion of the two-loop ${\cal O}(\alpha_t^2)$ corrections in the
so-called ``MSSM limit'' (i.e., {$v_s\rightarrow \infty$} with
$\lambda\,v_s$ {and $\kappa \,v_s$} held fixed). The uncertainty of
the calculation was estimated by varying the renormalization scheme of
the top/stop parameters between OS and $\drbar$, and also by varying
the scale in the \drbar\ calculation by a factor of 2 above and below
the default value. It was found in ref.~\cite{Dao:2019qaz} that the
inclusion of the ${\cal O}(\alpha_t^2)$ corrections actually worsens
the estimated uncertainty of the Higgs-mass calculation, raising it to
$5\!-\!6\%$ in the considered scenarios. To explain this seemingly
counter-intuitive finding, it was argued in ref.~\cite{Dao:2019qaz}
that the small scheme- and scale-dependence of the calculation that
includes only the two-loop ${\cal O}(\alpha_t\alpha_s)$ corrections is
due to accidental cancellations, and that the additional inclusion of
the ${\cal O}(\alpha_t^2)$ corrections makes the uncertainty estimate
sensitive to different classes of higher-order terms, for which these
cancellations do not occur.

The uncertainty estimate introduced in ref.~\cite{Athron:2016fuq} for
the FO calculation of the Higgs masses implemented in \FS\ was in turn
applied also to models beyond the MSSM. In NMSSM scenarios with
vanishing stop mixing and $\tan\beta = 5$, ref.~\cite{Athron:2016fuq}
estimated a theory uncertainty of about $\pm 6$~GeV for $\MS \approx
30$~TeV, where the prediction for the light Higgs mass can be in the
vicinity of $125$~GeV. In an E$_6$SSM scenario, also with vanishing
stop mixing and $\tan\beta = 5$, a suitable prediction for the Higgs
mass can be obtained for $\MS \approx 10$~TeV, but the estimated
uncertainty of the FO calculation is as large as $\pm 10$~GeV in that
point. Similar results were found for the MRSSM, where the uncertainty
estimate of ref.~\cite{Athron:2016fuq} was applied to the FO
calculation implemented in \SA. In all these cases, the comparison
with the hybrid calculation implemented in \FE\ highlighted the
importance of resumming the large logarithmic corrections in scenarios
with heavy SUSY particles.

\subsubsection{Uncertainty of the hybrid calculations}
\label{sec:hybUnc}

Just as the hybrid calculation of the Higgs mass combines an EFT
component and a FO component, its uncertainty estimate stems from the
combination of the uncertainties of the two components. A number of
techniques employed in public codes to estimate the latter have been
described in sections~\ref{sec:EFTunc} and~\ref{sec:FOunc}. We stress
that the EFT and FO {\em components} of the hybrid calculation
implemented in a given code do not necessarily coincide with the
stand-alone EFT and FO calculations that may also be provided by the
same code (e.g., due to the presence of subtraction terms), thus a
dedicated estimate of the uncertainty of the hybrid calculation
remains in order. This said, it is legitimate to expect that such
estimate will be comparable to or lower than the individual estimates
of the stand-alone calculations. In particular, the uncertainty of the
hybrid calculation should agree with the one of the pure EFT
calculation in scenarios with very heavy SUSY particles, and with the
one of the pure FO calculation in (now experimentally challenged)
scenarios where all SUSY particles are at the EW scale.

In the following we describe the uncertainty estimates that have been
developed for the three hybrid approaches described in
sections~\ref{sec:hybridFH}--\ref{sec:hybridHIM}. 

\vspace{-1mm}
\paragraph{Hybrid approach of \FH:}
The uncertainty estimate of the hybrid calculation implemented in
\FH\ was described in ref.~\cite{Bahl:2019hmm}. In its latest
implementation, the hybrid prediction of \FH\ for the mass of the
SM-like Higgs boson can be viewed as a complete EFT calculation, at
full NLL and ``gaugeless'' NNLL order, supplemented with a FO
calculation of the effects suppressed by powers of $v^2/\MS^2$, at
full one-loop and ``gaugeless'' two-loop order. Accordingly,
\FH\ obtains the uncertainty of the hybrid result by combining the
uncertainty of the EFT component, estimated as described in
section~\ref{sec:EFTunc} {\em minus} the contribution stemming from
the ${\cal O}( v^2/\MS^2)$ terms, with the uncertainty of the FO
calculation of the ${\cal O}( v^2/\MS^2)$ terms, estimated as
described in section~\ref{sec:FOunc}. In particular, the ``SM
uncertainty'' of the EFT component is estimated by changing the
definition of the top mass entering the determination of the pole
Higgs mass and by switching on the three-loop QCD corrections in the
extraction of the top Yukawa coupling from the top mass; the ``SUSY
uncertainty'' is estimated by varying the matching scale by a factor
of $2$ above and below the central value $\MS$, and by changing the
definition of the top Yukawa coupling entering the threshold
corrections to the quartic Higgs coupling; the ``EFT uncertainty''
is{,} however{,} omitted, because the ${\cal O}(
v^2/\MS^2)$ terms are accounted for by the FO component of the hybrid
calculation. The uncertainty of the FO calculation of the ${\cal O}(
v^2/\MS^2)$ terms is in turn estimated by changing the definition of
the top mass, by multiplying the ${\cal O}(\alpha_t\alpha_s)$ part by
a factor $1\pm\alpha_s/(4\pi)\ln(\MS^2/M_t^2)$, and by switching off
the resummation of the $\Delta_b$ terms. All of the above-listed
sources of uncertainty are summed linearly in absolute value.

As discussed in section~\ref{sec:hybridFH}, the hybrid calculation of
\FH\ involves additional sources of uncertainty when the input
parameters that determine the stop masses and mixing are defined in
the OS scheme. In that case, the stop mixing parameter $X_t$ is
converted to the $\drbar$ scheme, see eq.~(\ref{eq:convXt}), before
being passed to the EFT component of the calculation. Moreover, the FO
component contains two-loop counterterm contributions that do not
vanish when $ v^2/\MS^2 \rightarrow 0$, but are not canceled out by
the subtraction term introduced to avoid double counting, see
eq.~(\ref{eq:hybridFH}). Indeed, {these} contributions do not have an
equivalent in the EFT component, where all parameters are defined as
$\drbar$. Both the uncertainty associated {with} the
OS--\drbar\ conversion of $X_t$ and the uncertainty associated {with}
the counterterm contributions in the FO component are estimated by
switching between different definitions of the top mass, and by
multiplying the strong gauge coupling $\alpha_s$ by a factor
$1\pm\alpha_s/(4\pi)\ln(\MS^2/M_t^2)$. We remark that the latter
procedure introduces ${\cal O}(\alpha_t\alpha_s^2)$ terms enhanced by
$\ln(\MS^2/M_t^2)$ -- i.e., terms that are formally of ``gaugeless''
NNLL order -- in the uncertainty estimate. This reflects the fact
that, when the stop masses and mixing are defined in the OS scheme,
the hybrid procedure implemented in \FH\ provides only an incomplete
resummation of the ``gaugeless'' NNLL corrections. As a result, the
estimated uncertainty turns out to be somewhat larger than in the case
in which the input parameters in the stop sector are defined directly
in the \drbar\ scheme. It is, however, important to note that
additional sources of uncertainty must be considered if the $\drbar$
input parameters are extracted from (so-far hypothetical) measured
quantities, such as, e.g., the stop pole masses and decay widths.
  
The study of theory uncertainties presented in
ref.~\cite{Bahl:2019hmm} focused on the hybrid setup in which, in the
EFT part of the calculation, the heavier Higgs doublet is decoupled
together with the heavy SUSY particles at the scale $\MS$. In
scenarios where both Higgs doublets are light, this neglects the
resummation of the corrections enhanced by $\ln(\MS^2/\MA^2)$. It was
shown in ref.~\cite{Bahl:2019hmm} that the use of an inappropriate EFT
in the hybrid calculation is indeed reflected in an increase of the
estimated uncertainty at low $\MA$, compatible with the differences
found in ref.~\cite{Bahl:2018jom} between the calculation using the SM
as EFT and the calculation using the \thdm\ as EFT.

\vspace{-1mm}
\paragraph{Hybrid approach of \FE:}
As described in section~\ref{sec:hybridFE}, the hybrid calculation of
the SM-like Higgs mass implemented in \FE\ is organized in a way
similar to a {pure} EFT calculation. The only difference with respect
to the pure EFT case is that the corrections to the Higgs mass
suppressed by powers of $v^2/\MS^2$ are absorbed in the boundary
condition for the quartic Higgs coupling, via the requirement that a
FO computation of the pole Higgs mass give the same result above and
below the matching scale, see eq.~(\ref{eq:FEFTH}). Accordingly, the
estimate of the theory uncertainty of the hybrid result is organized
in a way similar to the one described in section~\ref{sec:EFTunc} for
the EFT calculation, omitting{,} however{,} the contribution that
stems from missing ${\cal O}( v^2/\MS^2)$ terms. In the latest
implementation of the \FE\ approach, described in
ref.~\cite{Kwasnitza:2020wli}, the ``SUSY uncertainty'' is taken as
the largest of two estimates: the first considers the effect of
varying the matching scale by a factor of $2$ above and below the
central value $\MS$, while the second considers the effects of several
non-logarithmic higher-order terms generated by changing the
definition of the parameters that enter the known part of the boundary
condition. We stress that, in this approach, the estimate of the
``SUSY uncertainty'' probes both the higher-order terms that do not
vanish when $ v^2/\MS^2 \rightarrow 0$ and those suppressed by powers
of $v^2/\MS^2$\,. The ``SM uncertainty'' is in turn taken as the
largest of two estimates, namely the effect of varying the scale at
which the pole Higgs mass is computed by a factor of $2$ above and
below the central value $Q=M_t$, and the effect of including
higher-order terms in the extraction of the top Yukawa coupling from
the top mass. The resulting estimates of the ``SUSY'' and ``SM''
uncertainties are then added linearly in the total uncertainty.
  
\vspace{-1mm}
\paragraph{Third hybrid approach:}
As detailed in section~\ref{sec:hybridFE}, the hybrid approach
proposed in ref.~\cite{Harlander:2019dge} combines a pure EFT
calculation of the SM-like Higgs mass with a separate calculation of
the corrections suppressed by powers of $v^2/\MS^2$\,. In
ref.~\cite{Harlander:2019dge} the hybrid result for the Higgs mass was
compared with the pure EFT result provided by \HS\ and with the FO
(namely, three-loop) result obtained from the combination of \FS\ and
\HIM, showing the expected agreement with one or the other in the
appropriate limits. The proposal of ref.~\cite{Harlander:2019dge} for
the uncertainty estimate of the hybrid result thus consisted in simply
taking, in each point of the parameter space, the lowest uncertainty
estimate between the one of the FO result and the one of the pure EFT
result. These estimates are described in sections~\ref{sec:EFTunc}
and~\ref{sec:FOunc}, respectively.

\subsubsection{Comparing the uncertainties:}

\begin{figure}[t]
  \includegraphics[width=0.52\textwidth]
                  {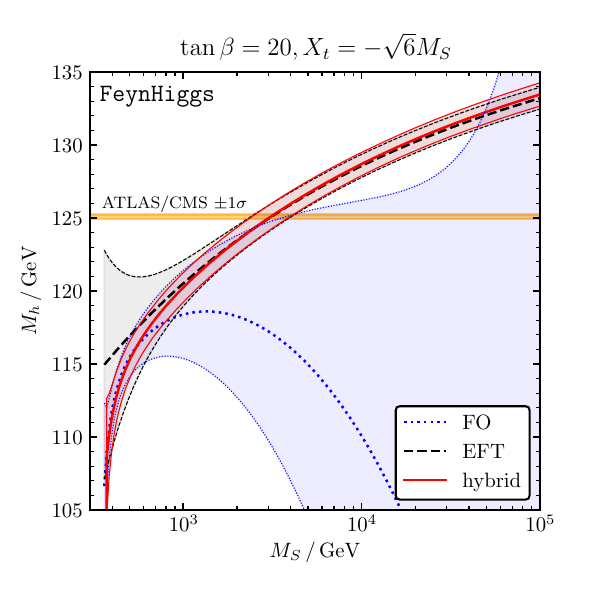}\hfill
                  \includegraphics[width=0.52\textwidth]
                                  {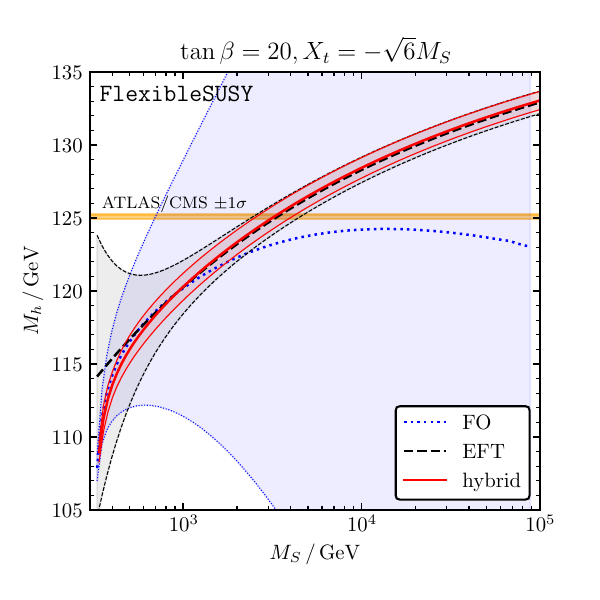}
  \caption{\em Comparison between the pure FO, pure EFT and hybrid
    calculations of the Higgs mass, with the corresponding estimates
    of the theory uncertainty, in an MSSM scenario with degenerate
    SUSY masses. The sfermion mass and mixing parameters are defined
    in the $\drbar$ scheme at the scale $Q=\MS$. The left plot is
    produced with {\rm \FH}, the right one with different modules of
    {\rm \FS.}}
\label{fig:uncertainties}
\end{figure}

To illustrate the estimates described in
sections~\ref{sec:EFTunc}--\ref{sec:hybUnc}, we compare in
figure~\ref{fig:uncertainties} the uncertainties of the hybrid
calculations implemented in \FH\ and in \FE\ with those of their FO
and EFT counterparts, in the same MSSM scenario as the one considered
in section~\ref{sec:hybridplot}. We recall that the parameters that
determine the stop masses and mixing are here defined in the $\drbar$
scheme at the scale $Q=\MS$. In view of the large stop mixing that
characterizes this scenario, the predictions of \FS, \HS\ and
\FE\ omit the three-loop corrections, as they rely on an expansion in
the ratio $X_t/\MS\,$. Consequently, in both the left and right plots
of figure~\ref{fig:uncertainties} the FO calculations include full
one-loop and gaugeless two-loop corrections, and the EFT calculations
provide a full NLL and gaugeless NNLL resummation of the logarithmic
corrections. In each plot, the dotted-blue, dashed-black and full-red
lines represent the predictions for the SM-like Higgs mass of the FO,
EFT and hybrid calculation, respectively, and the shaded bands around
each line represent the uncertainty estimates.

The shapes of the blue-shaded bands in the left and right plots of
figure~\ref{fig:uncertainties} show that, in this scenario, the FO
calculations of \FH\ and \FS\ start losing accuracy as soon as $\MS
\gtrsim 1$~TeV. As mentioned in the description of
figure~\ref{fig:hybrid}, there is a rather dramatic difference between
the FO predictions of the two codes at large $\MS$, due to the
different definitions of the top mass and the strong gauge coupling
entering the radiative corrections.\footnote{{We recall
    that the FO prediction of \FH\ has a milder dependence on $\MS$ in
    scenarios defined in terms of OS parameters for the stop sector
    (the latter is the recommended choice for FO predictions in
    \FH).}}$^,$\footnote{The difference between the blue-dotted curve
  for \FS\ in the right plot of figure~\ref{fig:hybrid} and the one in
  the right plot of figure~\ref{fig:uncertainties} is instead due to
  the omission of the three-loop corrections in the latter.}
~However, the respective uncertainty bands are so large that they
comfortably overlap. This highlights once again the inadequacy of the
FO calculation in scenarios with multi-TeV SUSY masses.

The shape of the grey-shaded bands in figure~\ref{fig:uncertainties}
shows that the EFT calculation is expected to be less accurate at low
$\MS$, where the missing ${\cal O}( v^2/\MS^2)$ effects are most
relevant. However, for values of $\MS$ large enough that the EFT
prediction for the SM-like Higgs mass is compatible with the LHC
measurement, the grey-shaded bands have already shrunk to an
almost-constant width, meaning that the ${\cal O}( v^2/\MS^2)$ effects
are essentially negligible.

Finally, the comparison between the three shaded bands in each plot of
figure~\ref{fig:uncertainties} shows that, for both codes, the
uncertainty estimate of the hybrid calculation in this scenario
{essentially} coincides with the one of the EFT
calculation as soon as $\MS\gtrsim 2$~TeV.
For low values of $\MS$, the uncertainty estimate of the hybrid
calculation of \FH\ {essentially} coincides with the one
of the FO calculation in the same code as soon as $\MS\lesssim
500$~GeV, whereas the uncertainty estimate of the hybrid calculation
of \FE\ remains smaller than the one of the FO calculation of
\FS\ (the latter is obtained following ref.~\cite{Harlander:2019dge},
but taking into account the omission of the three-loop corrections).
In the intermediate range of $\MS$, the uncertainty of the hybrid
calculation is, for both codes, smaller than the uncertainties of both
the EFT and FO calculations, underscoring the fact that the hybrid
calculation combines the advantages of the two approaches, while
avoiding their drawbacks. For $\MS\approx 2\!-\!4$~{TeV}, where the
uncertainty band of the theoretical prediction intersects the one of
the experimental measurement, the theory uncertainty of the hybrid
calculation in both codes is about $\pm1$~GeV {for this
  scenario}.

\subsection{The role of the parametric uncertainties}

\label{sec:parunc}
In addition to the theory uncertainty stemming from uncomputed
higher-order terms, the prediction for the Higgs masses in SUSY models
is subject to a ``parametric'' uncertainty, associated {with} the
experimental {uncertainty} with which the input parameters -- e.g.,
masses and couplings of the particles in the loops -- are known. This
uncertainty can be straightforwardly determined as the maximal shift
in the prediction for the Higgs masses obtained by varying the input
parameters within their experimental ranges. In phenomenological
analyses of SUSY models, it is customary to discuss the two kinds of
uncertainty without combining them, as they come from completely
different and independent sources.

Due to the relatively large size of the top Yukawa coupling, and to
the fact that it enters the top/stop contribution to the one-loop
corrections at the fourth power, the uncertainty of the top-mass
measurement is the one to which the parametric uncertainty of the
Higgs-mass prediction is most sensitive. A well-known ``rule of
thumb''~\cite{Heinemeyer:1999zf} (see also
refs.~\cite{Degrassi:2002fi, Allanach:2004rh}), which holds for MSSM
scenarios with moderate to large $\tan\beta$ and TeV-scale stops,
states that each GeV of variation in $M_t$ results in roughly one GeV
of variation in the prediction for the SM-like Higgs mass. For
example, in 2015 ref.~\cite{Vega:2015fna} showed that, in such
scenarios, a variation in the pole top mass of $1.5$~GeV -- i.e.,
$2\sigma$ according to the first combination of Tevatron and Run-1 LHC
measurements in ref.~\cite{ATLAS:2014wva} -- does indeed induce a
shift of about $1.3$~GeV in the prediction for $M_h$. For low values
of $\tan\beta$, where a larger contribution from the top/stop loops is
needed to obtain a prediction for the Higgs mass around $125$~GeV, the
induced shift can reach $2.5$~GeV. This parametric uncertainty would
have been comparable to, or even larger than, the estimated theory
uncertainty of the EFT and hybrid calculations, at least in simplified
MSSM scenarios with a realistic Higgs-mass prediction (see
e.g.~figure~\ref{fig:uncertainties}). However, the inclusion of Run-2
LHC data already brings the $2\sigma$ uncertainty of the top-mass
measurement down to $0.6$~GeV~\cite{Zyla:2020zbs}.  It should be
noted, on the other hand, that the question of how to relate the mass
parameter that is measured at hadron colliders to a theoretically
well-defined top-quark mass, suitable as input parameter for
higher-order calculations, is still the subject of debate -- see
e.g.~refs.\cite{Hoang:2008xm, Butenschoen:2016lpz, Nason:2017cxd}. The
related uncertainty should be taken into account as an additional
source of parametric uncertainty in the Higgs-mass predictions.
Future measurements at $e^+ e^-$ colliders running at the $t \bar t$
threshold would benefit from the fact that the relation between the
measured mass parameter and a theoretically well-defined
short-distance mass is well understood, and are expected to further
reduce the $2\sigma$ uncertainty to about
$0.1$~GeV~\cite{Fujii:2019zll}.


Until superparticles are discovered, it is of course pointless to
associate a parametric uncertainty to the input values of their masses
and couplings. Even in the felicitous event of a discovery, it is
unlikely that all of the SUSY parameters relevant to the Higgs-mass
prediction will be independently measured in the medium term.
Instead, as we illustrated in section~\ref{sec:general:radcor}, the
mass and couplings of the SM-like Higgs boson will serve as precision
observables to constrain the SUSY parameters that are not directly
accessible by experiment.

\subsection{Prospects}

As long as the dominant classes of higher-order terms are properly
identified and simulated, the margin of improvement for a given
estimate of the theory uncertainty of the Higgs-mass calculation is not
large, because of the unavoidable subjectiveness involved in the
choice of numerical coefficients, in the range of scale variation and
so on. Only the explicit computation of the dominant missing terms can
tell whether their estimate was too optimistic or too
pessimistic. When the accuracy of the Higgs-mass calculation is thus
improved, the existing uncertainty estimate must be adapted so that it
simulates the dominant terms among those that remained uncomputed.

Rather than the development of new, more-refined techniques to
estimate the theory uncertainty of the Higgs-mass calculation, the
most natural direction of development in this domain is likely to be
the application of the existing techniques to models or scenarios for
which an uncertainty estimate is not currently available. For example,
the uncertainty estimates discussed in sections~\ref{sec:EFTunc} and
\ref{sec:hybUnc} for the EFT and hybrid calculations all refer to
scenarios in which there is only one Higgs doublet below the SUSY
scale. It is fair to expect that they will soon be extended to
scenarios with two light Higgs doublets and, beyond the MSSM, to
scenarios in which the EFT features an even more complicated Higgs
sector. {In keeping with the trend towards an
  ``automation'' of the Higgs-mass calculations, it can also be
  expected that, in the medium term, estimates of the theory
  uncertainty similar to those described in this section will also be
  developed for the case of a general renormalizable theory, to be
  implemented in packages like \SA.}

{Finally,} it is worth noting that, in phenomenological analyses of
SUSY models, the spectrum of superparticle masses is generally more
complicated than in the simplified scenarios considered for
illustration in this report (e.g., it might arise as the output of a
spectrum generator).  In {realistic} SUSY scenarios, the question{s}
of which among the available EFTs {(or EFT towers)} better describes a
given mass spectrum {and of what is the optimal choice for the
  matching scale (or scales)} might not be clear-cut. In this case, a
comparison between the theory uncertainties associated {with}
different EFT calculations can be used to {investigate which one} is
most appropriate {for} the considered point of the parameter space.

\vfill
\newpage

\section{Outlook}
\label{sec:outlook}
\vspace*{3mm}

Resolving the underlying dynamics of electroweak symmetry breaking is
one of the main goals of particle physics, and the predictions for the
Higgs masses that characterize the SUSY extensions of the SM play a
crucial role in this context. Indeed, the mass of the Higgs boson
discovered at the LHC can now be considered an electroweak precision
observable: much like, before the LHC era, the $W$ mass and the
$Z$-pole observables provided constraints on the Higgs mass within the
SM, comparing the precisely-measured mass of the observed Higgs boson
with the corresponding theoretical prediction places very sensitive
constraints on the parameter space of the considered SUSY model.

The accuracy of the theoretical predictions for the Higgs masses in
SUSY models has improved very significantly over the years that
followed the Higgs discovery at the LHC. The progress in the
calculations has been discussed in a series of (so far) \kutsnr\ KUTS
meetings, and is summarized in this report. The major lines of
development included:
\begin{itemize}
\renewcommand\labelitemi{--}
\item
The improvement of the FO predictions for the Higgs {masses} in the
simplest SUSY extension of the SM, the MSSM, with new calculations of
two-loop corrections beyond the ``gaugeless'' and vanishing-momentum
approximations, as well as of the dominant three-loop effects;
\item
The precise calculation of the Higgs-mass spectrum in a number of
non-minimal SUSY extensions of the SM, including among others the
NMSSM, models with Dirac gauginos, and models with right-handed
neutrinos;
\item
A renewed focus on the all-order resummation of large logarithmic
corrections even in scenarios with moderately heavy superparticles,
through the development of EFT calculations of the Higgs masses and
their combination with the existing FO calculations in several
``hybrid'' approaches.  This led to the important result that somewhat
larger stop masses than previously thought are needed in the MSSM to
reproduce the observed value of the Higgs mass;
\item
A new effort to assess, for each considered choice of SUSY parameters,
the theory uncertainty of the Higgs-mass prediction that stems from
uncomputed higher-order corrections. This showed that widely used
``one size fits all'' estimates of the uncertainty could be viewed as
too optimistic or too pessimistic, depending on the considered regions
of the parameter space.
\end{itemize}

Despite all of these developments, the quest for high-precision
predictions for the Higgs masses in SUSY models is not by any means
concluded. As described at the end of
sections \ref{sec:fo}--\ref{sec:uncertainties} of this report, efforts
aimed at improving and extending the calculations, and KUTS meetings
to discuss them, are bound to continue. An obvious question in this
context is what should be the target for the precision of the
theoretical predictions. Ideally, to fully exploit the potential of
the Higgs-mass measurement in constraining the parameter space of SUSY
models, one would need to bring the theory uncertainty of the
prediction for the mass {of the SM-like Higgs boson} below the level
of the current (and future) experimental precision. This would however
require a reduction of the theory uncertainty by more than a factor of
$10$ compared to the current level, which is probably too ambitious a
target in the medium term. A more realistic goal is that, in the
coming years, the theory uncertainty of the Higgs-mass prediction be
reduced by a factor of about $2\!-\!3$, down to the level of the
current ``parametric'' uncertainty that stems from the experimental
uncertainty with which the SM input parameters are known. As discussed
in section~\ref{sec:parunc}, the dominant contribution to this
parametric uncertainty comes from the value of the top mass measured
at hadron colliders, whose relation with the theoretically
well-defined {top-mass} parameter that is needed as input for the
Higgs-mass calculation is an additional source of uncertainty. It is
therefore unlikely that further reductions of the theory uncertainty
would bring substantial benefits, at least until {an improved}
measurement of the top mass, e.g.~at future $e^+ e^-$ {colliders},
reduces the associated parametric uncertainty down to the level of the
experimental precision of the Higgs-mass measurement.

\vfill
\newpage

The urgency of further improvements in the accuracy of the Higgs-mass
predictions in SUSY models will also depend on the experimental
developments concerning the properties of the observed Higgs boson,
the electroweak precision observables and the direct searches for BSM
particles. In the MSSM, the minimal values of the stop masses that
lead to a prediction for the Higgs mass compatible with the measured
value lie typically above the current bounds from direct stop searches
at the LHC. Therefore, the scenario of a SM-like Higgs boson with mass
around $125$~GeV and no hints for additional particles from direct BSM
searches can still be considered a fully consistent realization of the
MSSM. If no deviation from the SM is detected in the coming years, a
SUSY model with superparticle masses beyond the kinematic reach of the
LHC {-- or even of future hadron colliders such as the FCC-hh --} will
continue to be a viable possibility (one could invoke fine-tuning
arguments to favor or disfavor certain classes of models). In this
case, the requirement that the prediction for the mass of the SM-like
Higgs boson agree -- within the uncertainties -- with the measured
value will place constraints on the multi-dimensional space of
experimentally inaccessible parameters of the considered model.
{For example, as can be inferred from
figure~\ref{fig:first} in section~\ref{sec:general:radcor}, a lower
bound on the stop masses of ${\cal O}(10~{\rm TeV})$ from future
searches at the FCC-hh would constrain the region of the MSSM
parameter space with large $\tan\beta$, which is consistent with a
possible SUSY explanation of the $(g-2)_{\mu}$ anomaly.}
{We also remark that, in the absence of new discoveries,
the} benefits of any possible improvement in the
calculation { of the Higgs masses} will have to be assessed
on a case-by-case basis. For example, in the simplified MSSM scenario
with a common SUSY scale and a fixed value of $\tan\beta$, the
correlation between $M_S$ and $X_t$ discussed in
section~\ref{sec:general:radcor} (see figure~\ref{fig:second} there)
illustrates the ultimate sensitivity on the unknown SUSY parameters
that could be reached in the idealized situation where experimental
and theory uncertainties are negligible. Even in that idealized
situation, however, all correlations get blurred when more SUSY
parameters are allowed to vary or non-minimal models are considered.

If, on the other hand, {any} significant deviation{s} from the
predictions of the SM {are} detected {(e.g., with definitive
confirmations of the lepton-flavor anomalies recently observed by
LHCb, or of the long-standing $(g-2)_{\mu}$ anomaly)} and/or {any} BSM
particles are discovered, future investigations on the theory side
will obviously focus on the classes of models that can accommodate the
observed phenomenology.  If SUSY models belong to this class, the
techniques and results discussed in this report will be crucial to
obtain accurate predictions for the Higgs masses, and the case for
improving the calculations until the theory uncertainty matches the
experimental accuracy of the Higgs-mass measurement will be even
stronger.  It is also worth stressing that, while these techniques
were developed in the context of SUSY, they can be applied more
generally to any BSM model that involves some kind of prediction for
the quartic scalar couplings. In combination with direct experimental
evidence for BSM physics, the Higgs-mass predictions will be a
powerful tool for unraveling the nature of the new phenomena.

\vfill
\newpage

\section*{Acknowledgements}
The authors of this report wish to thank all of the colleagues who,
directly or indirectly, contributed over the past seven years to the
progress of the Higgs-mass calculations in SUSY models and to the
success of the KUTS initiative.

\bigskip

%
The work of P.S.~and {M.Go.}~has been supported in part by the French
``Agence Nationale de la Recherche'' (ANR), in the context of the
LABEX ILP (ANR-11-IDEX-0004-02, ANR-10-LABX-63) and of the grant
``HiggsAutomator'' (ANR-15-CE31-0002).
The work of S.H.~has been supported in part by the MEINCOP (Spain) under
contract FPA2016-78022-P and under contract PID2019-110058GB-C21, in
part by the Spanish Agencia Estatal de Investigaci\'on (AEI), in part
by the EU Fondo Europeo de Desarrollo Regional (FEDER) through the
project FPA2016-78645-P, in part by the ``Spanish Red Consolider
MultiDark'' FPA2017-90566-REDC, and in part by the AEI through the
grant IFT Centro de Excelencia Severo Ochoa SEV-2016-0597.
%
%
H.B., T.B., J.B., P.D., D.M., I.S.~and G.W.~acknowledge support by the
Deutsche Forschungsgemeinschaft (DFG, German Research Foundation)
under Germany's Excellence Strategy -- EXC 2121 ``Quantum Universe''
-- 390833306.
{The work of H.E.H.~has been partially supported by the U.S.~Department
of Energy Grant~\uppercase{DE-SC}0010107.}
{The work of R.H., M.M., A.V., J.K.~and M.Ga.~has
been supported by the DFG Grant 396021762 -- TRR 257 ``Particle Physics
Phenomenology after the Higgs Discovery''.}
The work of R.H.~and S.P.~has {also} been supported by the
BMBF Grant 05H18PACC2. 
G.L.~acknowledges support by the Samsung Science \& Technology Foundation
under Project Number SSTF-BA1601-07, a Korea University Grant, and the
support of the U.S. National Science Foundation through grant PHY-1719877.
G.L.~is grateful to the Technion--Israel Institute of Technology and the
University of Toronto for partial support during completion of this work.
The work of H.R.~{and N.M.}~was partially funded by the
Danish National Research Foundation, grant number DNRF90. H.R.~is also
supported by the German Federal Ministry for Education and Research
(BMBF) under contract number 05H18VFCA1. Part of this work was
supported by a STSM Grant from COST Action CA16201 PARTICLEFACE.
D.S.~and T.K.~acknowledge support by the DFG grant STO 876/2-2.
C.E.M.W.~is supported in part by the U.S.~Department of Energy under
contracts DEAC02-06CH11357 at Argonne and by the DOE grant
DE-SC0013642 at the University of Chicago.
%
%
The work of B.C.A~has been partially supported by STFC Consolidated
HEP theory grants ST/P000681/1 and ST/T000694/1.
The work of S.B.~has been supported in part by the Research Executive
Agency (REA) of the European Union through the ERC Advanced Grants
MC@NNLO (340983) and MathAm (395568).
The work of T.N.D.~is funded by the Vietnam National Foundation for
Science and Technology Development (NAFOSTED) under grant number
103.01-2020.17.
F.D.~acknowledges support of the DFG grant SFB CRC-110 ``Symmetries
and the Emergence of Structure in QCD''.

\vfill
\newpage

\appendix
\section{Survey of public codes for the Higgs-mass calculation}
\label{sec:codes}
In the body of the report we mentioned several public codes that
provide a precise calculation of the Higgs masses in SUSY models. Most
of these codes compute in addition the full spectrum of SUSY particle
masses, as well as a number of other observables such as, e.g., the
decays widths of the Higgs bosons. They are often denoted as {\em
``spectrum generators''}. Moreover, there exist {\em
``spectrum-generator generators''} that can compute the mass spectrum
of a general renormalizable theory, and produce stand-alone codes
dedicated to specific models.

\newcommand\itemURL{\item[\small\texttt{URL:}]}
\newcommand\itemModel{\item[\small\textit{Model(s):}]}
\newcommand\itemCPV{\item[\small\textit{CPV/FLV/RPV:}]}
\newcommand\itemInputs{\item[\small\textit{Inputs:}]}
\newcommand\itemOutputs{\item[\small\textit{Outputs:}]}
\newcommand\itemSLHA{\item[\small\textit{SLHA\ format:}]}
\newcommand\itemScale{\item[\small\textit{Input scale:}]}
\newcommand\itemExtendability{\item[\small\textit{Extendability:}]}
\newcommand\itemStrategy{\item[\small\textit{Strategy:}]}
\newcommand\itemCorrections{\item[\small\textit{Corrections:}]}
\newcommand\itemOther{\item[\small\textit{Other\ features:}]}
\newcommand\itemCite{\item[\small\textit{What\ to\ cite:}]}
\newcommand\itemLanguage{\item[\small\textit{Language:}]}

In this appendix we provide a survey of public codes for the
Higgs-mass calculation, grouping them~\footnote{ Within
each group, the codes are listed in alphabetic order.} according to
whether their results apply to the MSSM, to some non-minimal SUSY
extension of the SM, or to a general renormalizable theory.  For each
code, we structure the description in the following fields:

\bigskip

\begin{itemize}[leftmargin=8em]

\itemURL The address of the code's web page.

\itemModel The model (or models) for which the code computes the mass
spectrum.

\itemInputs General description of the input parameters required for
the calculation. We specify whether the code accepts input parameters
at some ``high scale'' (e.g., the GUT scale) from which they are
evolved down with appropriate RGEs, or instead it accepts only
``low-scale'' inputs. The latter typically consist of OS parameters or
running parameters at a scale comparable to the SUSY masses. Note that
the codes also require a set of SM input parameters, which we refrain
from listing here.

\itemOutputs
General description of the results of the Higgs-mass calculation.

\itemCPV
Common approximations in the Higgs-mass calculation consist in
neglecting CP violation (CPV), flavor mixing in the sfermion mass
matrices (FLV) and \Rpar\ violation (RPV). We specify whether, and to
what extent, these effects can be included in the calculation.

\itemSLHA We specify whether the code accepts the SLHA
format for the input and/or output parameters. In particular, the
SLHA1 format~\cite{Skands:2003cj} is restricted to the ``vanilla''
MSSM, whereas the SLHA2 format~\cite{Allanach:2008qq} covers also a
number of extensions.

\itemStrategy The strategy adopted in the calculation of the Higgs
masses. In particular, we specify whether the calculation follows a
FO, EFT or hybrid approach.

\itemCorrections Details on the radiative corrections included in the
Higgs-mass calculation. We recall that the ``gaugeless limit''
consists in neglecting all corrections that involve the EW gauge
couplings. The ``MSSM limit'' of the NMSSM corresponds to taking
$v_s\rightarrow\infty$, with both $\lambda v_s$ and $\kappa v_s$ held
fixed.

\itemOther Information on additional outputs of the code beyond
the Higgs-mass calculation.

\itemExtendability When relevant, we mention how the code can be extended to
cover different models, include additional corrections and so on.

\itemLanguage Programming language, user interfaces and other technical
information on the code.

\itemCite A list of references to cite when the results of the code
are included in a publication (sometimes as a series of options
depending on which features are {employed}).

\end{itemize}

\vfill
\newpage

\subsection{MSSM codes}
\label{sec:mssmhmcodes}

The MSSM is the simplest and certainly the most-studied SUSY extension
of the SM. Indeed, some of the codes collected in this section include
corrections to the Higgs masses that are currently not available for
any other BSM model, namely two-loop momentum-dependent corrections
and three-loop corrections. We include in this section also codes that
deal with high-energy extensions of the MSSM (namely, seesaw models
for the generation of the neutrino masses) since the particle content
at the low scale -- and thus the calculation of the Higgs masses -- is
the same as in the MSSM. {Finally, we remark that both of
the ``spectrum-generator generators'' listed in
section~\ref{sec:generic} can be used to produce codes that compute
the Higgs mass spectrum of the MSSM.}

\subsubsection{CPSuperH}
\label{sec:cpsuperh}

\texttt{CPSuperH} performs a FO calculation of the masses and mixing matrices
of the Higgs bosons in the MSSM with complex parameters. It also
computes a number of CPV and flavor observables.

\begin{itemize}[leftmargin=8em]

\itemURL \url{http://www.hep.man.ac.uk/u/jslee/CPsuperH.html}

\itemModel MSSM.

\itemInputs
Soft SUSY-breaking parameters, $\mu$, $\tan\beta$ and $M_{H^\pm}$. The
SUSY-breaking parameters in the stop sector are defined in the \msbar\
scheme, at a scale of the order of the SUSY masses. High-scale
boundary conditions are not supported.

\itemOutputs
Pole Higgs masses and mixing matrix in the Higgs sector.  

\itemCPV
Complex parameters are supported. FLV and RPV are not supported.

\itemSLHA Not supported.

\itemStrategy FO calculation of the Higgs masses and mixing.

\itemCorrections
Full one-loop corrections, leading-logarithmic two-loop effects in the
gaugeless limit.

\itemOther
Higgs couplings and branching ratios.  Muon, electron, thallium,
neutron, mercury, radium, and deuteron electric dipole moments (EDMs).
Flavor observables such as $B\to X_s \gamma$, $B_s\to\mu\mu$,
$B_u\to\tau\nu$, $B_d\to\tau\tau$, CP asymmetry in $B\to X_s\gamma$,
and the SUSY contributions to the $B_d$ and $B_s$ mixings. The
anomalous magnetic dipole moment $(g-2)$ of the muon.

\itemExtendability
A beta version with EFT resummation of the large logarithmic
corrections, allowing to push the SUSY masses to arbitrary values, is
available upon request.

\itemLanguage
\texttt{CPSuperH} is written in Fortran.

\itemCite
Standard: refs.~\cite{Lee:2003nta,Lee:2007gn,Lee:2012wa}. EFT
improvement: ref.~\cite{Carena:2015uoe}.

\end{itemize}

\subsubsection{FeynHiggs}
\label{sec:feynhiggs}

\FH\ performs a hybrid calculation of the pole masses and mixing matrices
of the Higgs bosons in the MSSM with complex parameters, accounting
for the possibility of flavor mixing in the sfermion mass matrices. It
also computes a number of additional Higgs-related observables, as
well as EW observables and EDMs.

\begin{itemize}[leftmargin=8em]

\itemURL
\url{http://www.feynhiggs.de}

\itemModel
MSSM.

\itemInputs
Soft SUSY-breaking parameters, $\mu$, $\tan\beta$ and either $M_A$ or
$M_{H^\pm}$. The SUSY-breaking parameters in the stop sector can be
either OS or \drbar, at a scale of the order of the SUSY masses.
High-scale boundary conditions are not supported.

\itemOutputs
Pole Higgs masses, effective mixing matrix and ``$Z$-factors'' in the
Higgs sector. Uncertainty estimate of the Higgs-mass predictions.

\itemCPV
Complex parameters and flavor-mixing effects are supported. RPV is not
supported.

\itemSLHA
SLHA conventions (SLHA1 and SLHA2) for input and output.

\itemStrategy
FO calculation combined with an EFT calculation (``hybrid''
code{, see section~\ref{sec:hybridFH}}). The EFT
calculation allows for either one or two light Higgs doublets, with
several intermediate steps for the heavier Higgs, gaugino/higgsino and
gluino mass scales.

\itemCorrections
FO calculation: full one-loop corrections, including the effects of
flavor mixing in the sfermion sector. Two-loop corrections in the
vanishing-momentum and gaugeless limits, neglecting flavor mixing as
well as the tau Yukawa coupling. Optionally, two-loop ${\cal
O}(\at\as)$ momentum-dependent effects. EFT calculation: full NLL
resummation of the large logarithmic corrections, NNLL resummation in
the gaugeless limit, and three-loop ${\cal O}(y_t^4\,g_s^4)$ threshold
corrections to $\lambdaSM$ (the latter from \HIM).

\itemOther
Calculation of Higgs widths and branching ratios (SM and MSSM),
approximation for the LHC production cross sections, flavor
observables, muon $(g-2)$, $\Delta\rho$, $M_W$, $\sin^2
{\theta_{\text{w}}}$, and EDMs. {\FH\ can be directly linked to
\HB~\cite{Bechtle:2008jh, Bechtle:2011sb, Bechtle:2013wla,
Bechtle:2020pkv} and \HSi~\cite{Bechtle:2013xfa, Bechtle:2020uwn} to
check for collider constraints on the MSSM Higgs sector via
a \FH-specific compiler option in those codes.}

\itemExtendability
An extension of \FH\ to the NMSSM, based on
ref.~\cite{Drechsel:2016jdg}, is currently under construction.

\itemLanguage
\FH\ is written in Fortran. It has four modes of operation: as a
command-line tool, as a library for use with Fortran or C/C++, as a
Mathematica program, and as a web page
at \url{http://feynhiggs.de/fhucc}.

\itemCite
Standard: refs.~\cite{Heinemeyer:1998yj, Heinemeyer:1998np,
Degrassi:2002fi, Frank:2006yh, Hahn:2013ria, Bahl:2016brp,
Bahl:2017aev, Bahl:2018qog}. Additional references depending on usage:
for the momentum-dependent two-loop corrections,
refs.~\cite{Borowka:2014wla,Borowka:2015ura}; in case of complex
parameters, refs.~{\cite{Heinemeyer:2007aq, Hollik:2014wea,
Hollik:2014bua, Bahl:2020tuq}}; in case of the \thdm\ as low-energy
model, {refs.~\cite{Bahl:2018jom, Bahl:2020jaq,
Bahl:2020mjy}}; in {the} case {of several Higgs bosons close in mass},
ref.~\cite{Bahl:2018ykj}; for the estimate of the theory uncertainty,
ref.~\cite{Bahl:2019hmm}.

\end{itemize}
 
\subsubsection{H3m}
\label{sec:h3m}

\Htm\
computes the mass of the lighter CP-even Higgs boson in the MSSM,
including the three-loop ${\cal O}(\at\as^2)$ corrections from
refs.~\cite{Harlander:2008ju,Kant:2010tf}.

\begin{itemize}[leftmargin=8em]

\itemURL {
\url{https://www.ttp.kit.edu/Progdata/ttp10/ttp10-23}}
  
\itemModel MSSM.

\itemInputs
Soft SUSY-breaking parameters, $\mu$, $\tan\beta$ and $\MA$. The
SUSY-breaking parameters are defined in the \drbar\ scheme, at a scale
$\MS$ of the order of the SUSY masses. High-scale boundary conditions
are not supported.

\itemOutputs
Pole mass of the lighter CP-even Higgs boson.

\itemCPV Not supported.

\itemSLHA Input file in the SLHA1 format.

\itemStrategy
FO calculation of the lighter Higgs mass. The three-loop corrections
are available in the form of asymptotic expansions for a number of
mass hierarchies. The code determines the hierarchy that best fits the
considered point of the MSSM parameter space. The one-and two-loop
corrections are obtained via a call to \FH, and those involving
top/stop loops are adapted to the \drbar\ scheme.

\itemCorrections Full one-loop corrections, two-loop corrections in the
vanishing-momentum and gaugeless limits (neglecting also the tau
Yukawa coupling), three-loop ${\cal O}(\at\as^2)$ corrections from
refs.~\cite{Harlander:2008ju,Kant:2010tf}.

\itemLanguage \Htm\ is a Mathematica program.

\itemCite Refs.~\cite{Harlander:2008ju, Kant:2010tf, Kunz:2014gya}.

\end{itemize}

\subsubsection{Himalaya}
\label{sec:himalaya}

\HIM\
is a library of functions allowing for the inclusion of the three-loop
corrections of refs.~\cite{Harlander:2008ju, Kant:2010tf} in existing
FO or EFT calculations of the lighter CP-even Higgs mass that adopt
the \drbar\ renormalization scheme for the SUSY parameters.

\begin{itemize}[leftmargin=8em]

\itemURL \url{https://github.com/Himalaya-Library/Himalaya}

\itemModel MSSM.

\itemInputs
Soft SUSY-breaking parameters, $\mu$, $\tan\beta$ and $\MA$. The
SUSY-breaking parameters are defined in the \drbar\ scheme, at a scale
$\MS$ of the order of the SUSY masses. High-scale boundary conditions
are not supported. The code also requires the gauge and Yukawa
couplings of the MSSM, as well as the parameter $v$, all defined in
the \drbar\ scheme at the scale $\MS$.

\itemOutputs
One-, two- and three-loop corrections to $\Mh$ for the FO calculation
  in the \drbar\ scheme. One-, two- and three-loop contributions to
  $\Delta\lambda$ for the EFT calculation. Estimates of the
  uncertainty associated with the expansion in mass ratios of the
  three-loop corrections.

\itemCPV Not supported.

\itemSLHA Not supported.

\itemStrategy
FO calculation: the three-loop corrections to $\Mh$ from
refs.~\cite{Harlander:2008ju, Kant:2010tf} are available in the form
of asymptotic expansions for a number of mass hierarchies. The code
determines the hierarchy that best fits the considered point of the
MSSM parameter space. EFT calculation: the three-loop ${\cal O}(y_t^4
g_s^4)$ contribution to $\Delta\lambda$ is extracted from the ${\cal
O}(\at\as^2)$ correction to $\Mh$ as described in
ref.~\cite{Harlander:2018yhj}.

\itemCorrections
FO calculation: full one-loop, gaugeless and vanishing-momentum
two-loop, and ${\cal O}(\at\as^2)$ three-loop contributions to
$\Delta \Mh^2$. EFT calculation: full one-loop, gaugeless two-loop and
${\cal O}(y_t^4 g_s^4)$ three-loop contributions to $\Delta \lambda$.

\itemLanguage
\HIM\ is a library of C++ functions.  A Mathematica interface is provided.

\itemCite Refs.~\cite{Harlander:2017kuc,Harlander:2018yhj} for the code and 
refs.~\cite{Harlander:2008ju,Kant:2010tf} for the original three-loop
calculation.

\end{itemize}

\subsubsection{ISAJET}
\label{sec:isajet}

{\tt ISAJET} is a package that generates $e^+e^-$, $pp$, and $p\bar p$
collider events for a variety of processes, both in the SM and in a
number of BSM models which include the MSSM.  The SUSY mass spectrum
and the Higgs masses are calculated by the modules {\tt ISASUSY} and
{\tt ISASUGRA}.  Also provided is the module {\tt IsaTools}, which
computes several SUSY-related observables.

\begin{itemize}[leftmargin=8em]

\itemURL \url{http://www.nhn.ou.edu/~isajet}

\itemModel MSSM.

\itemInputs
The module {\tt ISASUSY} computes the mass spectrum of the MSSM
starting from Lagrangian parameters at the weak scale. It requires as
input the soft SUSY-breaking parameters, plus $\mu$, $\tan\beta$ and
$\MA$. The module {\tt ISASUGRA} obtains the soft SUSY-breaking
parameters via RG evolution starting from a set of high-scale boundary
conditions. In this case $\mu^2$ and $B_\mu$ are fixed by the minimum
conditions of the Higgs potential. Thirteen variants of high-scale
conditions are available, inspired by the gravity-mediated (SUGRA),
gauge-mediated (GMSB), anomaly-mediated (AMSB) and generalized
mirage-mediated mechanisms for SUSY breaking.

\itemOutputs
Loop-corrected masses and effective couplings of the Higgs bosons.

\itemCPV
Flavor-mixing effects in the RG evolution (but not in the Higgs-mass
calculation) are optionally accounted for by the module {\tt
RGEFLAV}. CPV and RPV are not supported.

\itemSLHA The SLHA1 format is supported for the output.

\itemStrategy
FO calculation of the Higgs masses and mixing in the effective
potential approach, expressing the corrections in terms of the running
couplings of the MSSM.

\itemCorrections
Two-loop RGEs for the MSSM. One-loop top/stop and bottom/sbottom
contributions to the neutral and charged Higgs masses in the
vanishing-momentum limit.

\itemOther
{\tt ISAJET} generates hadron and $e^+e^-$ collider events.  {\tt
ISASUSY} and {\tt ISASUGRA} compute the full spectrum of SUSY masses
and decay rates. {\tt ISASUGRA} also estimates, for a given choice of
high-scale boundary conditions, the degree of fine-tuning required to
maintain the weak scale around $100$~GeV. {\tt IsaTools} computes
various SUSY-related observables, namely: the relic density of
neutralino Dark Matter (DM); the cross sections for direct DM
detection; several $B$-decay rates; the muon $(g - 2)$.

\itemLanguage
{\tt ISAJET} and its various SUSY modules are written in Fortran.

\itemCite Ref.~\cite{Paige:2003mg}.

\end{itemize}

\subsubsection{MhEFT}
\label{sec:mheft}

\ME\
performs an EFT calculation of the CP-even Higgs masses in the
MSSM with heavy SUSY, covering also scenarios in which the mass of the
heavier Higgs doublet lies below the SUSY scale. The code also
computes the decay widths of the lighter CP-even Higgs boson.

\begin{itemize}[leftmargin=8em]

\itemURL \url{https://gabrlee.com/code}

\itemModel MSSM.

\itemInputs
Soft SUSY-breaking parameters, $\mu$, $\tan\beta$ and $\MA$. A common
mass parameter $\MS$ is assumed for all of the sfermions and for the
gluino. All parameters are renormalized in the \msbar\ scheme. The
SUSY parameters are defined at the scale $Q=\MS$, while $\tan\beta$
and $\MA$ are defined at the scale $Q=\MA$.

\itemOutputs
CP-even Higgs masses $\Mh$ and $\MH$. Effective mixing angles $\alpha$
and $\sin(\beta-\alpha)$. The latter are also used to obtain the
couplings of the lighter CP-even Higgs boson to third-generation
fermions (normalized to their SM values) and the $Hhh$ coupling.

\itemCPV Not supported.

\itemSLHA Not supported.

\itemStrategy
EFT calculation of the Higgs masses: SUSY particles decoupled at
$Q=\MS$, leaving the type-II \thdm\ as low-energy model, possibly
augmented with light higgsinos and EW gauginos. Intermediate threshold
for the heavy Higgs doublet at $Q=\MA$. Pole mass of the lighter
CP-even Higgs boson computed at $Q=M_t$.

\itemCorrections
Two-loop RGEs of the type-II \thdm, plus approximate one-loop effects
of higgsinos and gauginos, for the evolution of the couplings between
$Q=\MS$ and $Q=\MA$; three-loop RGEs of the SM below
$Q=\MA$. Threshold corrections $\Delta \lambda_i$ at the scale
$Q=\MS$: one-loop stop, sbottom and stau contributions; two-loop
${\cal O}(y_t^4 g_s^2)$ contributions; approximate inclusion of the
two-loop ${\cal O}(y_t^6)$ contributions in the limit
$\MA=\MS$. One-loop threshold corrections to the Yukawa couplings at
$Q=\MS$. One-loop correction to $\Mh$ computed in the SM at $Q=M_t$.

\itemOther 
Total width of the lighter CP-even Higgs boson. Partial widths,
branching ratios, and ratios to the SM values for the decays of the
lighter CP-even Higgs boson to third-generation fermions.

\itemLanguage \ME\ is a Mathematica program.

\itemCite Refs.~\cite{Draper:2013oza,Lee:2015uza}

\end{itemize}

\subsubsection{SOFTSUSY}
\label{sec:softsusyMSSM}

\SSY\
computes the mass spectrum of Higgs bosons and SUSY particles in the
MSSM and in the NMSSM, accounting for the possibility of \Rpar\
violation and of flavor mixing in the sfermion mass matrices. It accepts
the SUSY parameters at an arbitrary (high or low) input scale. It also
computes the widths and branching ratios for the decays of Higgs
bosons and SUSY particles.

\begin{itemize}[leftmargin=8em]

\itemURL \url{https://softsusy.hepforge.org}

\itemModel MSSM and NMSSM. For the
latter, see section~\ref{sec:softsusyNMSSM}.

\itemInputs
Soft SUSY-breaking parameters in the \drbar\ scheme at a user-defined
scale $\Qin$, spanning from the EW scale to the GUT scale (templates
for SUGRA, GMSB and AMSB boundary conditions are available). The
parameter $\tan\beta$ is given either at $\Qin$ or at $Q=\MZ$.  The
parameters $\mu^2$ and $B_\mu$ are fixed by the minimum conditions of
the Higgs potential. {Alternatively,} $\mu$ and either $B_\mu$ or the
pole mass $\MA$ are given as input, and the soft SUSY-breaking
parameters $m_{H_1}^2$ and $m_{H_2}^2$ are fixed by the minimum
conditions.

\itemOutputs
Pole Higgs masses, effective mixing angle $\alpha$ in the neutral
CP-even sector. Estimate of the theory uncertainty of the prediction
for $\Mh$.

\itemCPV
Flavor mixing in the sfermion mass matrices and \Rpar-violating
couplings are accounted for in the RG evolution of the parameters and
in the determination of the SUSY mass spectrum, but their effects are
not included in the radiative corrections to the Higgs masses. CPV is
not supported.

\itemSLHA
SLHA conventions (SLHA1 and SLHA2) for input and
output. Command-line input is also possible.

\itemStrategy
FO calculation of the Higgs masses. The gauge and Yukawa couplings of
the MSSM are extracted from the SM input parameters at the scale
$Q=\MZ$, while the soft SUSY-breaking parameters are given as input at
$\Qin$. All parameters are evolved with the RGEs of the MSSM to the
scale $\Qewsb$ where the minimum conditions of the potential are
imposed and the whole mass spectrum of the model is computed,
including the radiative corrections. This scale is set by default to
the geometric mean of the stop masses, but can be modified by the
user.

\itemCorrections
Two-loop RGEs for the MSSM, with optional inclusion of three-loop
effects.  The gauge and Yukawa couplings of the MSSM are extracted
from the SM inputs at one loop, with optional inclusion of two-loop
effects.  Full one-loop corrections to the Higgs masses. Two-loop
corrections to the neutral Higgs masses in the vanishing-momentum and
gaugeless limits. Optional inclusion of the three-loop
$\mathcal{O}(\at\as^2)$ corrections to $\Mh$ from \HIM.

\itemOther
Full mass spectrum for the SUSY particles at one loop, with optional
inclusion of two-loop SUSY-QCD corrections. In the presence of \Rpar\
violation, the neutrino masses and mixing are also computed. Decay
widths and branching ratios for SUSY particles and Higgs bosons are
computed in the $R$-parity-conserving case. \SSY\ also estimates, for
a given choice of high-scale boundary conditions, the degree of
fine-tuning required to maintain the weak scale around $100$~GeV.

\itemLanguage \SSY\ is written in C++. 

\itemCite
Standard: ref.~\cite{Allanach:2001kg}, which is the \SSY\ manual for
the \Rpar\ conserving MSSM. Additional references depending on usage:
for the effects of three-loop RGEs and two-loop threshold corrections
to gauge and Yukawa couplings, ref.~\cite{Allanach:2014nba}; for the
two-loop SUSY-QCD corrections to squark and gluino masses,
ref.~\cite{Allanach:2016rxd}; for the decay calculations,
ref.~\cite{Allanach:2017hcf}; for the $R$-parity-violating aspects,
ref.~\cite{Allanach:2009bv}; for the neutrino masses and mixing,
refs.~\cite{Allanach:2009bv, Allanach:2011de}.

\end{itemize}

\subsubsection{SPheno}
\label{sec:sphenoMSSM}
\SP\
computes the mass spectrum of Higgs bosons and SUSY particles in the
MSSM, accounting for the possibility of \Rpar- and
lepton-flavor-violating terms and of flavor mixing in the sfermion
mass matrices. It accepts the input SUSY parameters at an arbitrary
(high or low) scale. It also computes the widths and branching ratios
for the decays of Higgs bosons and SUSY particles, and a number of
flavor-related observables.

\begin{itemize}[leftmargin=8em]

\itemURL \url{https://spheno.hepforge.org}

\itemModel
The MSSM, plus several seesaw extensions with additional states at the
high scale. The Higgs-mass calculation in the seesaw extensions is the
same as in the MSSM.

\itemInputs
Soft SUSY-breaking parameters in the \drbar\ scheme at a user-defined
scale $\Qin$, spanning from the EW scale to the GUT scale (templates
for SUGRA, GMSB and AMSB boundary conditions are available). The
parameter $\tan\beta$ is given either at $\Qin$ or at $Q=\MZ$.  The
parameters $\mu^2$ and $B_\mu$ are fixed by the minimum conditions of
the Higgs potential. {Alternatively,} $\mu$ and either $B_\mu$ or the
pole mass $\MA$ are given as input, and the soft SUSY-breaking
parameters $m_{H_1}^2$ and $m_{H_2}^2$ are fixed by the minimum
conditions.
\itemOutputs
Pole Higgs masses, effective mixing angle $\alpha$ in the neutral
CP-even sector.

\itemCPV

Flavor mixing in the sfermion mass matrices and the subset
of \Rpar-violating couplings that also violate lepton flavor are
accounted for in the RG evolution of the parameters and in the
determination of the SUSY mass spectrum, but their effects are not
fully included in the radiative corrections to the Higgs masses.
Complex parameters are supported, but the mixing between neutral
CP-even and CP-odd Higgs bosons is not implemented.

\itemSLHA SLHA conventions (SLHA1 and SLHA2) for input and
output.

\itemStrategy
FO calculation of the Higgs masses. The gauge and Yukawa couplings of
the SM are extracted from the SM input parameters at the scale
$Q=\MZ$, evolved with the RGEs of the SM to a higher scale $\Qewsb$
and there converted into their MSSM counterparts. {Alternatively,}
they are converted directly at $Q=\MZ$ and then evolved up to $\Qewsb$
with the RGEs of the MSSM.  The soft SUSY-breaking parameters are
given as input at $\Qin$ and evolved with the RGEs of the MSSM to
$\Qewsb$. At this scale, which is set by default to the geometric mean
of the stop masses but can be modified by the user, the minimum
conditions of the potential are imposed and the whole mass spectrum of
the model is computed, including the radiative corrections. In
scenarios with heavy SUSY, a hybrid calculation of $\Mh$ is also
available, providing a LL resummation of the large logarithmic
corrections (see section~\ref{sec:hybridFE}).

\itemCorrections
Two-loop RGEs for the MSSM, three-loop RGEs for the SM. The gauge and
Yukawa couplings of the SM are extracted from the SM inputs at one
loop, but the extraction of the top Yukawa coupling includes also
two-loop QCD effects. Full one-loop corrections to the Higgs masses,
including the effects of flavor mixing in the sfermion
sector. Two-loop corrections to the neutral Higgs masses in the
vanishing-momentum and gaugeless limits, including only the
third-generation Yukawa couplings. In the RPV case the Higgs-mass
calculation includes only the one-loop top/stop and bottom/sbottom
corrections in the effective potential approach.

\itemOther
Full mass spectrum for the SUSY particles at one loop. In the RPV and
seesaw extensions, masses and mixing matrix of the neutrinos are also
provided.  Decay widths and branching ratios for SUSY particles and
Higgs bosons.  A large set of flavor observables (quark- and
lepton-flavor-violating processes), muon $(g-2)$, $\Delta\rho$ and
EDMs.

\itemExtendability
Versions of \SP\ for SUSY models beyond the MSSM can be produced
with \SA, see section~\ref{sec:sarah}.

\itemLanguage \SP\ is written in Fortran 90. 

\itemCite
Standard: refs.~\cite{Porod:2003um,Porod:2011nf}. For the hybrid calculation of
$\Mh$, ref.~\cite{Staub:2017jnp}.

\end{itemize}

\subsubsection{SuSeFLAV}
\label{sec:suseflav}

\SF\ computes the mass spectrum of Higgs bosons and
SUSY particles in the MSSM and in its seesaw extension, starting from
high-scale SUSY parameters. The code also computes a number of
low-energy observables and, in the seesaw extension, lepton-flavor
violating decays.

\begin{itemize}[leftmargin=8em]

\itemURL \url{https://github.com/debtosh/SuSeFLAV}

\itemModel
The MSSM and its seesaw extension with three heavy right-handed
neutrino superfields. The Higgs-mass calculation in the latter is the
same as in the MSSM.

\itemInputs
The soft SUSY-breaking parameters are given as input in the \drbar\
scheme at a high scale, namely the GUT scale for SUGRA boundary
conditions and the messenger scale for GMSB boundary conditions. The
parameter $\tan\beta$ is given at $Q=\MZ$.  The parameters $\mu^2$ and
$B_\mu$ are fixed by the minimum conditions of the Higgs potential.

\itemOutputs
Pole Higgs masses, effective mixing angle $\alpha$ in the neutral
CP-even sector.

\itemCPV
Flavor mixing in the sfermion mass matrices is accounted for in the RG
evolution of the parameters and in the determination of the SUSY mass
spectrum at tree level, but its effects are not included in the
radiative corrections. RPV and CPV are not supported.

\itemSLHA
SLHA2 conventions for input and output.

\itemStrategy
FO calculation of the Higgs masses. The gauge and Yukawa couplings of
the MSSM are extracted from the SM input parameters at the scale
$Q=\MZ$, while the soft SUSY-breaking parameters are given as input at
the high scale. In the seesaw extension, the three right-handed
neutrino superfields are decoupled from the RGEs at their respective
mass scales. All parameters are evolved to the scale $\Qewsb$, which
is fixed as the geometric mean of the stop masses. At this scale the
minimum conditions of the potential are imposed and the whole mass
spectrum of the model is computed, including the radiative
corrections.

\itemCorrections
Two-loop RGEs for the MSSM and for the seesaw extension.  The gauge
and Yukawa couplings of the MSSM are extracted from the SM inputs at
one loop, but the extraction of the top Yukawa coupling includes also
the SM part of the two-loop QCD effects. Full one-loop corrections to
the Higgs masses. Two-loop corrections to the neutral Higgs masses in
the vanishing-momentum and gaugeless limits.

\itemOther
Full mass spectrum for the SUSY particles at one loop. \SF\ also
computes several low-energy observables, namely $\Delta\rho$, the muon
$(g-2)$ and $B\to X_s\gamma$, and it estimates the degree of
fine-tuning required to maintain the weak scale around $100$~GeV. In
the seesaw extension, the code computes also a number of lepton-flavor
violating decays of the muon and the tau.

\itemExtendability
A new version is under development, covering scenarios where the
sfermions of the first two generations are heavy and also computing
the decay rate of the proton.

\itemLanguage \SF\ is written in Fortran 95. 

\itemCite
Ref.~\cite{Chowdhury:2011zr}.

\end{itemize}

\subsubsection{SuSpect}
\label{sec:suspect}

\Su\ computes the mass spectrum of Higgs bosons and SUSY particles in the
MSSM. It accepts the SUSY parameters at an arbitrary (high or low)
input scale.

\begin{itemize}[leftmargin=8em]

\itemURL \url{http://suspect.in2p3.fr}

\itemModel MSSM.

\itemInputs
Soft SUSY-breaking parameters in the \drbar\ scheme at a user-defined
scale $\Qin$, spanning from the EW scale to the GUT scale (templates
for SUGRA, GMSB and AMSB boundary conditions are available). The
parameter $\tan\beta$ is given as input at $Q=\MZ$.  The parameters
$\mu^2$ and $B_\mu$ are fixed by the minimum conditions of the Higgs
potential. {Alternatively,} $\MA$ (either pole or running) and $\mu$
are given as input, and the soft SUSY-breaking parameters $m_{H_1}^2$
and $m_{H_2}^2$ are fixed by the minimum conditions.

\itemOutputs
Pole Higgs masses, effective mixing angle $\alpha$ in the neutral
CP-even sector.

\itemCPV Not supported.

\itemSLHA SLHA1 conventions for input and output.

\itemStrategy
FO calculation of the Higgs masses. The gauge and Yukawa couplings of
the MSSM are extracted from the SM input parameters at the scale
$Q=\MZ$, while the soft SUSY-breaking parameters are given as input at
$\Qin$. All parameters are evolved with the RGEs of the MSSM to the
scale $\Qewsb$, where the minimum conditions of the potential are
imposed and the whole mass spectrum of the model is computed,
including the radiative corrections. This scale is set by default to
the geometric mean of the stop masses, but can be modified by the
user.

\itemCorrections
Two-loop RGEs for the MSSM.  The gauge and Yukawa couplings of the
MSSM are extracted from the SM inputs at one loop, but the extraction
of the top Yukawa coupling includes also the SM part of the two-loop
QCD effects.  Full one-loop corrections to the Higgs masses. Two-loop
corrections to the neutral Higgs masses in the vanishing-momentum and
gaugeless limits.

\itemOther
Full mass spectrum for the SUSY particles at one loop.  Check of
precision observables such as the muon $(g-2)$ and $B\to
X_s\gamma$. Estimate of the degree of fine-tuning required to maintain
the weak scale around $100$~GeV.

\itemLanguage
Version {\tt 2} of \Su\ is written in Fortran, while version {\tt 3},
available since 2014, is written in C++. The two versions are
developed in parallel.

\itemCite Standard: ref.~\cite{Djouadi:2002ze}. For version {\tt 3}, cite also
ref.~\cite{Brooijmans:2012yi}.\footnote{This is a preliminary reference, a
dedicated publication is in preparation.}

\end{itemize}

\subsubsection{SUSYHD}
\label{sec:susyhd}

\SH\
performs an EFT computation of the lighter CP-even Higgs mass of the
MSSM in heavy-SUSY scenarios where the SUSY particles and the heavier
Higgs doublet are decoupled at a common scale. The code also covers
Split-SUSY scenarios where higgsinos and gauginos are decoupled at a
lower scale than the heavy scalars.

\begin{itemize}[leftmargin=8em]

\itemURL \url{http://users.ictp.it/susyhd}

\itemModel MSSM. 

\itemInputs
Soft SUSY-breaking parameters, $\mu$, $\tan\beta$ and $\MA$ in the
$\drbar$ scheme at a user-defined input scale of the order of the SUSY
masses. An OS definition for the parameters in the stop sector is also
envisaged, but it leads to a large logarithmic term in the two-loop
part of $\Delta\lambda$.

\itemOutputs
Pole mass of the lighter CP-even Higgs boson, plus uncertainty estimate.

\itemCPV Not supported.

\itemSLHA Not supported.

\itemStrategy
Pure EFT calculation of $\Mh$ with full NLL resummation of the large
logarithmic corrections. The EFTs valid below the SUSY scale can be
either directly the SM, or Split SUSY followed by the SM. In the
former case, a partial NNLL resummation of the large logarithmic
corrections is also provided.

\itemCorrections
Three-loop RGEs for the SM and two-loop RGEs for Split SUSY,
neglecting the effects of the Yukawa couplings of the first two
generations. The threshold correction to the quartic Higgs coupling
includes full one-loop contributions, two-loop ${\cal O}(y_t^4 g_s^2)$
contributions and two-loop ${\cal O}(y_t^6)$ contributions, the latter
in the limit of degenerate masses for the stops and the heavy Higgs
doublet.  At the EW scale, the running gauge and Yukawa couplings of
the SM and the relation between $\lambdaSM$ and $\Mh$ are determined
at two loops from the interpolation formulas of
ref.~\cite{Buttazzo:2013uya}.

\itemLanguage \SH\ is a Mathematica program.

\itemCite Ref.~\cite{Vega:2015fna}

\end{itemize}

\subsection{Beyond-MSSM codes}
\label{sec:bmssm}

In this section we collect codes for the computation of the Higgs-mass
spectrum in SUSY models that feature an extended particle spectrum
with respect to the MSSM. In particular, three of the listed codes are
devoted to the NMSSM, and one to the $\mu\nu$SSM.  {We remark that
both of the ``spectrum-generator generators'' listed in
section~\ref{sec:generic} can be used to produce codes that compute
the Higgs mass spectrum of the NMSSM, and that \SA\ can also produce a
code for the $\mu\nu$SSM.}

\subsubsection{munuSSM}
\label{sec:munussm}
\vspace*{-1mm}

\munu\ computes the mass spectrum of the neutral scalars of the $\mu\nu$SSM
(namely, Higgs bosons and sneutrinos) with full one-loop
corrections. It relies on \FH\ to include also corrections beyond one
loop in the MSSM limit. {The code also computes the decays of all
non-colored (neutral and charged) scalars.}

\begin{itemize}[leftmargin=8em]

\itemURL \url{https://gitlab.com/thomas.biekoetter/munussm}

\itemModel The $\mu\nu$SSM.

\itemInputs The superpotential and soft-SUSY-breaking parameters,
 the sneutrino vevs and $\tan\beta$. The non-SM input parameters that
determine the tree-level masses of the neutral scalars are defined in
the \drbar\ scheme at a scale of the order of the SUSY masses. The
soft SUSY-breaking parameters in the stop and sbottom sectors can
optionally be defined in the OS scheme. High-scale boundary conditions
are not supported.

\itemOutputs
Pole masses of the eight CP-even and seven CP-odd neutral scalars of
the model. Effective couplings of all neutral scalars to SM particles.

\itemCPV
The Higgs and neutrino couplings of the $\mu\nu$SSM violate by
construction lepton flavor and \Rpar. CPV and flavor violation in the
squark sector are not supported.

\itemSLHA Not supported.

\itemStrategy
FO calculation of the neutral scalar masses in the $\mu\nu$SSM,
supplemented with higher-order corrections in the MSSM limit through
an interface with \FH.

\itemCorrections
Full one-loop corrections to the neutral scalar masses in the
$\mu\nu$SSM. Higher-order corrections included in the MSSM limit:
two-loop corrections in the gaugeless and vanishing-momentum limits,
and the resummation (full NLL, gaugeless NNLL) of the large
logarithmic corrections.

\itemOther
The package computes also the decay widths and branching ratios of all
non-colored (neutral and charged) scalars. An automated interface
to \HB~\cite{Bechtle:2008jh, Bechtle:2011sb, Bechtle:2013wla,
Bechtle:2020pkv} and \HSi~\cite{Bechtle:2013xfa, Bechtle:2020uwn}
tests the considered scenario against the bounds from Higgs searches
at colliders.

\itemLanguage
\munu\ is a Python package, but some routines for numerically involved
calculations are written in Fortran to allow for quadruple
floating-point precision.

\itemCite
Ref.~\cite{Biekotter:2020ehh} is the manual of \munu, and
refs.~\cite{Biekotter:2017xmf,Biekotter:2019gtq} describe the one-loop
calculation of the scalar masses in the $\mu\nu$SSM.

\end{itemize}

\subsubsection{NMSSMCALC}
\label{sec:nmssmcalc}

\NC\
performs a FO calculation of the Higgs masses in the $Z_3$-symmetric
NMSSM, allowing for the possibility of CP violation. The code also
computes the decay widths and branching ratios of the Higgs bosons,
and, in the CPV case, a number of EDMs.

\begin{itemize}[leftmargin=8em]

\itemURL \url{http://www.itp.kit.edu/~maggie/NMSSMCALC}

\itemModel The $Z_3$-conserving NMSSM.

\itemInputs
The input parameters for the Higgs sector are $\tan\beta$, $\lambda$,
$A_\lambda$, $\mu_{\mathrm{eff}}=\lambda v_s$, $\kappa$ and
$A_\kappa$, all defined in the \drbar\ scheme at a user-defined scale
$\Qin$. Optionally, $A_\lambda$ can be replaced by the pole mass
$\MHp$. In the CPV case the input parameters can be complex, but the
imaginary parts of $A_\lambda$ and $A_\kappa$ are fixed by the minimum
conditions of the potential; the phase of the vev $v_2$ must also be
supplied. The remaining input parameters are the soft SUSY-breaking
masses for gauginos and sfermions, and the trilinear Higgs-sfermion
couplings. The parameters in the stop sector can be defined either in
the \drbar\ scheme at the scale $\Qin$ or in the OS scheme. High-scale
boundary conditions are not supported.

\itemOutputs
Pole masses of the Higgs bosons; mixing matrices and ``$Z$-factors''
for the CP-even and CP-odd sectors in the CP-conserving case, or
$5\!\times\!5$ mixing matrix and ``$Z$-factors'' in the CPV case.

\itemCPV Complex parameters are supported. FLV and RPV are not supported.

\itemSLHA SLHA2 conventions for input and output.

\itemStrategy FO calculation of the Higgs masses and mixing.

\itemCorrections Full one-loop corrections to the Higgs masses. Two-loop  
$\mathcal{O}(\at\as)$ corrections to the Higgs masses in the
vanishing-momentum limit, plus two-loop $\mathcal{O}(\at^2)$
corrections in the vanishing-momentum and MSSM limits.

\itemOther Tree-level mass spectrum for the SUSY particles. 
Decay widths and branching ratios of the Higgs bosons into all
possible two-fermion final states, off-shell decays into heavy quarks
and gauge bosons. The two-body decays include all of the available QCD
corrections, and those to down-type fermions include also SUSY effects
through effective couplings.  Decay width and branching ratios of the
top quark (relevant in scenarios with light Higgs bosons). In the CPV
case, EDMs of electron, neutron, mercury and thallium. The code
produces also a table of Higgs couplings that can be passed
to \HB~\cite{Bechtle:2008jh, Bechtle:2011sb, Bechtle:2013wla,
Bechtle:2020pkv} and \HSi~\cite{Bechtle:2013xfa, Bechtle:2020uwn} to
test the considered scenario against the bounds from Higgs searches at
colliders.

\itemExtendability
An extension of the code featuring the full one-loop calculation of the
two-body decays is available at the {\tt URL}
\url{http://www.itp.kit.edu/~maggie/NMSSMCALCEW}\,.

\itemLanguage \NC\ is written in Fortran.

\itemCite
Standard: the code manual, ref.~\cite{Baglio:2013iia}, plus
refs.~\cite{Ender:2011qh,Graf:2012hh} for the one-loop corrections to
the Higgs masses and refs.~\cite{Muhlleitner:2014vsa,Dao:2019qaz} for
the two-loop corrections. Additional references depending on usage:
for the decay calculation in \NC,
refs.~\cite{Djouadi:1997yw,Djouadi:2018xqq}; for the decay calculation
in {\tt NMSSMCALCEW}, also refs.~\cite{Baglio:2019nlc, Dao:2020dfb};
for the EDMs, ref.~\cite{King:2015oxa}.

\end{itemize}

\subsubsection{NMSSMTools}
\label{sec:nmssmtools}

\NT\
is a suite of codes for calculating the mass spectrum and the decays
of Higgs bosons and SUSY particles in the general NMSSM, with input
parameters given either at the SUSY scale or at a high (namely, GUT or
messenger) scale.

\begin{itemize}[leftmargin=8em]

\itemURL \url{https://www.lupm.univ-montp2.fr/users/nmssm}

\itemModel General NMSSM, with or without $Z_3$ symmetry.

\itemInputs SUSY-scale mode: the (possibly complex) superpotential and
soft SUSY-breaking parameters, as well as $\tan\beta$ and $v_s$, are
given as input in the \drbar\ scheme at a user-defined scale $\Qin$ of
the order of the SUSY masses. Exceptions are the soft SUSY-breaking
masses of the Higgs doublets and of the singlet, which are determined
by the minimum condition of the Higgs potential. High-scale mode: the
parameters are given as input at the GUT scale or (for GMSB) at the
messenger scale. In this case the three parameters that are chosen to
be fixed by the minimum conditions of the potential depend on the
considered SUSY-breaking mechanism at the high scale.

\itemOutputs
Pole masses of the Higgs bosons; mixing matrices for the CP-even and
CP-odd sectors in the CP-conserving case, or a single $5\!\times\!5$
mixing matrix in the CPV case.

\itemCPV
Complex parameters are supported. FLV and RPV are not supported.

\itemSLHA  SLHA2 conventions for input and output.

\itemStrategy

FO calculation of the Higgs masses and mixing. The EW gauge couplings
and $v$ are extracted from the SM inputs directly at the SUSY scale,
whereas the Yukawa and strong-gauge couplings are evolved up from the
EW scale.

\itemCorrections
The two-loop RGEs of the NMSSM are used in the case of high-scale
inputs. The extraction of the \drbar\ parameters $(g, g^\prime, v)$ is
performed at one loop. The extraction of the top and bottom Yukawa
couplings includes also two-loop QCD corrections.  Full one-loop
corrections to the Higgs masses, including the effect of
$Z_3$-violating couplings.  In the CP-conserving case, two-loop ${\cal
O}(\alpha_t\alpha_s + \alpha_b\alpha_s)$ corrections to the neutral
Higgs masses in the vanishing-momentum limit. The two-loop corrections
that involve only the third-family Yukawa couplings are included in the
vanishing-momentum and MSSM limits. In the CPV case the two-loop
corrections are included only at the leading-logarithmic order.

\itemOther
Tree-level mass spectrum for the SUSY particles. Decay widths and
branching ratios for the Higgs bosons as well as the SUSY particles.
The package includes modules to perform automatic scans of the
parameter space, and to check whether a given point is constrained by
measured Higgs couplings, collider searches for Higgs bosons and for
SUSY particles, $K$- and $B$-physics observables, muon $(g-2)$, and
EDMs (in the CPV case).  It can also be linked to {\tt
micrOMEGAs}~\cite{Belanger:2001fz, Belanger:2006is, Belanger:2018mqt},
which was adapted to the NMSSM in ref.~\cite{Belanger:2005kh}, to
compute the relic density and the direct-detection cross sections of
Dark Matter.

\itemLanguage \NT\ is written in Fortran.

\itemCite
Standard: refs.~\cite{Ellwanger:2004xm, Ellwanger:2005dv}.  Additional
references depending on usage: in the CPV case,
ref.~\cite{Domingo:2015qaa}; for scenarios with GUT-scale inputs,
ref.~\cite{Ellwanger:2006rn}; for SUSY particle decays,
ref.~\cite{Das:2011dg}, which was based on
ref.~\cite{Muhlleitner:2003vg}; for Dark Matter,
ref.~\cite{Belanger:2005kh}.

\end{itemize}

\subsubsection{SOFTSUSY}
\label{sec:softsusyNMSSM}
\SSY, whose general features are described in
section~\ref{sec:softsusyMSSM}, also allows for the computation of the
mass spectrum in the general NMSSM. We describe here the features that
are specific to the NMSSM calculation.

\begin{itemize}[leftmargin=8em]

\itemURL \url{https://softsusy.hepforge.org}

\itemModel General NMSSM, with or without $Z_3$ symmetry.

\itemInputs
In addition to the parameters that are in common with the MSSM, \SSY\
requires as input at the scale $\Qin$ the superpotential and soft
SUSY-breaking parameters that involve the singlet. However, three
among the Lagrangian parameters are fixed by the minimum conditions of
the Higgs potential at the scale $\Qewsb$ (e.g., in the
$Z_3$-conserving NMSSM these are chosen as $v_s$, $\kappa$ and
$m_S^2$).

\itemOutputs
Pole Higgs masses, effective $3\!\times\!3$ mixing matrix in the
CP-even sector and effective mixing angle in the CP-odd sector.

\itemCPV
Not supported.

\itemSLHA
SLHA2 conventions for input and output. Command-line input is also
possible.

\itemStrategy
Same as for the MSSM calculation, see section~\ref{sec:softsusyMSSM}.

\itemCorrections
Two-loop RGEs for the NMSSM.  The gauge and Yukawa couplings are
extracted from the SM inputs at one loop, with optional inclusion of
two-loop effects in the MSSM limit.  The EW parameter $v$ is extracted
at one loop. Full one-loop corrections to the Higgs masses, including
the effect of $Z_3$-violating couplings.  Two-loop ${\cal
O}(\alpha_t\alpha_s + \alpha_b\alpha_s)$ corrections to the neutral
Higgs masses in the vanishing-momentum limit. The two-loop corrections
that involve only the third-family Yukawa couplings are included
{in} the vanishing-momentum and MSSM limits.

\itemOther 
Full mass spectrum for the SUSY particles at one loop, with optional
inclusion of two-loop SUSY-QCD corrections.  Decay widths and
branching ratios for SUSY particles and Higgs bosons are also
computed.

\itemCite
Refs.~\cite{Allanach:2001kg, Allanach:2013kza}. 

\end{itemize}

\subsection{Generic codes}
\label{sec:generic}

In this section we describe codes that aim to work for \emph{any}
model, implementing general quantum field theory calculations and
adapting them to the model under consideration.  These are known as
``metacodes'' or ``spectrum-generator generators'', because they
produce as output a stand-alone code for the computation of the mass
spectrum of the model that was defined as input by the user. For each
metacode, we describe here the Higgs-mass calculation implemented in
the generated codes.

\subsubsection{FlexibleSUSY}
\label{sec:flexiblesusy}

\FS\
is a package that uses the analytic results for RGEs, tadpoles and
self-energies provided by \SA, as well as numerical routines
from \SSY, to generate stand-alone spectrum generators for any SUSY
(or non-SUSY) model defined by the user. When available, higher-order
corrections beyond those provided by \SA\ are included for specific
models. In heavy-SUSY scenarios where the EFT valid below the SUSY
scale is the SM, the package also provides a hybrid calculation of the
SM-like Higgs mass following the \FE\ approach (see
section~\ref{sec:hybridFE}). The module {\tt FlexibleBSM} generates
stand-alone codes for the pure EFT calculation of the Higgs masses,
with user-supplied boundary conditions.  Finally, the package contains
pre-generated spectrum generators for several SUSY models, namely
MSSM, NMSSM, E$_6$SSM and MRSSM. Below we list the general features of
the spectrum generators produced by \FS, focusing on the case of SUSY
models.

\begin{itemize}[leftmargin=8em]

\itemURL \url{https://flexiblesusy.hepforge.org}

\itemModel
Any renormalizable extension of the SM whose Lagrangian does not
contain 3- or 4-tensor interactions.

\itemInputs
Superpotential parameters and soft SUSY-breaking parameters in
the \drbar\ scheme at a user-defined scale $\Qin$, spanning from the
EW scale to the GUT scale. The parameter $\tan\beta$ and all scalar
vevs other than $v$ are also required as input at the scale
$\Qin$. For each of the scalars that acquire a vev, one of the
parameters that contribute to the mass is fixed through the minimum
conditions of the scalar potential.

\itemOutputs
Pole Higgs masses, effective mixing matrices in the Higgs sector.
Estimate of the theory uncertainty of the prediction for $\Mh$.

\itemCPV CPV and FLV are supported. RPV is not supported.

\itemSLHA SLHA conventions (SLHA1 and SLHA2) for input and output.

\itemStrategy
FO calculation: The gauge and Yukawa couplings of the SUSY model are
extracted from the SM input parameters at a user-defined low-energy
scale $\QSM$, while the soft SUSY-breaking parameters are given as
input at $\Qin$. All parameters are evolved with the RGEs of the SUSY
model to the scale $\Qewsb$, where the minimum conditions of the
potential are imposed and the whole mass spectrum of the model is
computed, including the radiative corrections. This scale is set by
default to the geometric mean of the stop masses, but can be modified
by the user.

Hybrid calculation: Only for scenarios where the EFT valid below the
SUSY scale is the SM.  The FO calculation of the Higgs masses is
performed at a scale $Q=\MS$ of the order of the SUSY masses, starting
from an ansatz for the gauge and Yukawa couplings and the parameter
$v$ of the SUSY model, which we collectively denote as $P^i(\MS)$. A
matching condition for $\lambdaSM(\MS)$ is thus extracted from the
results of the FO calculation as described in
section~\ref{sec:hybridFE}, and $P^i(\MS)$ are converted into their SM
counterparts $P^i_\smallSM(\MS)$. All parameters are then evolved down
to a user-defined scale $\QSM$, where the pole Higgs mass is computed
within the SM. The ansatz for $P^i(\MS)$ is adjusted until the values
of $P^i_\smallSM(\QSM)$ obtained via RG evolution coincide with those
extracted from a set of experimental observables.

EFT calculation: Standard multi-scale approach. The matching
conditions for the couplings of the EFTs at each of the scales where
some heavy particles are integrated out must be supplied by the user.

\itemCorrections

FO calculation: Two-loop RGEs from \SA\ for any model. For the MSSM,
optional inclusion of the three-loop RGEs from
refs.~\cite{Jack:2003sx, Jack:2004ch}. The gauge and Yukawa couplings
and the parameter $v$ of the SUSY model are extracted from the SM
inputs at one loop, but the extraction of the top-Yukawa and
strong-gauge couplings includes also the SM part of the two- and
three-loop QCD effects (for the MSSM, the two-loop SUSY-QCD effects
can also be included). Full one-loop corrections to the Higgs masses
from \SA\ for any model. For the MSSM, two-loop corrections to the
neutral Higgs masses in the gaugeless and vanishing-momentum limits,
and optional inclusion of the three-loop $\mathcal{O}(\at\as^2)$
corrections to $\Mh$ from \HIM. For the NMSSM, two-loop ${\cal
O}(\alpha_t\alpha_s + \alpha_b\alpha_s)$ corrections to the neutral
Higgs masses in the vanishing-momentum limit, remaining gaugeless
two-loop corrections in the vanishing-momentum and MSSM limits.

Hybrid calculation: For any model, the one-loop corrections to all of
the Higgs masses are included as in the FO calculation, and the large
logarithmic corrections to $\Mh$ are resummed at the NLL order. For
the MSSM, where the FO calculation includes also two- and three-loop
corrections, the resummation is extended to the NNLL order in the
gaugeless limit, and it optionally includes also the N$^3$LL terms
that involve the highest powers of the strong gauge coupling.

EFT calculation: For any model, \FS\ requires as input the one-loop
matching conditions, and generates a stand-alone code for the EFT
calculation of the Higgs masses at the NLL order. Pre-generated codes
for various MSSM scenarios with hierarchical mass spectra are included
in the package (see section~\ref{sec:EFTcodes}). In the simplest MSSM
scenario where the EFT valid below the SUSY scale is the SM, the
resummation of the large logarithmic corrections to $\Mh$ is performed
at the same order as in the hybrid calculation described above. For
the other MSSM scenarios the resummation is performed at the NLL
order, but the known two-loop contributions to the matching conditions
for the quartic Higgs coupling(s) are included.

\itemOther
Mass spectrum and mixing matrices at one loop for all BSM scalars and
fermions. Effective Higgs couplings to photons and to
gluons. Predictions for $\MW$ and {the running weak mixing
angle}, the muon $(g-2)$ and the EDMs of quarks and leptons. For the
models with high-scale boundary conditions, the package includes a
semi-analytic solver of the boundary-value problem (BVP). This allows
for the study of models that are highly constrained (such as the
CNMSSM or the CE$_6$SSM) or in which the BVP has multiple solutions.

\itemLanguage
\FS\
is a Mathematica package. The spectrum generators produced by \FS\ are
written in C++.  A Mathematica interface for the spectrum generators
is also provided.

\itemCite
Standard: refs.~\cite{Athron:2014yba, Athron:2017fvs}, mentioning that
the package includes numerical routines
from \SSY~\cite{Allanach:2001kg,Allanach:2013kza} and analytic results
from \SA~\cite{Staub:2008uz, Staub:2009bi, Staub:2010jh, Staub:2012pb,
Staub:2013tta}. Additional references depending on usage: for the
hybrid calculation of \FE, ref.~\cite{Athron:2016fuq}{, as
well as ref.~\cite{Kwasnitza:2020wli} in the case of the MSSM}; for
the three-loop corrections to the MSSM Higgs mass from \HIM,
refs.~\cite{Harlander:2008ju, Kant:2010tf, Harlander:2017kuc,
Harlander:2018yhj}; for the semi-analytic solver of the BVP,
ref.~\cite{Athron:2009bs}.

\end{itemize}

\subsubsection{SARAH}
\label{sec:sarah}

\SA\ is a multipurpose tool for any renormalizable model.
To define a model, the user specifies the superfield content, the
superpotential (or, for non-SUSY models, directly the field content
and the Lagrangian), and the way gauge symmetries are broken and
fields mix.  \SA\ then determines all mass matrices and interaction
vertices, and uses them to adapt to the model under consideration the
general formulas for the two-loop RGEs, for the one- and two-loop
tadpoles and self-energies of the scalars, and for the one-loop
self-energies of fermions and vector bosons. All of these results are
given as output in analytical form and can be used by other codes such
as, e.g., \FS\ (see section~\ref{sec:flexiblesusy}).  Most relevant
here is the ability to produce stand-alone spectrum generators,
similar to \SP\ (see section~\ref{sec:sphenoMSSM}) and relying on that
code's library of routines. The Higgs-mass calculation can
alternatively adopt the FO, EFT and hybrid approaches.  Below we list
the features of these spectrum generators, focusing on the case of
SUSY models.

\begin{itemize}[leftmargin=8em]

\itemURL \url{https://sarah.hepforge.org}

\itemModel
Any renormalizable extension of the SM, modulo some restrictions for
certain features.  A library of pre-defined models, both SUSY and
non-SUSY, is provided in the package.

\itemInputs
Superpotential parameters and soft SUSY-breaking parameters in
the \drbar\ scheme at a user-defined scale $\Qin$, spanning from the
EW scale to the GUT scale. The parameter $\tan\beta$ and all scalar
vevs other than $v$ are also required as input at the scale
$\Qin$. For each of the scalars that acquire a vev, one of the
parameters that contribute to the mass is fixed through the minimum
conditions of the scalar potential.

\itemOutputs
Pole Higgs masses, effective mixing matrices in the Higgs sector.

\itemCPV All supported.

\itemSLHA SLHA conventions (SLHA1 and SLHA2) for input and output.

\itemStrategy 
FO calculation: The gauge and Yukawa couplings and the parameter $v$
of the SM are determined at the scale $Q=\MZ$, then evolved to a
user-defined scale $\Qewsb$ where they are converted into their SUSY
counterparts. {Alternatively,} the gauge and Yukawa couplings and the
parameter $v$ of the SUSY model are extracted from the SM inputs
directly at the scale $\Qewsb$. The remaining parameters of the SUSY
model are in turn evolved from $\Qin$ to $\Qewsb$. At this scale,
which is set by default to the geometric mean of the stop masses but
can be modified by the user, the minimum conditions of the scalar
potential are imposed and the whole mass spectrum is computed,
including the radiative corrections.

Hybrid calculation: Only for scenarios where the EFT valid below the
SUSY scale is the SM. The gauge and Yukawa couplings and the parameter
$v$ of the SM are extracted from a set of experimental observables at
the scale $Q=\MZ$, then evolved to a scale $Q=\MS$ of the order of the
SUSY masses, where they are converted into their SUSY counterparts and
used for the FO calculation of the Higgs masses. A matching condition
for $\lambdaSM(\MS)$ is thus extracted from the results of the FO
calculation as described in section~\ref{sec:hybridFE}, and it is used
for an EFT calculation of $\Mh$.

Finally, the code allows for pure EFT calculations of the Higgs masses
in which the matching conditions at each of the scales where some
heavy particles are integrated out are obtained from the general
analytic results of refs.~\cite{Braathen:2018htl, Gabelmann:2018axh}.

\itemCorrections
FO calculation: Two-loop RGEs for any model.  The gauge and Yukawa
couplings and the parameter $v$ of the SUSY model are extracted from
the SM inputs at one loop. Full one-loop corrections to the Higgs
masses. Two-loop corrections to the neutral Higgs masses in the
gaugeless and vanishing-momentum limits. For models beyond the MSSM,
possible singularities in the two-loop corrections associated with the
``Goldstone Boson Catastrophe'' -- see section~\ref{sec:nmssmfixed} --
are addressed as described in refs.~\cite{Braathen:2016cqe,
Braathen:2017izn}, requiring the partial inclusion of
momentum-dependent effects.

Hybrid calculation: The one- and two-loop corrections to all of the
Higgs masses are included as in the FO calculation, and the large
logarithmic corrections to $\Mh$ are resummed at the LL order.

EFT calculation: The general formulas for two-loop RGEs and one-loop
matching conditions allow for the NLL resummation of the large
logarithmic corrections to the Higgs masses in any SUSY model with a
hierarchical mass spectrum. The package contains pre-defined model
files for several hierarchical scenarios in the MSSM and in the NMSSM
(see section~\ref{sec:EFTcodes}).

\itemOther
Mass spectrum and decay widths at one loop for all BSM scalars and
fermions. Unitarity constraints on the scalar couplings.  Model files
for Monte Carlo tools (\texttt{UFO}~\cite{Degrande:2011ua},
\texttt{WHIZARD}~\cite{Kilian:2007gr},
\texttt{CalcHEP}~\cite{Pukhov:2004ca, Belyaev:2012qa}).
Model files for other tools:
\texttt{Vevacious} \cite{Camargo-Molina:2013qva},
\texttt{FeynArts}~\cite{Hahn:2000kx}.
Flavor observables through
\texttt{FlavorKit}~\cite{Vicente:2014xda}. \LaTeX\ output.

\itemLanguage \SA\
is a Mathematica package. The \SP-like spectrum generators produced
by \SA\ are written in Fortran 90.

\itemCite 
Refs.~\cite{Staub:2008uz, Staub:2009bi, Staub:2010jh, Staub:2012pb,
  Staub:2013tta} for the core functions, refs.~\cite{Goodsell:2014bna,
  Goodsell:2015ira, Braathen:2017izn} for the two-loop corrections to
  the scalar masses.  Additional references depending on usage: for
  the hybrid calculation, ref.~\cite{Staub:2017jnp}; for the automated
  EFT calculation, ref.~\cite{Gabelmann:2018axh}; for the decay 
  widths, ref.\cite{Goodsell:2017pdq}; for the unitarity constraints,
  {refs.~\cite{Goodsell:2018tti,Goodsell:2020rfu}} .

\end{itemize}

\vfill
\newpage

\bibliographystyle{utphys}
\bibliography{kuts}

\end{document}